\documentclass[prb,twocolumn,floatfix]{revtex4}

\usepackage{verbatim}   

\usepackage{amsmath}
\usepackage[dvips]{graphicx}
\usepackage{subfigure}
\usepackage{latexsym}
\usepackage{bm}
\usepackage{wasysym}
\usepackage{amssymb}
\usepackage{longtable}

\def\etal{~\textit{et~al.}} 	
\def\ra{\rangle} 			
\def\la{\langle} 			
\def\z2{$\mathbb{Z}_2$}
\def\dd{\textrm{d}}
\def\bh{\boldsymbol{\chi}}
\def\bx{\boldsymbol{\xi}}
\def\bm{\boldsymbol{\mu}}
\def\bC{\boldsymbol{C}}
\def\br{\boldsymbol{r}}

\def\mD{\mathcal{D}}

\def\bkp{\boldsymbol{\kappa}}
\def\rv{\boldsymbol{r}}
\def\drv{\Delta \boldsymbol{r}}

\newcommand{\ket}[1]{
\begin{tabular}{@{} r @{} c @{} l @{}}
$\Bigl|$ & #1 & $\Bigr\ra$
\end{tabular}}

\newcommand{\bra}[1]{
\begin{tabular}{@{} r @{} c @{} l @{}}
$\Bigl\la$ & #1 & $\Bigr|$
\end{tabular}}

\newcommand{\bond}[1]{\la #1 \ra}

\newcommand{\plaq}[2]{
\includegraphics[width=#2cm,height=#2cm,viewport=0 140 300 -100]{#1}}
\newcommand{\DPL}{
\includegraphics[width=0.3cm]{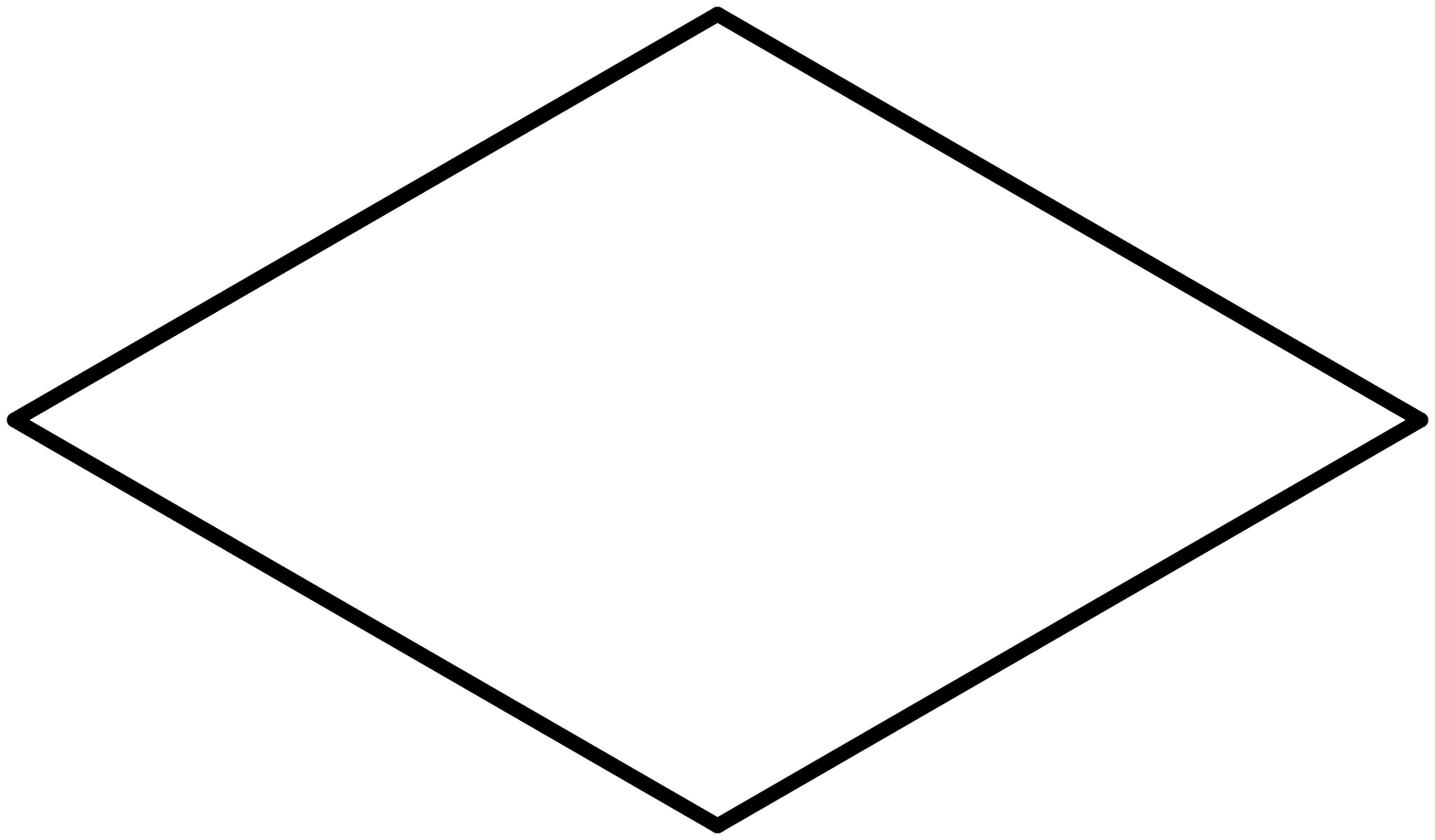}}
\newcommand{\CDPL}{
\includegraphics[width=0.4cm]{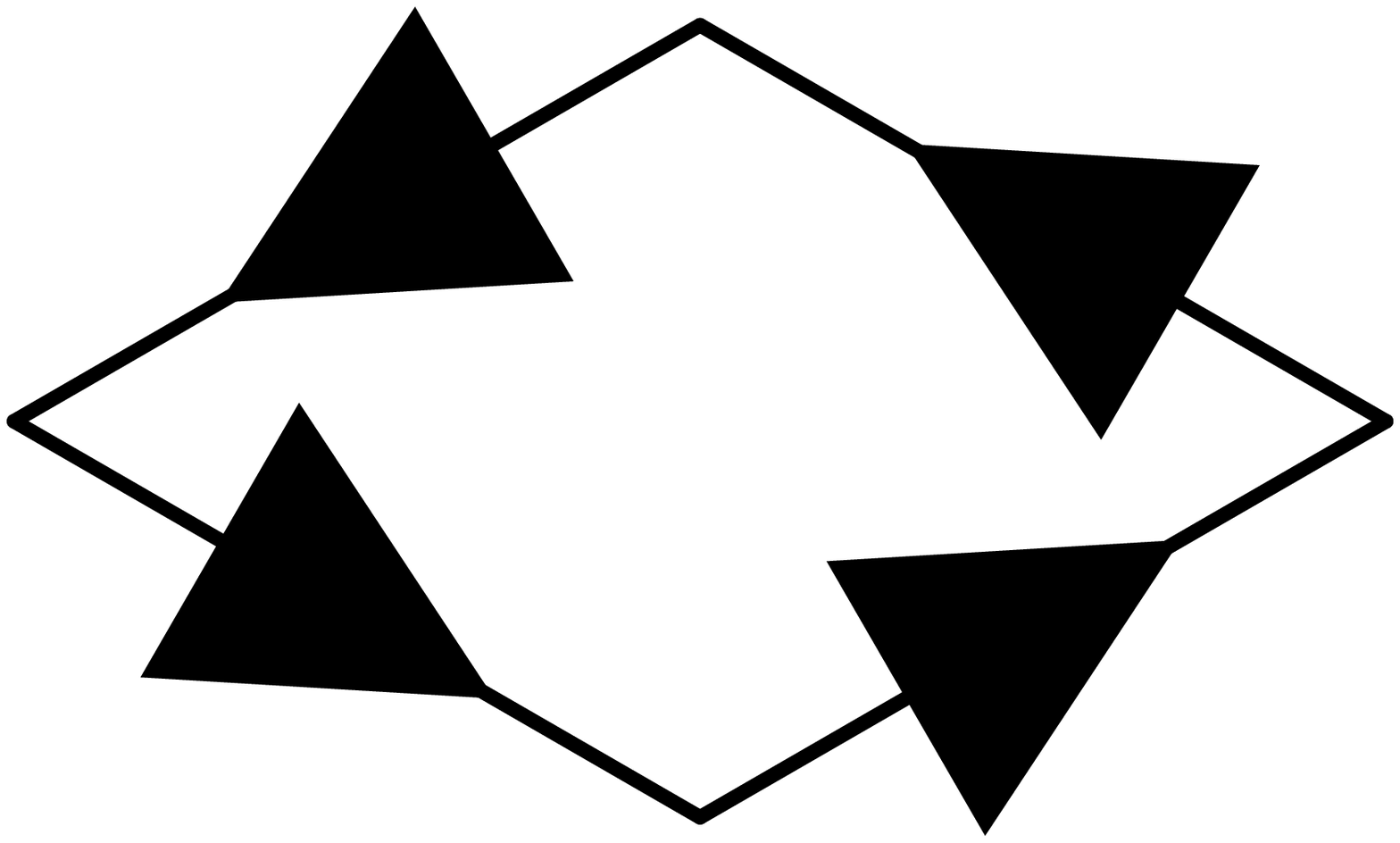}}
\newcommand{\CTRI}{
\includegraphics[width=0.3cm]{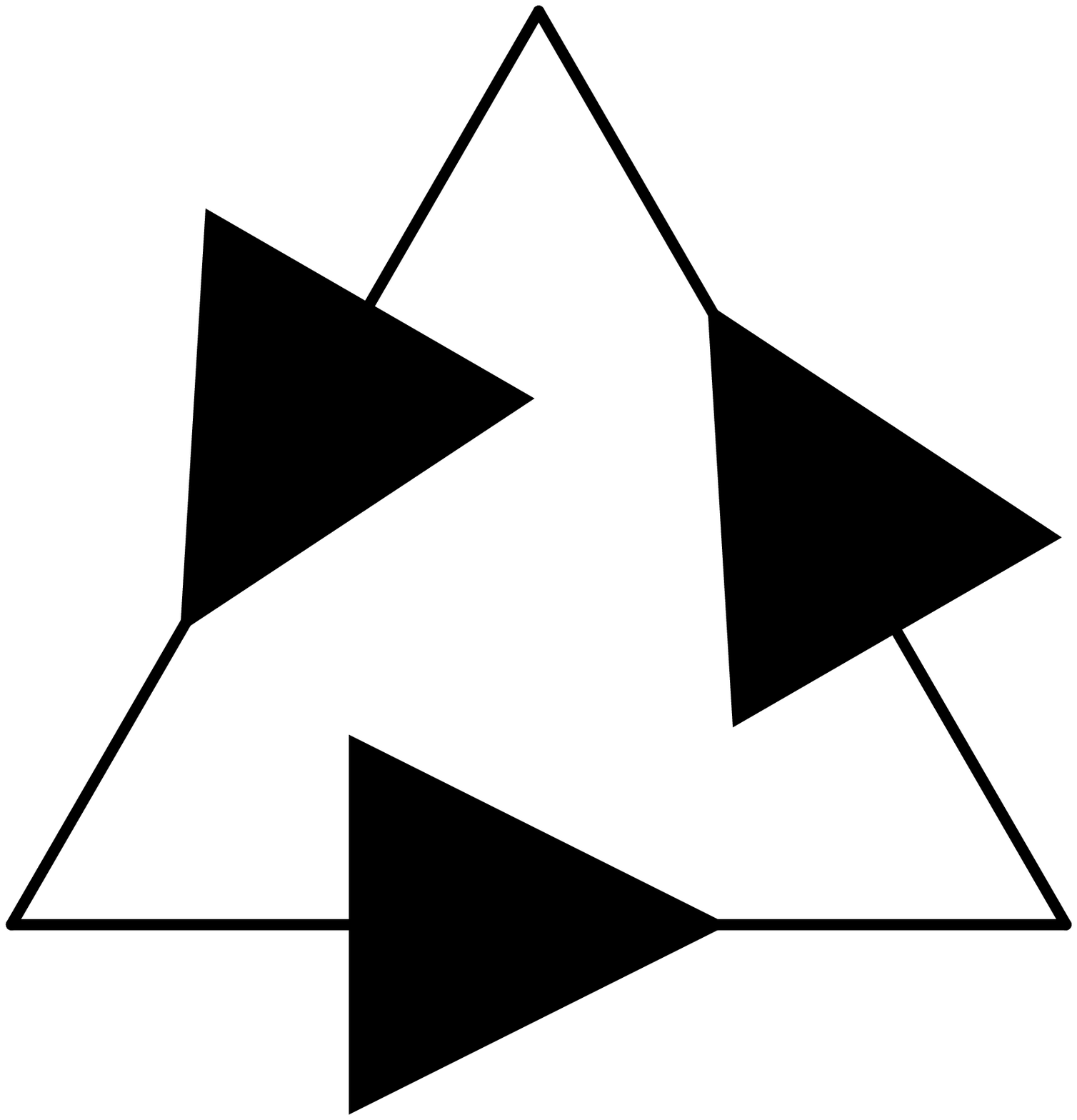}}
\newcommand{\CHEX}{
\includegraphics[width=0.4cm]{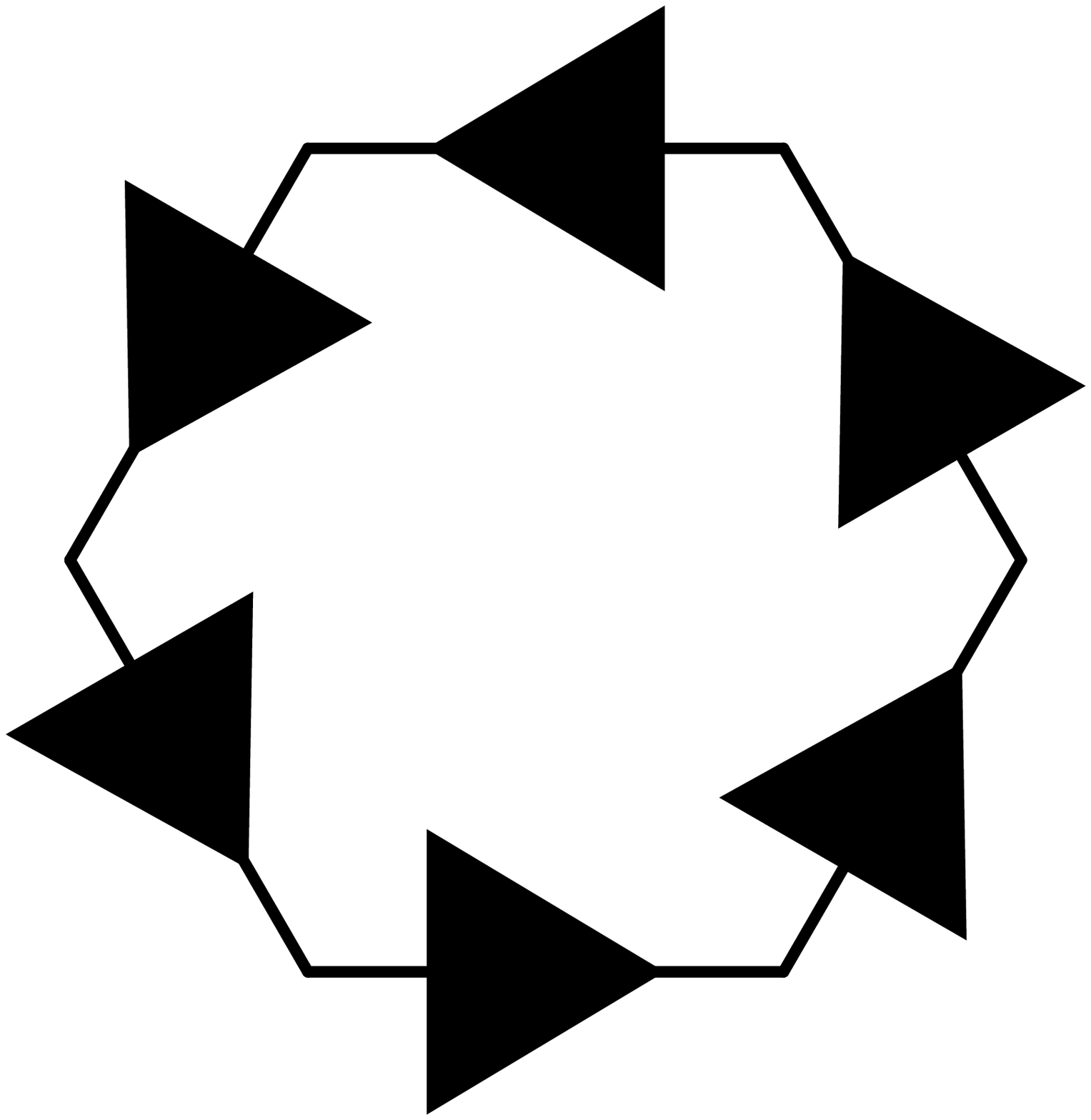}}
\newcommand{\DDPL}{
\includegraphics[width=0.4cm]{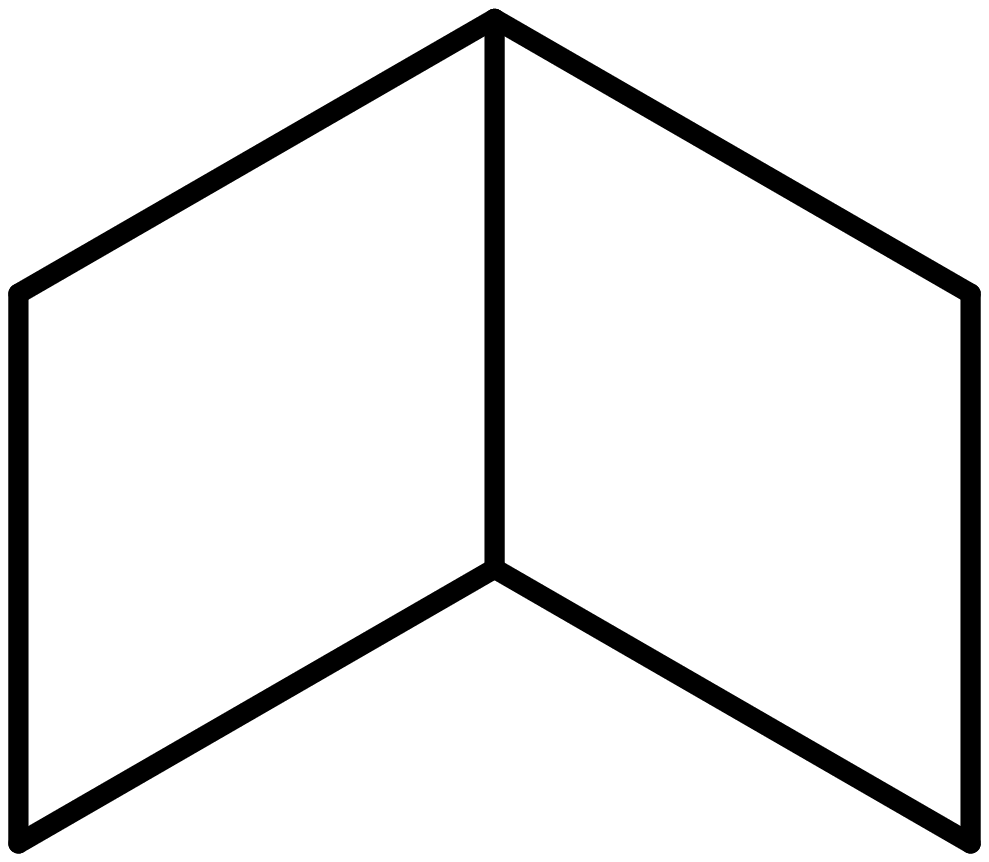}}
\newcommand{\DDPLA}{
\includegraphics[width=0.4cm]{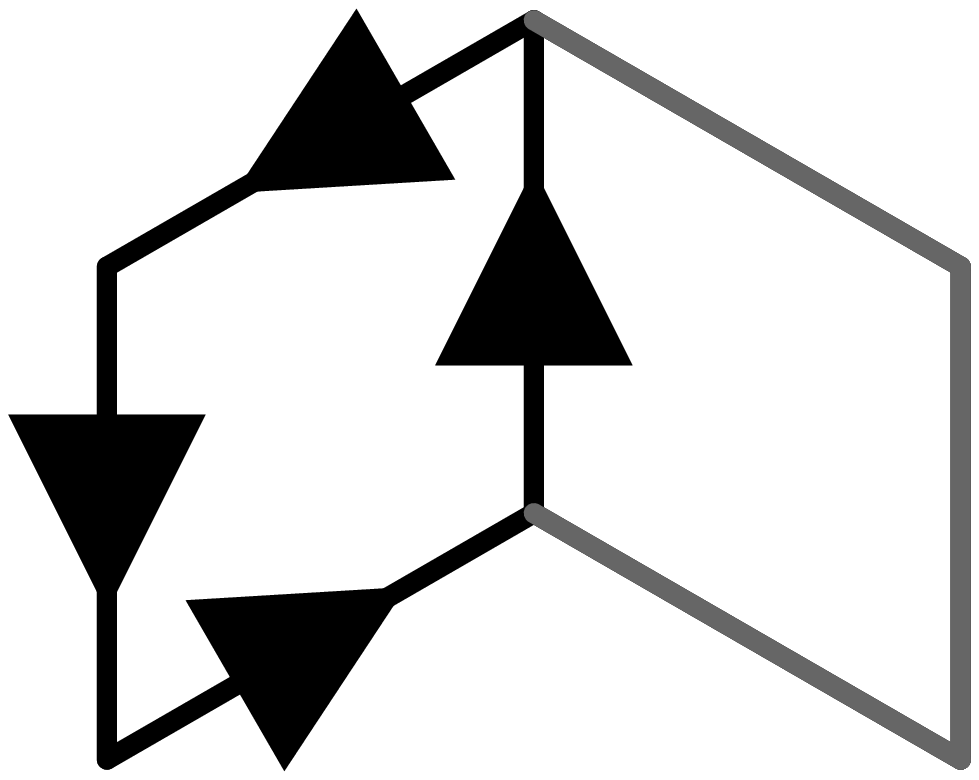}}
\newcommand{\DDPLB}{
\includegraphics[width=0.4cm]{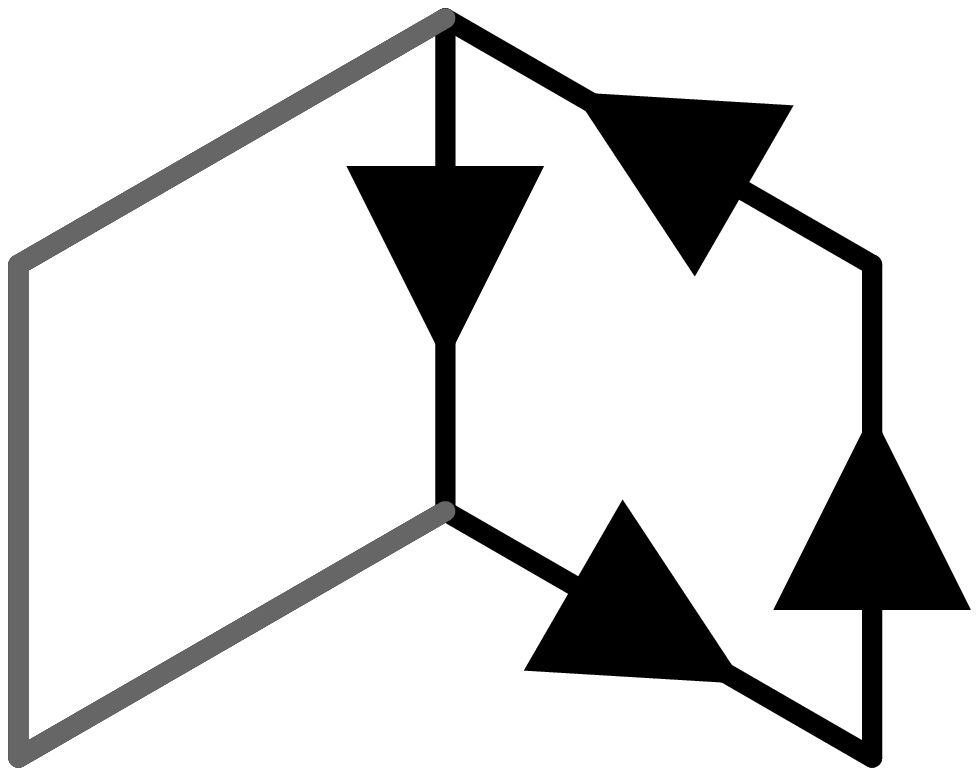}}

\newcommand{\DPLa}{\plaq{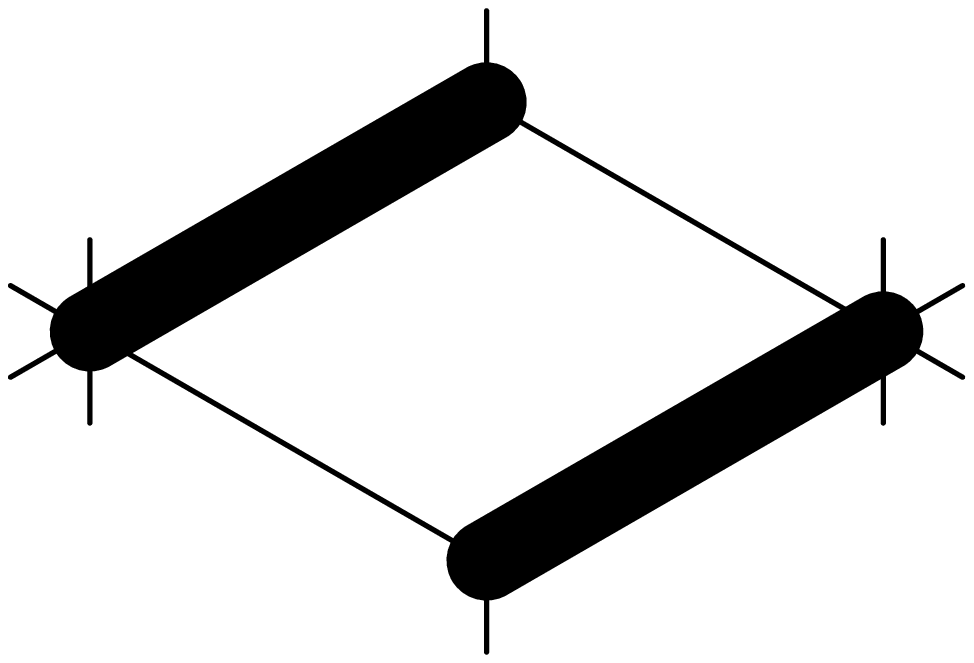}{0.5}}
\newcommand{\DPLb}{\plaq{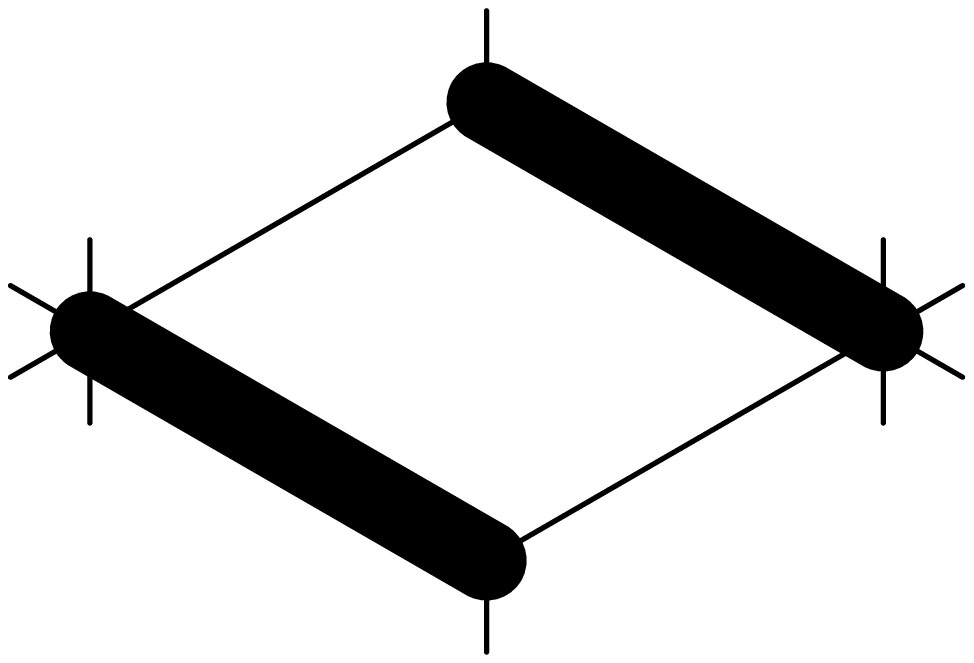}{0.5}}
\newcommand{\DPLc}{\plaq{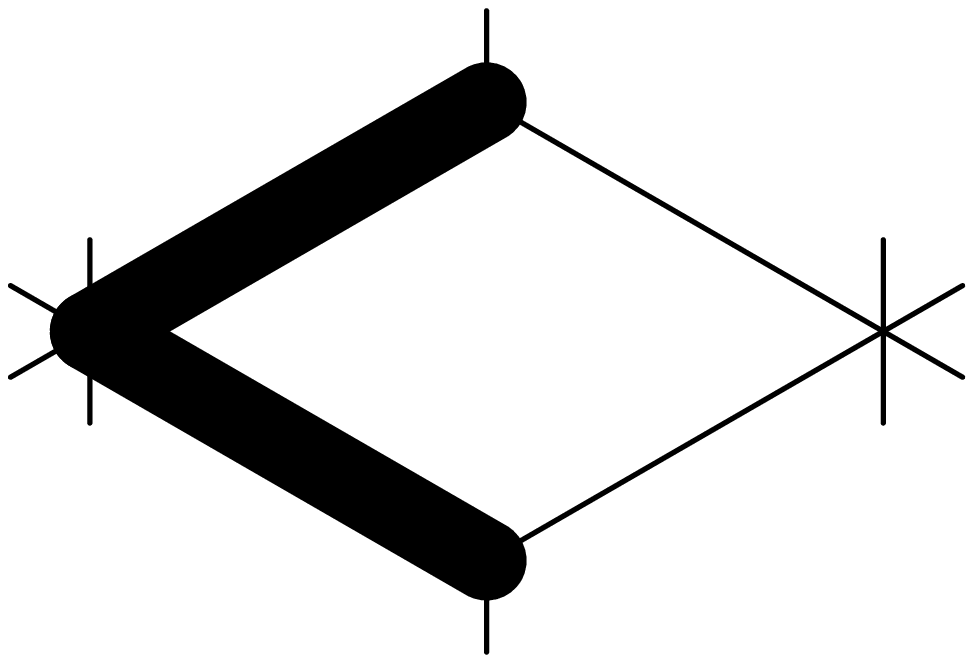}{0.5}}
\newcommand{\DPLd}{\plaq{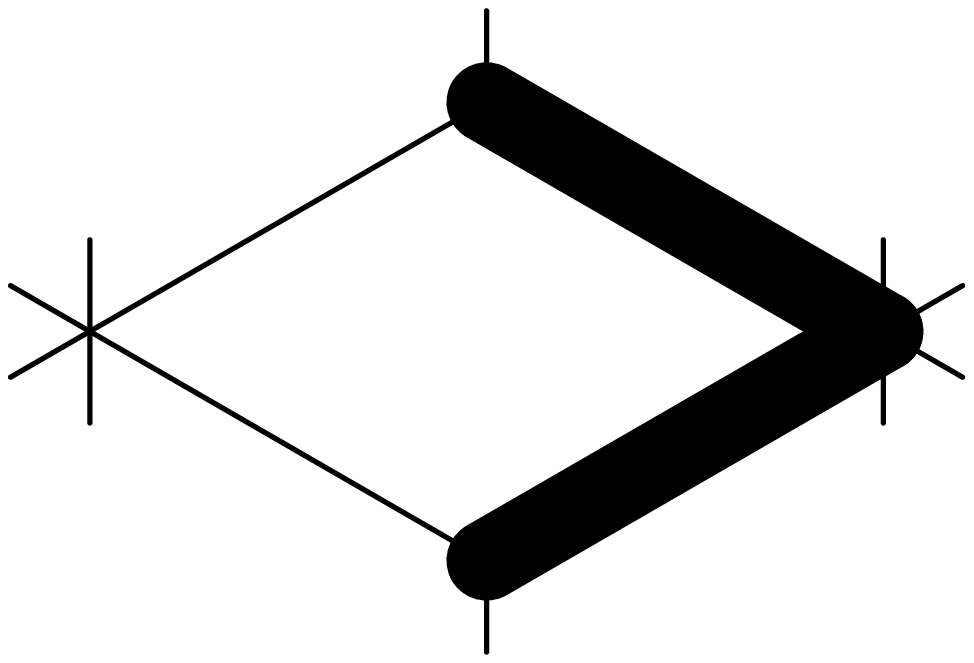}{0.5}}

\begin{document}

\title{Disordered, Spin Liquid and Valence-Bond Ordered Phases of the Kagome Lattice Quantum Ising Models With Transverse Field and XXZ Dynamics}
\author{P. Nikoli\'c}
\affiliation{Massachusetts Institute of Technology, 77 Massachusetts Ave.,  Cambridge, MA 02139}
\date{\today}

\begin{abstract}

General conditions in which disordered, spin liquid, and valence-bond ordered phases occur in quantum Ising antiferromagnets are studied using the prototype Kagome lattice spin models. A range of quantum dynamical processes in the Ising model, with and without total Ising spin conserved, are analytically shown to yield all three characteristic quantum paramagnetic phases in the Kagome system. Special emphasis is given to the XXZ model that can be sensibly compared to the Kagome lattice Heisenberg antiferromagnet. It is explicitly demonstrated that the total-spin-conserving dynamics can yield a resonant valence bond (RVB) liquid phase with very short-ranged correlations, but also a valence-bond ordered phase compatible with the one proposed to explain the seemingly gapless singlet states of the Heisenberg antiferromagnet on the Kagome lattice. Likely consequences for generic spin models are discussed. The analysis combines compact U(1) gauge theory, duality transformations, lattice-field-theoretical methods, and variational approach.

\end{abstract}

\maketitle

\section{Introduction}

The quest for a quantum spin liquid has been a major pursuit in condensed matter physics ever since the Anderson's proposal for its existence and relevance to the cuprates. \cite{RVB} Since then, some new materials with geometric frustration emerged as promising candidates for the spin liquid physics at low temperatures. Among two-dimensional systems the Mott insulators $\textrm{Cs}_2 \textrm{CuCl}_4$ (Ref.~\cite{Coldea}) and $\kappa - \textrm{(BEDT-TTF)}_2 \textrm{Cu}_2 (\textrm{CN})_3$ (Ref.~\cite{BEDTsl}), based on the triangular lattice, experimentally exhibit unconventional magnetic behavior in certain circumstances, without detectable symmetry breaking. Numerical studies also reveal possible spin liquid in realistic systems, ranging from a Wigner crystal near melting, \cite{Bernu} to various antiferromagnets with multiple-spin exchange. \cite{TriRing1, TriRing2} However, there is still no unambiguous experimental evidence that a spin liquid is found in any of these or similar cases. Even theoretically, much more is known about general properties of the spin liquid \cite{ReSaSpN, SaRe, SLWen, SLFradkin, Z2} then about fundamental and microscopic circumstances needed for its realization. Gaining more microscopic insight can only be useful for understanding the unconventional magnetic behavior found in various experiments, and perhaps also for clarifying available options for quantum computing. \cite{QuantComp1, QuantComp2}

In this paper we attempt to learn general lessons on the role played by the lattice structure, symmetries and type of dynamics in shaping the phases of the frustrated quantum magnets. The present analysis is focused on the prototype Kagome lattice Ising antiferromagnets, but the range of explored models admits all characteristic quantum paramagnetic phases: disordered, spin liquid, and valence-bond solid. Through connecting these outcomes with both the fundamental and microscopic properties of the models, and with information that emerges from calculations, we can deduce some conclusions of broader significance for the spin models on other lattices.

The Kagome lattice is an excellent choice for this pursuit because it is one of the few simple spin systems (with only nearest-neighbor interactions) where disordered and spin liquid phases are believed to exist. On the other frequently studied lattices it usually takes further-neighbor and multiple-spin exchange to destabilize the zero temperature Neel order. The situation is somewhat better when the frustrated Ising antiferromagnets are concerned, but even then the quantum dynamics typically leads to a paramagnetic ground state that breaks lattice symmetries (such as on the triangular and fully frustrated square lattices).

The Kagome antiferromagnets are unique among two-dimensional spin systems in that they not only exhibit a promising spin liquid-like behavior, but also sometimes hold additional surprises. The experimental research on the spin $S=\frac{3}{2}$ Kagome layered materials $\textrm{SrCr}_{9p} \textrm{Ga}_{12-9p} \textrm{O}_{19}$ (SCGO) \cite{SCGOmuon, SCGOHeatCap} and $\textrm{Ba}_2 \textrm{Sn}_2 \textrm{ZnGa}_3 \textrm{Cr}_7 \textrm{O}_{22}$ (QS ferrite) \cite{QSfer} discovered a spin-glass together with a heat capacity that is not thermally activated, and largely not dependent on the magnetic field at low temperatures. The numerical exact diagonalization studies of the $S=\frac{1}{2}$ Heisenberg antiferromagnet on small Kagome samples \cite{KNum, KMagnTherm} revealed a disordered ground state, and a seemingly gapless band of numerous singlet excitation that fill the spectrum below the finite spin-gap. In comparison to the other quantum paramagnets, \cite{Lhlong, Lhshort} this one appeared qualitatively different, and called for classification as a new and exotic kind of spin liquid. Such kind of spectrum, in which there seems to be no gap in a completely disordered phase, is still not understood. The physical pictures proposed by various theoretical efforts favored a spin liquid, \cite{Elser, KagMila, KagSpN, KagRK, KagSpecDimer} but some new ideas open up a possibility of a valence bond crystal with very large unit-cell that gives rise to an extremely low energy scale for the singlet degrees of freedom. \cite{KagSUN, KagVBC}

The transverse field Ising model on the Kagome lattice is another example of how special this lattice is. The Monte-Carlo simulations \cite{FrIsing1, FrIsing2} pointed out that unlike the other Ising systems, Kagome prefers not to order even for small transverse fields. One perspective in explaining the reason for this has been taken in the Ref.~\cite{KagIsing} In this paper, a different perspective will be given. The present approach is also extended to the XXZ model, which can be regarded as the Heisenberg model with easy-axis anisotropy.

\section{Models and Overview}

\begin{figure}
\includegraphics[width=2.1in,angle=90]{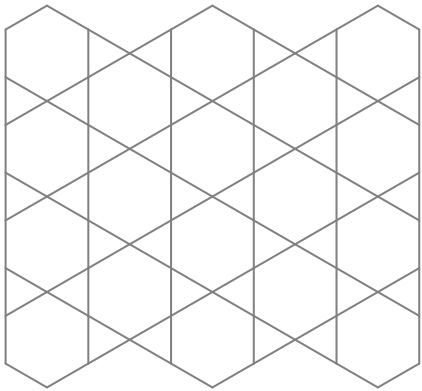}
\vskip -2mm
\caption{\label{KGrid}Kagome lattice is a corner-sharing two dimensional lattice. The frustrated units are triangular plaquettes, and they are only minimally connected into the lattice.}
\end{figure}

In this paper we analyze the nearest-neighbor spin $S=\frac{1}{2}$ quantum Ising antiferromagnets on the Kagome lattice (Fig.~\ref{KGrid}). Two kinds of spin dynamics will be explored, represented by the following simple models:
\begin{itemize}
\item transverse field Ising model (TFIM):
\begin{equation}\label{TFIM}
H = J_z \sum_{\bond{ij}} S^z_i S^z_j - \Gamma \sum_i S^x_i \ ;
\end{equation}
\item Heisenberg model with easy-axis anisotropy (XXZ):
\begin{equation}\label{EAHM}
H = J_z \sum_{\bond{ij}} S^z_i S^z_j + J_{\bot} 
  \sum_{\bond{ij}} \bigl( S^x_i S^x_j + S^y_i S^y_j \bigr) \ .
\end{equation}
\end{itemize}
In contrast to the transverse field case, the XXZ dynamics preserves total Ising magnetization, making the Hamiltonian ~(\ref{EAHM}) symmetric under global spin-flip. Furthermore, the transverse field gives rise to the most local kind of spin dynamics, while the XXZ dynamics involves pairs of spins, and thus introduces some correlation. It will become apparent that these two fundamental differences yield very different low energy physics. The consequent analysis will also admit introduction of other dynamical processes, spatially extended to larger clusters of spins, but consistent with the symmetries of these two basic models. 

The calculations in this paper are restricted to weak dynamical perturbations of the pure Ising model: $\Gamma, J_{\bot} \ll J_z$. This limit is a combination of analytical convenience, and essential physical interest in the context of frustrated magnetism. The main question being asked is how the quantum fluctuations (created by weak dynamical perturbations) lift the degeneracy of the pure Ising model. Is the ground state ordered like in many other Ising systems (order-by-disorder on triangular and fully frustrated square lattices)? Under what general circumstances a completely disordered ground state is possible, with or without non-trivial topology? In attempt to answer these questions, we will formulate a lattice field theory and apply to it a technique specialized for frustrated systems, but otherwise analogous to the usual mean-field approach in the unfrustrated problems. Namely, instead of finding the mean-field solutions that minimize energy in various parameter regimes, we will seek solutions that maximize ``entropy'' of quantum fluctuations. \cite{DimerSG} When needed, those solutions will be subject to a verification of stability. This will provide a reliable picture of the phases that exist in our models. 

Physics of the TFIM is trivial when the transverse field $\Gamma$ is large, while in the limit $\Gamma \ll J_z$ the quantum dynamics, as a matter of principle, has a chance to yield interesting valence bond ordered or disordered ground states after lifting the huge degeneracy of the pure Ising model. Even though this issue has been already understood for the Kagome TFIM, \cite{KagIsing} the following approach is going to bring some more insight: it will allow us to propose certain variational wavefunctions. Of course, much closer to the true challenge posed by the Kagome Heisenberg antiferromagnet is the XXZ model, which can be regarded as its anisotropic version. In the limit $J_{\bot} \ll J_z$ that we are going to explore the easy-axis anisotropy is strong, but sensible comparison with the isotropic case will become apparent. In particular, it will be shown by variational arguments that the short-range valence bond states arise most naturally in the easy-axis XXZ models on the corner-sharing lattices, which together with the lattice-field-theoretical calculations will lead to the valence bond ordered state already proposed for the isotropic Kagome antiferromagnet. \cite{KagVBC} A spin liquid phase will also appear as a possibility in closely related models.

This paper is organized in two main sections, each divided into several subsections. The TFIM model is discussed in the section ~\ref{tfim}, while the XXZ model is studied in the section ~\ref{eahm}. Initial discussion of the XXZ model relies heavily on the definitions and ideas introduced in the TFIM section (sections ~\ref{tfim} through ~\ref{TFIMfluct}). Readers interested only in the physical nature of the valence-bond ordered and spin liquid phases of the XXZ models can skip all calculations and go directly to the section ~\ref{EAHMvar}. All conclusions are summarized and bigger perspective is taken in the Discussion.

\newpage


\section{Transverse Field Ising Model}\label{tfim}

\begin{figure}[b]
\includegraphics[width=2.2in]{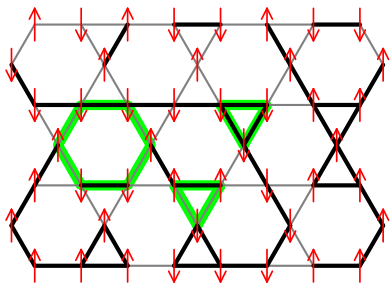}
\vskip -2mm
\caption{\label{KagDimer}Every loop holds an even number of unfrustrated bonds. Dimers represent the frustrated bonds, or pairs of aligned spins.}
\end{figure}

We start from the Kagome lattice Ising model in a weak transverse field $\Gamma \ll J_z$:
\begin{equation}\label{TFIM2}
H = J_z \sum_{\bond{ij}} S^z_i S^z_j - \Gamma \sum_i S^x_i \ .
\end{equation}
Let us first understand the ground states of the pure Ising Hamiltonian ($\Gamma=0$). They are the \emph{least frustrated} states in which the number of \emph{frustrated} bonds (two aligned spins) is minimized. If every frustrated bond is visualized by a dimer, then every appropriate dimer covering determines a spin configuration up to a global spin flip. Consider a loop on the Kagome lattice (Fig.~\ref{KagDimer}). The unfrustrated bonds on the loop mark locations where the two neighboring spins on the loop have different orientation. When going one full circle around the loop one ends at the same spin from which one started, so that the number of times the spin orientation is changed must be even. Therefore, every loop contains an even number of unfrustrated bonds, and the parity of the number of dimers on the loop is determined by the loop size. The number of dimers on the triangular (hexagonal) Kagome plaquettes must be odd (even). This is the only constraint for the Kagome lattice dimer coverings that correspond to arbitrary spin states.

\begin{figure}
\includegraphics[width=2.6in]{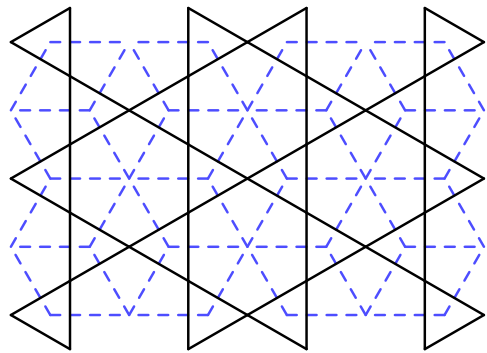}
\vskip -2mm
\caption{\label{KagDice}Duality between the Kagome (solid line) and dice (dashed) lattices. Duality transforms a Kagome site into the dice plaquette inside which it sits, and vice versa. The 3-coordinated dice sites and Kagome triangles transform into each other, as well as the 6-coordinated dice sites and Kagome hexagons. Every Kagome bond intersects one dual dice bond.}
\end{figure}

\begin{figure}
\includegraphics[width=1.9in]{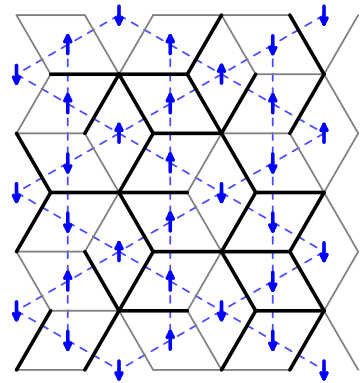}
\vskip -2mm
\caption{\label{DiceDimer}Frustrated bonds represented by dimers on the dice lattice.}
\end{figure}

It will be convenient to immediately switch to the dual picture. Duality between the Kagome and dice lattices is depicted in the Fig.~\ref{KagDice}. Since every Kagome bond corresponds to one dice lattice bond, the frustrated bonds can be represented by dimers on either lattice. An example is shown in the Fig.~\ref{DiceDimer}. There must be an odd (even) number of dimers emanating from every 3-coordinated (6-coordinated) dice lattice site. The number of dimers (and thus frustration) is minimized if there is exactly one dimer emanating from every 3-coordinated dice lattice site. This condition fixes the number of dimers in the least frustrated states, since the dice lattice is bipartite. The degeneracy of the least frustrated states is apparently huge.

Small dynamical perturbations will mix the least frustrated states and lift their immense degeneracy. In principle, one can perturbatively derive an effective theory that describes dynamics at energy scales well below $J_z$. This effective theory lives in the Hilbert space spanned only by the least frustrated states. Therefore it takes form of a soft-core quantum dimer model on the dice lattice, where exactly one dimer emanates from every 3-coordinated dice site, while an arbitrary even number of dimers emanate from every 6-coordinated site. For our purposes it will be sufficient to concentrate just to the first order of degenerate perturbation theory: 
\begin{eqnarray}\label{DiceDimerModel}
H_{\textrm{eff}} = 
  & - & \frac{\Gamma}{2} \sum_{\DPL_{}} 
        \left(\ket{\DPLa}\bra{\DPLb} + h.c. \right) \nonumber \\
  & - & \frac{\Gamma}{2} \sum_{\DPL_{}}
        \left(\ket{\DPLc}\bra{\DPLd} + h.c. \right) \ .
\end{eqnarray}
The dimer dynamics consists of two different flips on the dice plaquettes that are consistent with minimum frustration (see Fig.~\ref{DimerFlip1}). Note that these two processes involve flipping of only one spin on the Kagome lattice. For the purposes of simplicity and staying close to the original spin dynamics, we will not consider a more general dimer model with different energy scales for the two types of flips. Since the dice lattice is bipartite, it is possible to apply the standard techniques and cast this dimer model as a compact U(1) gauge theory. \cite{DimerFT} From that point on, duality transformations and lattice-field-theoretical methods are at disposal to study the possible phases. After understanding the phases that the lattice theory yields, it will be apparent that sufficiently small higher order perturbations cannot destabilize them.

Some comments are in order before proceeding. First, the dimer representation is insensitive to the global spin flip. This is of no concern for the models studied here, since magnetically ordered phases will not be found. Secondly, the dice lattice dimer model in its own right has non-trivial topology, just like the other quantum dimer models. However, only one of its topological sectors corresponds to the physical spin states on a torus. Therefore, the original spin model need not have any interesting topology.

\subsection{Compact U(1) Gauge Theory}\label{TFIMU1}

Calculations in this paper rely heavily on duality between the Kagome and dice lattices (see Fig.~\ref{KagDice}). In order to facilitate mathematical manipulations, we will treat both lattices on the same footing, and regard the pairs of objects related by duality as identical. The notation that we will use from now on is summarized in the Table \ref{Notation1}. Note that, according to this principle, any quantity that lives on a Kagome bond equivalently lives on the dual dice bond, and may be labeled by either Kagome, or dice bond labels. Also, we will apply the following convention: if an equation shows a relationship between expressions defined on different lattices, the dual lattice objects are always implied (for example, Kagome site $\Leftrightarrow$ dual dice plaquette, Kagome bond $\Leftrightarrow$ dual dice bond\dots).

\begin{table}
\begin{tabular}{|l|l|}
\hline
$i,j\dots$ & Kagome lattice sites, or dual dice plaquettes \\
$p,q\dots$ & dual dice lattice sites, or Kagome plaquettes \\
$p_3$        & 3-coordinated dice sites, or Kagome triangles \\
$p_6$        & 6-coordinated dice sites, or Kagome hexagons \\
$\bond{ij}$  & Kagome lattice bonds \\
$\bond{pq}$  & dual dice lattice bonds \\
\hline
\end{tabular}
\caption{\label{Notation1}Notation for the Kagome and dual dice lattices}
\end{table} 

\begin{figure}
\includegraphics[width=2.6in]{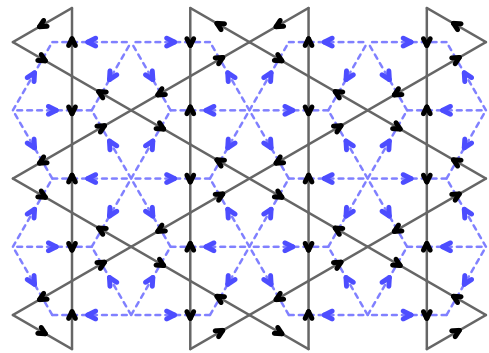}
\vskip -2mm
\caption{\label{BondOrientations}The reference bond orientations of the Kagome and dice lattices. The bipartite dice lattice bonds are oriented from the 6-coordinated site to the 3-coordinated site. Every Kagome bond orientation is locked to the orientation of the dual dice bond by the ``right hand rule''. Note that the Kagome orientations circulate around triangles and hexagons in the different directions.}
\end{figure}

Let us also introduce a vector notation. We will distinguish vectors $R_{pq}$ from the corresponding bond scalars $R_{\bond{pq}}$ in that the vectors will change sign if the bond orientation is reversed: $R_{pq}=-R_{qp}$, while the scalars will not: $R_{\bond{pq}}=R_{\bond{qp}}$. In order to establish a formal connection between the bond vectors and scalars, we assign orientation to the lattice bonds. Let the vector $\eta_{pq}$ equal $+1$ if the bond $\bond{pq}$ is oriented from $p$ to $q$, and $-1$ otherwise. Then, the vectors and corresponding bond scalars are related by $R_{pq} = \eta_{pq} R_{\bond{pq}}$. These relations are applicable to both the dice and Kagome lattices. The bond orientation will transform by duality according to the ``right hand rule''. Since the dice lattice is bipartite, we will orient its bonds in a natural way, and chose the orientation to be from the 6-coordinated to the 3-coordinated site on every bond. This fixes orientation of the Kagome bonds as well, and we show both in the Fig.~\ref{BondOrientations}.

Now we can define the electric field $E_{pq}$ as a vector corresponding to the scalar bond energy $E_{\bond{pq}}$:
\begin{equation}
E_{\bond{pq}} = \bigg \lbrace
\begin{array}{l@{,}l}
$0 \ $ & \textrm{$\ $ vacancy (unfrustrated bond $\bond{pq}$)} \\
$1 \ $ & \textrm{$\ $ dimer (frustrated bond $\bond{pq}$)}
\end{array} \nonumber
\end{equation}
\begin{equation}\label{U1E}
E_{pq} = \eta_{pq} E_{\bond{pq}} \ .
\end{equation}
The Hilbert space of the least frustrated states has restrictions that are easily expressed in the form of the Gauss' Law. The number of dimers $E_{pq}=\eta_{pq}$ emanating from any 3-coordinated site is one, and from any 6-coordinated site an even number $(2n_{p_6})$. We use the convention that every dice bond is oriented from the 6-coordinated to the 3-coordinated site, and write:
\begin{eqnarray}\label{U1constr}
(\forall p_3) \quad \sum_{q \in p_3} E_{p_3q} = -1 \ , \nonumber \\
(\forall p_6) \quad \sum_{q \in p_6} E_{p_6q} = 2n_{p_6} \ .
\end{eqnarray}
The interpretation of this Gauss' Law is that there is a fixed background charge $-1$ on every 3-coordinated dice site, and a number $0 \leqslant n_{p_6} \leqslant 3$ of charge $2$ bosons on every 6-coordinated site. The charged bosons are independent degrees of freedom living on the 6-coordinated dice sites. Formally, they emerge because the dice lattice dimer model is not hard-core.

Dynamics of the fields and particles can be easily formulated if the Hilbert space is expanded to allow arbitrary integer strength of the electric field, and arbitrary particle occupation. Promoting $E_{pq}$ and $n_{p_6}$ into free integers between $-\infty$ and $+\infty$ makes it easy to write the creation and annihilation operators: $\exp(\pm i \mathcal{A}_{pq})$ for the field lines, and $\exp(\pm i \varphi_{p_6})$ for the particles. The vector potential $\mathcal{A}_{pq}$ and the boson phase $\varphi_{p_6}$ are conjugate angle operators to the electric field $E_{pq}$ and particle number $n_{p_6}$ respectively:
\begin{equation}
\lbrack \mathcal{A}_{pq} , E_{pq} \rbrack = 
\lbrack \varphi_{p_6} , n_{p_6} \rbrack = i \ .
\end{equation}
After the Hilbert space has been expanded, we must at least introduce a large energy cost to all ``unphysical'' states, so that the low energy physics will still correspond to the dimer model ~(\ref{DiceDimerModel}). This is achieved in the large $U$ limit through the following new term in the Hamiltonian:
\begin{equation}\label{HU}
H_u = U \sum_{\bond{pq}} \Bigl( E_{\bond{pq}} - \frac{1}{2} \Bigr) ^ 2 =
      U \sum_{\bond{pq}} E_{pq}^2 + \textrm{const.} \ .
\end{equation}
The term linear in electric field is a global constant, since it expresses the fixed total number of dimers on the dice lattice (in the least frustrated states).

\begin{figure}
\subfigure[{}]{\includegraphics[width=3in]{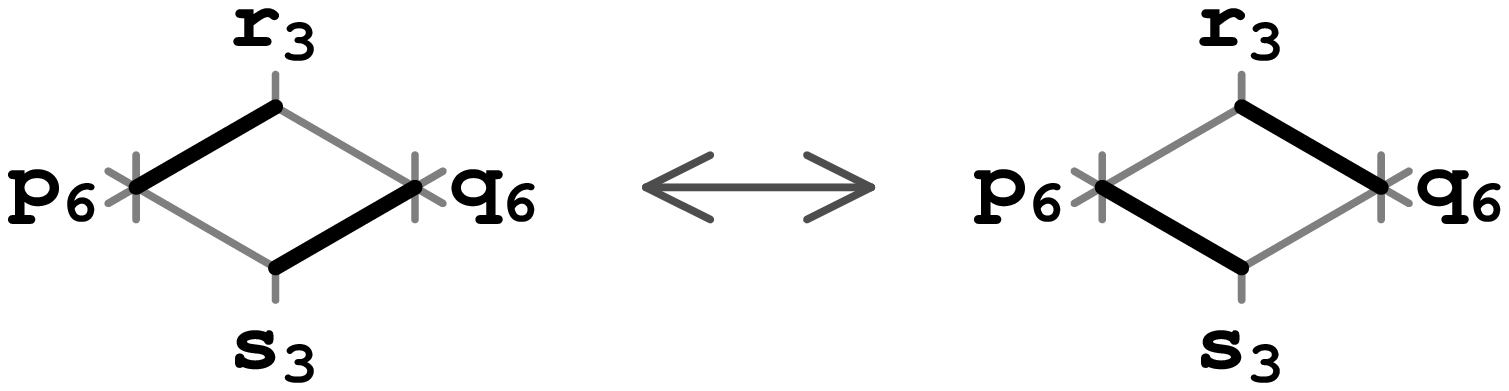}}
\vskip -2.3mm
\subfigure[{}]{\includegraphics[width=3in]{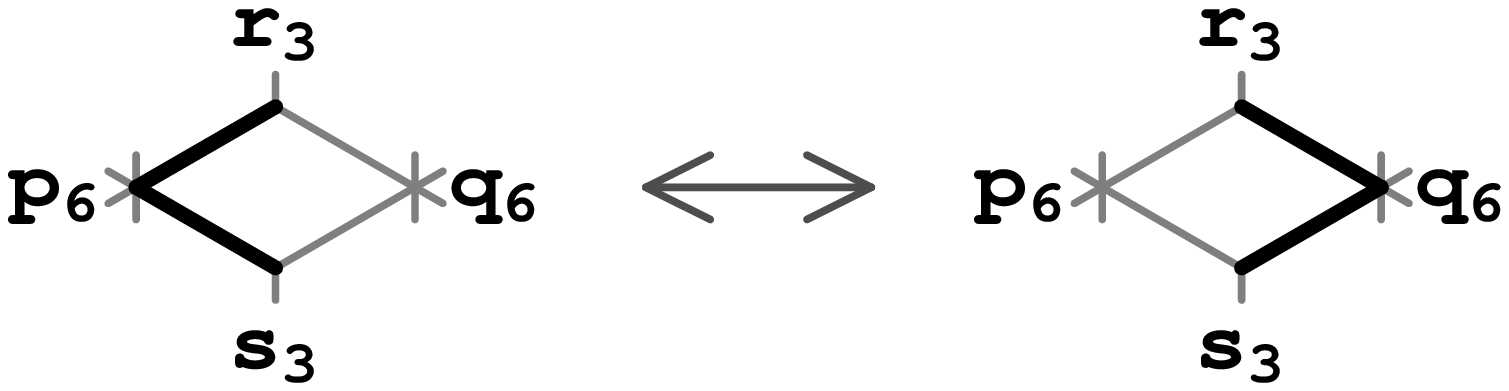}}
\vskip -2.3mm
\caption{\label{DimerFlip1}Elementary processes on a dice plaquette that preserve the minimum frustration. (a) The number of dimers emanating from every site is preserved; (b) a pair of dimers is exchanged between two 6-coordinated sites.}
\end{figure}

\begin{figure}
\includegraphics[width=2.6in]{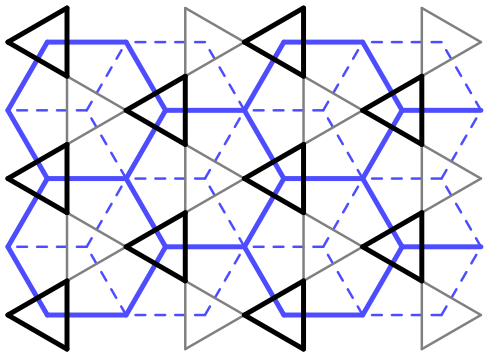}
\vskip -2mm
\caption{\label{Epsilon}Special bond signs on the Kagome and dice lattices. $\varepsilon_{\bond{ij}} \equiv \varepsilon_{\bond{pq}}$ take the value $-1$ on the emphasized bonds, and $+1$ on all other bonds.}
\end{figure}

\begin{figure}
\subfigure[{}]{\includegraphics[width=2.6in]{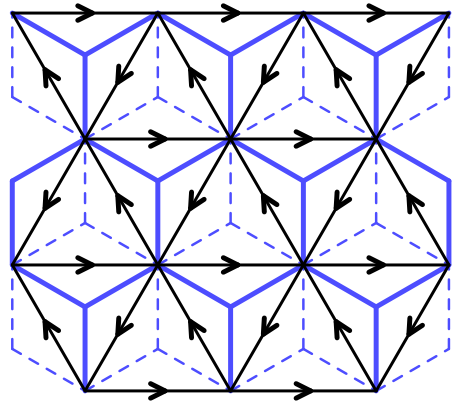}}
\subfigure[{}]{\includegraphics[width=2.6in]{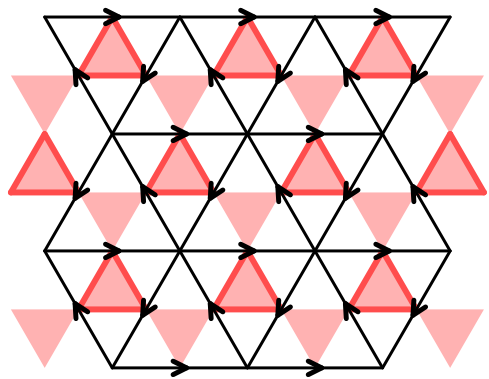}}
\vskip -2mm
\caption{\label{TriangularEta}Bond orientations $\eta_{p_6q_6}$ of the triangular lattice formed by the 6-coordinated dice sites, or equivalently the Kagome hexagon centers. The dice and Kagome bonds with $\varepsilon_{\bond{pq}} \equiv \varepsilon_{\bond{ij}} = -1$ are emphasized. Notice in (b) that every triangular lattice bond contains one Kagome site.}
\end{figure}

Now we formulate dynamics of ~(\ref{DiceDimerModel}) in the U(1) language. The two processes of interest are shown in the Fig.~\ref{DimerFlip1}. Recall that every dimer means $E_{\bond{pq}}=1$, and every vacancy $E_{\bond{pq}}=0$, and that $n_{p_6}$ is the number of dimer pairs emanating from a 6-coordinated site. Therefore, we can easily exploit the dice bond orientations, and arrange the creation and annihilation operators to describe the allowed dimer flip processes (a) and (b) shown in the Fig.~\ref{DimerFlip1}:
\begin{eqnarray}
\label{DimerFlipA}
(a) & \sim &   \exp(i\mathcal{A}_{\bond{p_6s_3}}) 
               \exp(-i\mathcal{A}_{\bond{q_6s_3}}) \times \nonumber \\
    & & \times \exp(i\mathcal{A}_{\bond{q_6r_3}})
               \exp(-i\mathcal{A}_{\bond{p_6r_3}}) + h.c. \\
    & = & 2 \cos ( \mathcal{A}_{p_6s_3} + \mathcal{A}_{s_3q_6} +
                   \mathcal{A}_{q_6r_3} + \mathcal{A}_{r_3p_6} ) \nonumber \\
    & = & 2 \cos \Biggl( \sum_{pq}^{\CDPL_{}} \mathcal{A}_{pq} \Biggr) \ , \nonumber
\\ 
\label{DimerFlipB}
(b) & \sim &   \exp(i\varphi_{q_6}) \exp(-i\varphi_{p_6}) \times \nonumber \\
    & & \times \exp(-i\mathcal{A}_{\bond{p_6s_3}}) 
               \exp(i\mathcal{A}_{\bond{q_6s_3}}) \times \\
    & & \times \exp(i\mathcal{A}_{\bond{q_6r_3}})
               \exp(-i\mathcal{A}_{\bond{p_6r_3}}) + h.c. \nonumber \\
    & = & 2 \cos ( \varphi_{q_6} - \varphi_{p_6} -
                   \mathcal{A}_{p_6s_3} - \mathcal{A}_{s_3q_6} +
                   \mathcal{A}_{q_6r_3} + \mathcal{A}_{r_3p_6} ) \nonumber \\
    & = & 2 \cos \Biggl( \varphi_{q_6} - \varphi_{p_6} +
                         \eta_{p_6q_6} \sum_{pq}^{\CDPL_{}} 
                         \varepsilon_{\bond{pq}} \mathcal{A}_{pq} \Biggr) \ . \nonumber
\end{eqnarray}
In the last lines of these expressions the sums are taken around a plaquette in the counter-clockwise sense; this is the lattice circulation, or the curl. The expression ~(\ref{DimerFlipA}) is the usual ``magnetic'' energy, while the expression ~(\ref{DimerFlipB}) is the boson hopping between the neighboring 6-coordinated sites. In this paper we will not use a more conventional form of particle hopping that involves an ``integral'' of the vector potential along only one path between the two sites. In the equation ~(\ref{DimerFlipB}) we have introduced two new symbols: $\varepsilon_{\bond{pq}}$ and $\eta_{p_6q_6}$. The former is needed to correct the signs of $\mathcal{A}_{pq}$ that appear in the circulation. Note that the signs have been altered with respect to the ordinary circulation in ~(\ref{DimerFlipA}) only on one of the two paths that connect the two 6-coordinated sites. This allows us to choose $\varepsilon_{\bond{pq}}$ as shown in the Fig.~\ref{Epsilon}. The other symbol, $\eta_{p_6q_6}$ is needed to ensure that the expression inside the cosine of ~(\ref{DimerFlipB}) transforms properly when the sites $p_6$ and $q_6$ are exchanged. This is a new vector, defined on the triangular lattice formed by the 6-coordinated dice sites, or equivalently the centers of the Kagome hexagons. Since it takes the values $\pm 1$, it defines the bond orientations shown in the Fig.~\ref{TriangularEta}. Notice that $\eta_{p_6q_6}$ must be related to $\varepsilon_{\bond{pq}}$: if one takes a closer look at the cosines in ~(\ref{DimerFlipB}), one can see that when the boson hops from $p_6$ to $q_6$, the circulation starting from $p_6$ must first go through $\varepsilon_{\bond{pq}}=-1$ bonds. When this is satisfied for the counter-clockwise circulation, $\eta_{p_6q_6}$ should be $+1$, otherwise it should be $-1$. Exactly this is achieved by relating $\eta$ and $\varepsilon$ vectors as shown in the Fig.~\ref{TriangularEta}.

Finally, we can summarize the compact U(1) gauge theory on the dice lattice. The Hamiltonian is:
\begin{eqnarray}\label{U1Hamiltonian}
H & = & U \sum_{\bond{pq}} E_{pq}^2 
        - \Gamma\sum_{\DPL_{}} \Biggl\lbrack
           \cos \Biggl( \sum_{pq}^{\CDPL_{}} \mathcal{A}_{pq} \Biggr) \nonumber \\
  & + & \cos \Biggl( \varphi_{q_6} - \varphi_{p_6} +
             \eta_{p_6q_6} \sum_{pq}^{\CDPL_{}} 
             \varepsilon_{\bond{pq}} \mathcal{A}_{pq} \Biggr) \Biggr\rbrack \ ,
\end{eqnarray}
and the Hilbert space is constrained by the Gauss' Law:
\begin{eqnarray}\label{U1Gauss}
(\forall p_3) \quad \sum_{q \in p_3} E_{p_3q} = -1 \ , \nonumber \\
(\forall p_6) \quad \sum_{q \in p_6} E_{p_6q} = 2n_{p_6} + 2 \ .
\end{eqnarray}
For convenience that will become apparent later, we have shifted $n_{p_6}$ by one in the bottom expression (this sets to zero the total background charge on the lattice). In the limit of $U \to \infty$ this Hamiltonian is an exact rewriting of the effective theory ~(\ref{DiceDimerModel}). For finite and large $U$, one perturbatively obtains a theory as a $\Gamma / U$ expansion that introduces dimer flip processes on larger loops. This physically corresponds to further-neighbor and multiple-spin exchange that would be also generated by not so small $\Gamma$ in the original spin model ~(\ref{TFIM2}). Owing to this approximate correspondence between finite $U$ and larger $\Gamma$, we might have means to qualitatively see some trends beyond very small $\Gamma / J_z$.

\subsection{Lattice Field Theory}\label{TFIMft}

The path-integral corresponding to ~(\ref{U1Hamiltonian}) describes a (2+1)D electrodynamics. All fluctuations are constrained by ~(\ref{U1Gauss}). The action will contain a usual Berry's phase (we will omit the time index):
\begin{equation}\label{U1Berry}
S_B = -i \sum_{\tau} \Biggl( \sum_{\bond{pq}} \mathcal{A}_{pq} \Delta_{\tau} E_{pq}
           + \sum_{p_6} \varphi_{p_6} \Delta_{\tau} n_{p_6} \Biggr) \ ,
\end{equation}
and a potential energy part:
\begin{equation}\label{U1pot}
S_{\textrm{pot}} = 
  U\delta\tau \sum_{\tau}\sum_{\bond{pq}} E_{pq}^2 \ ,
\end{equation}
where $\delta\tau$ is the imaginary time increment. The kinetic energy, which involves the cosines in ~(\ref{U1Hamiltonian}), can be brought to a more tractable form by applying the Villain's approximation. Two new fields will appear and play a significant role: the magnetic field scalar $B_i$ that lives on the Kagome sites dual to the dice plaquettes, and the particle current $j_{p_6q_6}$ that lives as a vector on the triangular lattice bonds. Both will be integer valued, reflecting the compactness of the U(1) gauge theory, and the fluctuations of both will be suppressed by the scale $g=|\log(\Gamma\delta\tau / 2)|$. They take part in the action as follows:
\begin{eqnarray}\label{U1kin}
S_{\textrm{kin}} & = & \sum_{\tau} \Biggl\lbrack
      g \Biggl(\sum_i B^2_i + \sum_{\bond{p_6q_6}} j^2_{p_6q_6} \Biggr) \\
   & + & i \sum_i B_i \sum_{pq}^{\CDPL_i} \mathcal{A}_{pq} \nonumber \\
   & + & i \sum_{\bond{p_6q_6}}
      j_{p_6q_6} \Biggl( \varphi_{q_6} - \varphi_{p_6} +
      \eta_{p_6q_6} \sum_{pq}^{\CDPL_{}} 
      \varepsilon_{\bond{pq}} \mathcal{A}_{pq} \Biggr) \Biggr\rbrack \ . \nonumber
\end{eqnarray}
After writing this, the angles $\mathcal{A}_{pq}$ and $\varphi_{p_6}$ can be formally integrated out. Fluctuations of the boson phases $\varphi_{p_6}$ will give rise to the particle current conservation law:
\begin{equation}\label{U1CurrCon}
(\forall p_6) \quad \Delta_{\tau} n_{p_6} + \sum_{q_6 \in p_6} j_{p_6q_6} = 0 \ .
\end{equation}
Fluctuations of the vector potential will give rise to the Maxwell's equation for the magnetic field curl: $\mathcal{A}_{pq}$ is coupled to the magnetic field $B_i$, and current $j_{p_6q_6}$ in ~(\ref{U1kin}), as well as the time derivative of the electric field $E_{pq}$ in ~(\ref{U1Berry}), which is the ``displacement'' current. However, this equation will take an unusual form, because the particle and displacement currents formally live on different lattices. The easyest way to derive it is to rewrite the terms in which $\mathcal{A}_{pq}$ appears using the Kagome lattice notation. For this purpose, let us note that the particle current $j_{p_6q_6}$ is related to the triangular lattice bond variable $j_{\bond{p_6q_6}}$ by: $j_{p_6q_6} = \eta_{p_6q_6} j_{\bond{p_6q_6}}$, which in turn can be regarded as actually living on the sites of the Kagome lattice (see Fig.~\ref{TriangularEta}). Therefore, we can label $j_{\bond{p_6q_6}}$ as $j_i$, where $i$ is the Kagome site that sits on the triangular bond $\bond{p_6q_6}$. From ~(\ref{U1Berry}) and ~(\ref{U1kin}) we have:
\begin{eqnarray}
\mathcal{A}_{pq} \Delta_{\tau} E_{pq} & \equiv & 
    \mathcal{A}_{ij} \Delta_{\tau} E_{ij} \nonumber \\
B_i \sum_{pq}^{\CDPL_i} \mathcal{A}_{pq} & \equiv &
    B_i \sum_{j \in i} \mathcal{A}_{ij} \\
j_{p_6q_6} \eta_{p_6q_6} \sum_{pq}^{\CDPL_i} 
    \varepsilon_{\bond{pq}} \mathcal{A}_{pq} & \equiv &
    j_i \sum_{j \in i} \varepsilon_{\bond{ij}} \mathcal{A}_{ij} \ . \nonumber
\end{eqnarray}
The vector potential fluctuations set to zero the sum of everything coupled to $\mathcal{A}_{ij}$ on every Kagome bond $\bond{ij}$. This is the Maxwell's equation:
\begin{equation}\label{U1Maxwell}
(\forall \bond{ij}) \quad \Delta_{\tau} E_{ij} =
   B_i - B_j + \varepsilon_{\bond{ij}} ( j_i - j_j ) \ .
\end{equation}
Once the phase and vector potential fluctuations have been integrated out, the remaining action contains only integer valued fields:
\begin{equation}\label{U1Action}
S = g \sum_{\tau} \Bigl\lbrack 
          \sum_i (B^2_i + j^2_i)
  +   \sum_{\bond{ij}} E_{ij}^2 
        \Bigr\rbrack \ ,
\end{equation}
whose fluctuations are subject to the constraints ~(\ref{U1Gauss}), ~(\ref{U1CurrCon}), and ~(\ref{U1Maxwell}). For convenience, we have chosen the imaginary time increment $\delta\tau$ such that $g=U\delta\tau$; this will not affect the low energy physics, at least for large $U$.

Now we can proceed by solving the constraints. To this end, we want to completely switch back to the Kagome lattice. The duality transformation that follows had been worked out in two typical cases. (a) If the compact U(1) gauge theory contains no charged particles, then the dual theory is an integer-valued height model. (b) If the particles and the electric field live on the same lattice, then the dual theory has a non-compact U(1) gauge structure, and a charged matter field. \cite{DimerFT} The case (a) emerges from the hard-core dimer models on bipartite lattices, while the case (b) has been proposed as an approximate description of the hard-core dimer models on non-bipartite lattices. Our case is somewhat in between. It turns out that the dual theory for our case resembles the height model.

\begin{figure}
\includegraphics[width=2.6in]{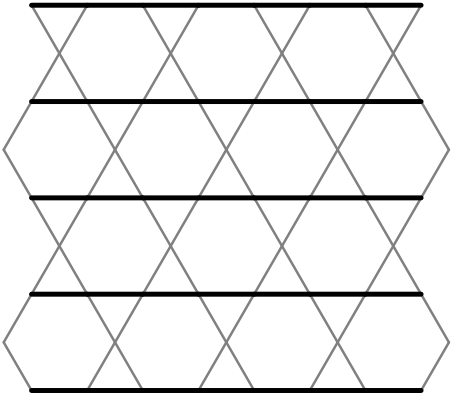}
\vskip -2mm
\caption{\label{ZetaFig}One characteristic and periodic configuration for $\zeta_{\bond{ij}}$. Every dimer represents $\zeta_{\bond{ij}}=1$, and vacancy $\zeta_{\bond{ij}}=0$.}
\end{figure}

It is convenient to redefine the magnetic field and current as time derivatives of two integer-valued ``height'' fields, $\chi_i$ and $\lambda_i$:
\begin{equation}\label{ChiLambda}
B_i = \Delta_{\tau} \chi_i \qquad , \qquad j_i = \Delta_{\tau} \lambda_i \ .
\end{equation}
Now, by introducing the $\lambda_{p_6q_6}$ vector analogous to $j_{p_6q_6}$, we can write solutions of the current conservation ~(\ref{U1CurrCon}) and Maxwell's equation ~(\ref{U1Maxwell}):
\begin{equation}\label{U1CurrConSol}
n_{p_6} + \sum_{q_6 \in p_6} \lambda_{p_6q_6} = 0 \ ,
\end{equation}
\begin{equation}\label{U1MaxwellSol}
E_{ij} = \chi_i - \chi_j + \varepsilon_{\bond{ij}} ( \lambda_i - \lambda_j ) +
   \zeta_{ij} \ ,
\end{equation}
where $\zeta_{ij}$ is an integer that does not vary with time, and will be determined by substituting this expression into the Gauss' Law ~(\ref{U1Gauss}). For consistency, let us first rewrite the Gauss' Law using the Kagome lattice labels:
\begin{eqnarray}\label{U1Gauss2}
(\forall \triangle_{p_3}) \quad \sum_{ij}^{\CTRI_{p_3}} E_{ij} = 1 \ , \nonumber \\
(\forall \hexagon_{p_6}) \quad \sum_{ij}^{\CHEX_{p_6}} E_{ij} = -2n_{p_6} - 2 \ .
\end{eqnarray}
The electric field divergence on the 3 and 6-coordinated dice sites in ~(\ref{U1Gauss}) transforms by duality into the lattice curl on the Kagome triangles and hexagons respectively. Taking curls of ~(\ref{U1MaxwellSol}) will annihilate $\chi_i$ on all Kagome plaquettes, as well as $\lambda_i$ on the Kagome triangles, since $\varepsilon_{\bond{ij}}$ is fixed on every triangle (see Fig.~\ref{Epsilon}). However, $\lambda_i$ will not be annihilated by the curls on the Kagome hexagons. One can easily show that:
\begin{equation}
\quad \sum_{ij}^{\CHEX_{p_6}}
   \varepsilon_{\bond{ij}} ( \lambda_i - \lambda_j ) = 
   2 \sum_{q_6 \in p_6} \lambda_{p_6q_6} \ ,
\end{equation}
which in turn is equal to $-2 n_{p_6}$ according to ~(\ref{U1CurrConSol}). Consequently, the equations that $\zeta_{ij}$ must satisfy are:
\begin{eqnarray}\label{Zeta}
(\forall \triangle_{p_3}) \quad \sum_{ij}^{\CTRI_{p_3}} \zeta_{ij} = 1 \ , \nonumber \\
(\forall \hexagon_{p_6}) \quad \sum_{ij}^{\CHEX_{p_6}} \zeta_{ij} = - 2 \ .
\end{eqnarray}
There are many possible choices. In order to reveal them more tractably, let us use the Kagome bond orientations in Fig.~\ref{BondOrientations}, and switch to the appropriate bond scalars $\zeta_{\bond{ij}}$ (notice that the counter-clockwise circulations coincide with the bond orientations on the triangles, but not on the hexagons):
\begin{eqnarray}\label{Zeta2}
(\forall \triangle_{p_3}) \quad \sum_{\bond{ij}}^{\triangle_{p_3}} 
    \zeta_{\bond{ij}} = 1 \ , \nonumber \\
(\forall \hexagon_{p_6}) \quad \sum_{\bond{ij}}^{\hexagon_{p_6}}
    \zeta_{\bond{ij}} = 2 \ .
\end{eqnarray}
If we decide to use only the values 0 and 1 for $\zeta_{\bond{ij}}$, and visualize the value 1 as a dimer, we see that every triangular plaquette must hold one dimer, and every hexagonal plaquette two dimers. One such configuration is depicted in the Fig.~\ref{ZetaFig}. The other possible configurations need not be periodic on the lattice, but ``breaking'' of the translational symmetry is unavoidable as long as $\zeta_{\bond{ij}}$ are integers.

This concludes the solution of all the constraints ~(\ref{U1Gauss}), ~(\ref{U1CurrCon}), and ~(\ref{U1Maxwell}). The final lattice field theory describes fluctuations of the two integer-valued height fields on the (2+1)D Kagome lattice. We obtain the action by substituting ~(\ref{ChiLambda}) and ~(\ref{U1MaxwellSol}) into ~(\ref{U1Action}):
\begin{eqnarray}\label{DHAction}
S & = & g \sum_{\tau} \Bigl\lbrack 
        \sum_i \Bigl( (\Delta_{\tau} \chi_i)^2 + (\Delta_{\tau} \lambda_i)^2 \Bigr) \\
  & & + \sum_{\bond{ij}} \Bigl( \chi_i - \chi_j +
             \varepsilon_{\bond{ij}} ( \lambda_i - \lambda_j ) +
             \zeta_{ij} \Bigr)^2
  \Bigr\rbrack \ . \nonumber      
\end{eqnarray}

\subsection{Important Properties}\label{TFIMprop}

Before proceeding with analysis of fluctuations in the lattice theory ~(\ref{DHAction}), we have to reveal several of its important properties. We begin by finding all configurations of integer-valued $\chi_i=\chi^{(0)}_i$ and $\lambda_i=\lambda^{(0)}_i$ that minimize the action. Clearly, $\chi^{(0)}_i$ and $\lambda^{(0)}_i$ should have no time dependence. However, in terms of spatial variations, there will be a large degeneracy. Let us define:
\begin{equation}\label{Xi}
\xi_{ij} = \chi^{(0)}_i - \chi^{(0)}_j +
           \varepsilon_{\bond{ij}} ( \lambda^{(0)}_i - \lambda^{(0)}_j ) +
           \zeta_{ij} \ .
\end{equation}
At a saddle-point, the action reduces to the sum of $\xi^2_{ij}$ on all Kagome bonds. The constraints that $\xi_{ij}$ must obey can be extracted by tracing back $\xi_{ij}$ to $E_{ij}$. Recall that this lattice field theory describes a particular soft-core dimer model in which $E_{\bond{ij}}=1$ represents a dimer. Therefore, the action will be minimized for all allowed dimer coverings of the Kagome lattice, by taking $\xi_{\bond{ij}}=1$ for a dimer, and $0$ for a vacancy. There will be one dimer on every Kagome triangle, and an arbitrary even number of dimers on every Kagome hexagon. Note that every action minimum corresponds to two least frustrated spin configurations of the Kagome Ising model ~(\ref{TFIM2}), which are related to each other by the global spin flip.

Shifting the height fields by $\chi^{(0)}_i$ and $\lambda^{(0)}_i$ allows us to study fluctuations about a particular saddle-point. The action takes a more general form:
\begin{eqnarray}\label{DHAction2}
S & = & g \sum_{\tau} \Bigl\lbrack 
        \sum_i \Bigl( (\Delta_{\tau} \chi_i)^2 + (\Delta_{\tau} \lambda_i)^2 \Bigr) \\
  & & + \sum_{\bond{ij}} \Bigl( \chi_i - \chi_j +
             \varepsilon_{\bond{ij}} ( \lambda_i - \lambda_j ) +
             \xi_{ij} \Bigr)^2
  \Bigr\rbrack \ . \nonumber      
\end{eqnarray}
After ``summation by parts'', we can write it in a matrix form (up to a constant):
\begin{eqnarray}\label{DHMatrAct}
\frac{S}{g} 
  & = & \sum_{\tau}\sum_i \Bigl\lbrack
          2\chi_i \sum_{j \in i} \xi_{ij}
        + 2\lambda_i \sum_{j \in i} \varepsilon_{\bond{ij}} \xi_{ij} \\
  & & + \chi_i \Bigl( 6\chi_i - (\chi_{i,\tau+1}+\chi_{i,\tau-1})
      -\sum_{j \in i} (\chi_j + \varepsilon_{\bond{ij}}\lambda_j) \Bigr) \nonumber \\
  & & + \lambda_i \Bigl( 6\lambda_i - (\lambda_{i,\tau+1}+\lambda_{i,\tau-1})
      -\sum_{j \in i} (\lambda_j + \varepsilon_{\bond{ij}}\chi_j) \Bigr) 
      \Bigr\rbrack \nonumber \\
  & = & \bh^T \bC \bh + ( \bh^T \bx + \bx^T \bh )
      \ , \nonumber      
\end{eqnarray}
where the vectors $\bh$ and $\bx$ are arranged as:
\begin{eqnarray}\label{Vect}
\bh & = & \lbrack \ \cdots \ ()_{i'} \ 
    (\chi_i,\lambda_i)_i \ ()_{i''} \ \cdots \ \rbrack^T \\
\bx & = & \Biggl\lbrack \ \cdots \ ()_{i'} \ 
    \Biggl( \sum_{j \in i} \xi_{ij}, 
    \sum_{j \in i} \varepsilon_{\bond{ij}} \xi_{ij} \Biggr)_i \
    ()_{i''} \ \cdots \ \Biggr\rbrack^T \ . \nonumber
\end{eqnarray}

It is crucially important to understand the properties of the saddle-point vectors $\bx$ and the coupling matrix $\bC$. They follow in a straight-forward manner from ~(\ref{DHMatrAct}) and ~(\ref{Vect}), but due to tediousness of algebra, we deffer derivation to the Appendix \ref{AlgebraApp}. Here we will only summarize the results:
\begin{itemize}
\item all saddle-point vectors $\bx$ have the same norm:
\begin{equation}\label{VectNorm}
\bx^T\bx = \textrm{const.} \ ;
\end{equation}
\item all saddle-point vectors $\bx$ are degenerate eigenvectors of the coupling matrix $\bC$:
\begin{equation}\label{EigenVect}
\bC\bx = 6\bx \ .
\end{equation}
\end{itemize}

\begin{figure}
\subfigure[{}]{\includegraphics[width=1.2in]{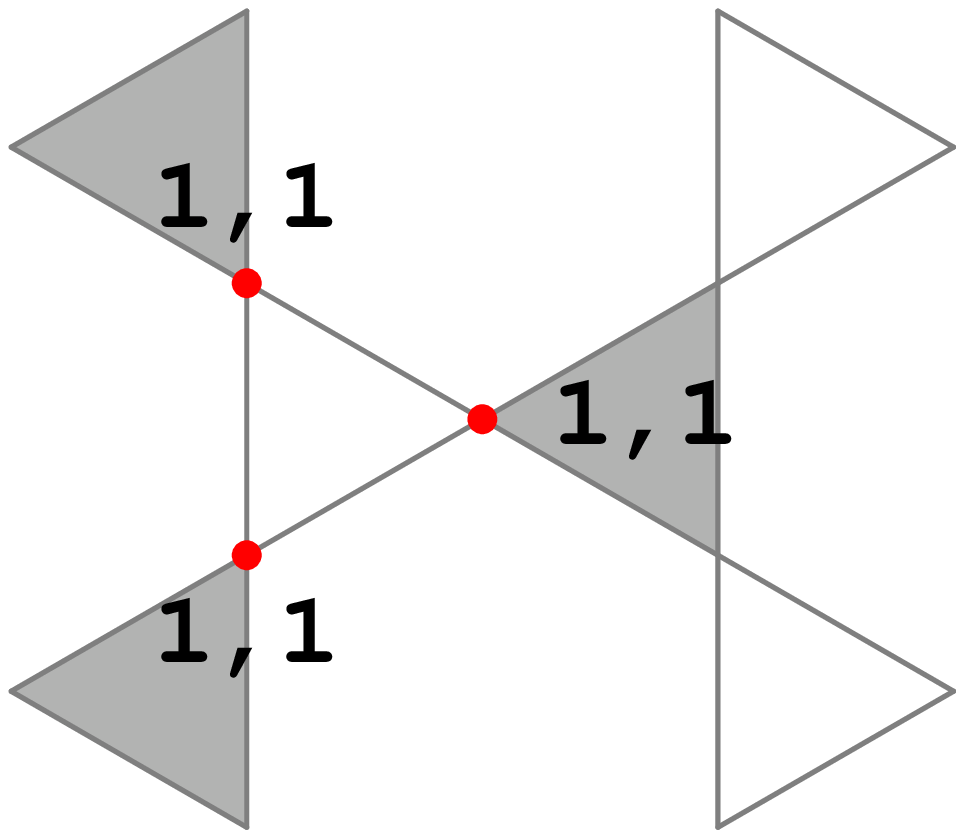}}
\subfigure[{}]{\includegraphics[width=1.2in]{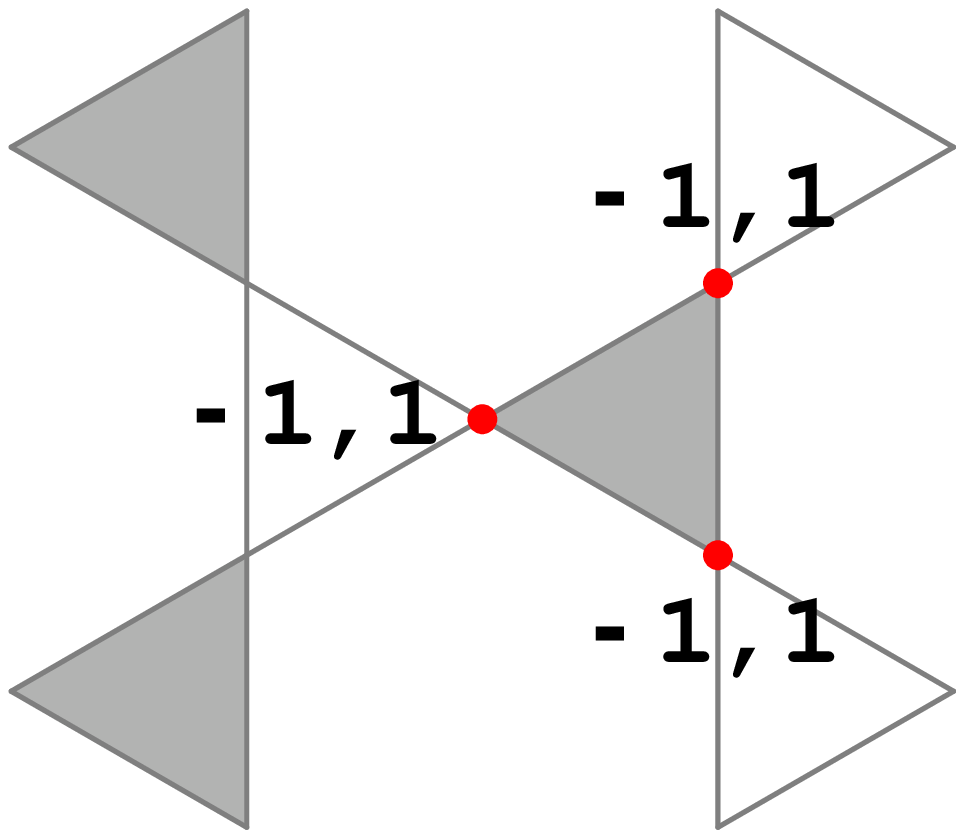}}
\vskip -4mm
\caption{\label{ZeroModes}Unphysical fluctuations in the lattice field theory ~(\ref{DHAction}). The pairs of numbers in the figure indicate all non-zero values of $(\chi_i, \lambda_i)$ that constitute a redundant integer-valued field configuration (up to a multiplicative integer constant). One can easily see from ~(\ref{U1MaxwellSol}) that arbitrary superposition of these configurations does not affect the bond energy $E_{ij}$ on any bond. $\varepsilon_{\bond{ij}}$ is $-1$ on the bonds of the shaded triangles, and $+1$ on the other bonds. }
\end{figure}

In fact, the coupling matrix $\bC$ is completely dispersionless: its eigenvalues have only frequency dependence, and no dependence on spatial wavevectors. If the height fields were real instead of integer numbers, there would be six localized bare modes per Kagome lattice unit-cell. Two would be ``gapless'' (zero eigenvalue at zero frequency) and degenerate, while the other four would be ``gapped'' and degenerate. The ``gapless'' bare modes are actually unphysical, and merely a redundancy of the state representation in terms of the height fields. Exciting them at arbitrary places and frequencies does not affect at all the only physical quantity in the problem, the bond energy $E_{ij}$ (see Fig.~\ref{ZeroModes}). Their existence can also be confirmed by counting arguments: there are six bond variables ($E_{ij}$), and two constraints on the Kagome triangles (Gauss' Law) per unit-cell, leaving only four independent variables per unit-cell. Also note that because of ~(\ref{EigenVect}) the gapless bare modes do not couple to the saddle-point vectors $\bx$ in ~(\ref{DHMatrAct}). In other words, they behave much like some ``gauge'' degrees of freedom.

\subsection{Effect of Fluctuations}\label{TFIMfluct}

The lattice theory ~(\ref{DHAction2}) that we want to analyze is very similar to the simple and well understood integer-valued height model. In $2+1$ dimensions the height model is known to order and give a ``smooth'' phase. This means that fluctuations of the height field $\chi_i$ are such that the average $\la (\chi_i - \chi_j)^2 \ra$ does not diverge as the sites $i$ and $j$ go far apart. The question that we now ask is whether the Kagome double-height theory ~(\ref{DHAction2}) also lives in a ``smooth'' phase, and what kind of lattice symmetry breaking (if any) is obtained. We will try to find answers in a fashion inspired by the Reference. \cite{DimerSG} First we investigate which microstates are most frequently visited by the system as it fluctuates. This is not a trivial problem for the frustrated magnets, since the microstates that minimize the action are macroscopically degenerate and only the entropy of quantum fluctuations can distinguish them. If we somehow found that the preferred microstates possessed long-range order, we would have to verify whether that order is truly stable against fluctuations. Note that the formal lack of dispersion in the coupling matrix $\bC$ does not automatically signal localization, and hence the ``rough'' phase: the bare modes are strongly interacting in order to give only integer-valued height fields on all sites.

Finding the microstates that are most frequently visited by the system is an extension of the standard mean-field approach, where such states simply minimize the energy. The appropriate quantity to minimize in general circumstances is a ``free energy''. Physically, when the ground state is a superposition of many different states (microstates), the ones with large probability amplitudes squared will be characterized by small ``free energy''. The probability amplitude of a state $\psi$ can be in principle calculated from the imaginary-time (or finite temperature) path-integral:
\begin{eqnarray}
|a_{\psi}|^2 & = & \textrm{tr} \Bigl( e^{-\int\dd\tau H} | \psi \ra \la \psi | \Bigr) \\
  & \propto & \sum_{\{ \psi(\tau) \}} | \la \psi | \psi(\tau_0) \ra |^2 e^{-S(\{\psi(\tau)\})} 
  \nonumber \ .
\end{eqnarray}
Here, $\psi(\tau)$ denotes entire microscopic field configuration at the imaginary-time instant $\tau$, which corresponds to a quantum state. The overlap $| \la \psi | \psi(\tau_0) \ra |^2$ is taken at an instant $\tau_0$ for which $| \psi \ra \la \psi |$ is specified in the Heisenberg picture. A fundamental property of the overlap is that it becomes larger when the state $| \psi(\tau_0) \ra$ is more ``similar'' to the state $|\psi\ra$ (and very sharply so if the diagonal representation is used to construct the path-integral). We will use this property in order to simplify calculation and only estimate the probability of the system being in a neighborhood of the state $\psi$. Having in mind that we are looking for a potential long-range order, which would unavoidably imply existence of a static order parameter, we can seek probability of the system spending an extended period of time in a neighborhood of the state $\psi$. The quantity that we will consider is:
\begin{equation}
e^{-F(\psi)} \propto \sum_{\{ \psi(\tau) \}} \prod_{\tau} \theta \bigl( \psi, \psi(\tau) 
  \bigr) e^{-S(\{\psi(\tau)\})} \ .
\end{equation}
The neighborhood is specified by the function $\theta(\psi, \psi')$ that becomes smaller when the states $\psi$ and $\psi'$ become more different. Physically, $F(\psi)$ is the free energy associated with the fluctuations from the vicinity of $\psi$. With a proper choice of the neighborhood function, the states with smaller free energy are more energetically and entropicaly favored.

We now turn to our specific problem given by the action ~(\ref{DHAction2}). Recall that the saddle-point vectors $\bx$ correspond to static spin states. Since the action is expanded about the saddle-point $\bx$ in the expression ~(\ref{DHAction2}), any non-zero values of the height fields mean moving away from that saddle-point. This allows us to define the neighborhood function $\theta$ in a soft, but controlled way:
\begin{equation}
\prod_{\tau} \theta(\psi, \psi') = \exp(-gm^2 \bh^T\bh) \ ,
\end{equation}
where $m^2$ is a tunable parameter controlling the neighborhood size. Then, the free energy of a state is:
\begin{eqnarray}\label{FreeEnergy}
e^{-F(\bx)} & = & \sum_{\bh} e^{-S'(\bh;\bx)} \\
  & = & \sum_{\bh} e^{ 
  -g\bigl[\bh^T \bC \bh + (\bh^T\bx + \bx^T\bh) + m^2 \bh^T\bh \bigr] } \ . \nonumber
\end{eqnarray}
Without the ``mass'' term $m^2$ the free energy would not depend on the saddle-points, because mere shifts of variables in the path integral that leave the action invariant would switch between them. This addition to the theory does not alter its fundamental properties. The physical bare modes are ``gapped'' to begin with, and the mass only changes the gap in the coupling matrix. On the other hand, the ``gapless'' unphysical bare modes do not couple to the physical degrees of freedom, and giving them mass is only a convenient way to handle them. Also note that the neighborhood function does not alter the nature of dispersion in the action, and thus cannot introduce unwanted effects at the neighborhood boundary.

We can make progress in two limits: very small and very large coupling constant $g$. For $g \ll 1$ the summation over integer fields $\bh$ in ~(\ref{FreeEnergy}) can be approximated by an integration:
\begin{eqnarray}\label{SmallG}
e^{-F(\bx)} & \approx & \int_{-\infty}^{\infty} \mD\bh e^{-S'(\bh;\bx)} \\
  & \propto & \exp \Bigl\lbrack g\bx^T(\bC+m^2)^{-1}\bx \Bigr\rbrack \ . \nonumber
\end{eqnarray}
However, since all the saddle-point vectors are degenerate eigenvectors of the coupling matrix ~(\ref{EigenVect}) and have the same normalization ~(\ref{VectNorm}), the free energy will have no dependence on the saddle-points. This limit together with large $m^2$ corresponds to the calculation from the Reference, \cite{DimerSG} only performed to all orders of $1/m^2$.

For $gm^2 \gg 1$ it is convenient to perform the Poisson resummation in ~(\ref{FreeEnergy}). Let us introduce a vector $\bm$ with integer components $\mu^{(k)}_i$ ($k=1,2$), and write:
\begin{eqnarray}\label{LargeG}
e^{-F(\bx)} & = & \sum_{\bm}\int\limits_{-\infty}^{\infty} \mD\bh 
  \exp \Bigl\lbrack -S'(\bh;\bx) - i\pi(\bm^T\bh+\bh^T\bm) \Bigr\rbrack \nonumber \\
  & \propto & \sum_{\bm} \exp \Bigl\lbrack
    g \Bigl( \bx+\frac{i\pi}{g}\bm \Bigr)^T (\bC+m^2)^{-1} 
      \Bigl( \bx+\frac{i\pi}{g}\bm \Bigr) \Bigr\rbrack 
    \nonumber 
\end{eqnarray}
This expression simplifies considerably due to ~(\ref{VectNorm}) and ~(\ref{EigenVect}):
\begin{eqnarray}\label{FreeEnergy2}
e^{-F(\bx)} & \propto & \sum_{\bm} \exp \Bigl\lbrack 
  -\frac{\pi^2}{g}\bm^T(\bC+m^2)^{-1}\bm \\
  & & +\frac{i\pi}{6+m^2}(\bm^T\bx+\bx^T\bm) \Bigr\rbrack \ . \nonumber
\end{eqnarray}
Note that we could also add an explicit ``vortex core'' term $u\bm^T\bm$ to the free energy. It would soften the integer-valued constraints for the height fields, and yield a sine-Gordon theory in the large $u$ limit. Doing this would be useful if we needed to discuss stability of phases, but for the purposes of present problem, this will prove to be unnecessary. The smallest eigenvalue of the matrix $\bC+m^2$ in the last equation is $m^2$. Therefore, in the limit of $gm^2 \gg 1$ (and $u \ll 1$), the only rapidly varying part of the exponent on the right-hand side is the purely imaginary part. We can easily understand the oscillatory effect that it induces as long as $m^2 \ll 1$. First, note that the components of the saddle-point vectors $\bx$ always have integer values: all possibilities are shown in the Table ~\ref{SPBowties}. Then, we can decompose the integer-valued components of the vector $\bm$ into two parts:
\begin{equation}
\mu^{(k)}_i = 6\mathcal{M}^{(k)}_i + \delta\mu^{(k)}_i \quad , \quad \delta\mu^{(k)}_i 
   \in \lbrace 0 \dots 5 \rbrace \ .
\end{equation}
Since the quadratic part of the exponent in ~(\ref{FreeEnergy2}) varies only very slowly, we can neglect fluctuations of $\delta\mu$ in it. Similarly, since $m^2 \ll 1$, we can neglect fluctuations of $\mathcal{M}$ in the oscillatory part (they approximately contribute a $2\pi\times\textrm{integer}$ phase). We approximately have:
\begin{eqnarray}\label{TFIMlargeG}
e^{-F(\bx)} & \propto & \sum_{\boldsymbol{\mu}} \exp \Bigl\lbrack -\frac{(6\pi)^2}{g}
  \boldsymbol{\mathcal{M}}^T(\bC+m^2)^{-1}\boldsymbol{\mathcal{M}} \nonumber \\
  & & +\frac{i\pi}{6+m^2}(\boldsymbol{\delta\mu}^T\bx+\bx^T\boldsymbol{\delta\mu})
  \Bigr\rbrack \ .
\end{eqnarray}
Clearly, the oscillatory part will give rise to a destructive interference for every non-zero component of the saddle-point vector $\bx$ (yielding factors of the order of $m^2$, or $(gm^2)^{-1}$ in the path-integral weight). In order to minimize the free energy $F(\bx)$, the saddle-point vector $\bx$ should have as many zero components as possible. The appropriate quantity to consider is:
\begin{equation}
n_p = 2n_c + n_a \ ,
\end{equation}
where $n_a$, $n_b$ and $n_c$ are respectively the total numbers of the A, B, and C type sites from the Table ~\ref{SPBowties} in a saddle-point vector. Configurations that maximize $n_p$ are preferred, and entropically selected by fluctuations.

\begin{figure}
\subfigure[{}]{\includegraphics[width=1.5in]{tfim-stripe.eps}}
\subfigure[{}]{\includegraphics[width=1.5in]{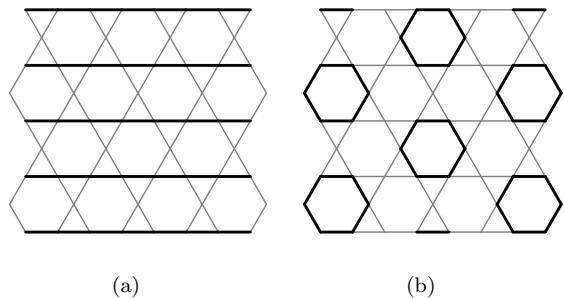}}
\vskip -2mm
\caption{\label{MaxFlip}Two periodic (not typical) preferred dimer configurations (with maximum flippability). Every site with two dimers emanating from it holds a flippable spin, and only every third site holds an unflippable spin.}
\end{figure}

Simple algebra can be worked out to find $n_p$. The total number of Kagome lattice sites is:
\begin{equation}\label{N1}
n_a + n_b + n_c = N \ ,
\end{equation}
while the total number of dimers in a least frustrated state is:
\begin{equation}\label{N2}
\frac{1}{2} (2n_a + n_b) = \frac{2N}{3} \ ,
\end{equation}
since one dimer sits on every Kagome triangle, and the triangles share corners instead of bonds. Combining these two equations we find:
\begin{equation}\label{N3}
n_a = \frac{N}{3} + n_c \ , \qquad n_b = \frac{2N}{3} - 2n_c \ .
\end{equation}
In order for all $n_a$, $n_b$, and $n_c$ to be positive and smaller than $N$, $n_c$ must be bounded between $0$ and $N/3$. We see that $n_p$ is maximized simply when the number of C-sites is maximized, which means: $n_a=2N/3$, $n_b=0$, $n_c=N/3$. Note that the preferred configurations also have the maximum number of \emph{flippable} spins, whose flipping costs no energy. Every A-site is a flippable spin, because the numbers of frustrated and unfrustrated bonds emanating from it are equal (flipping a spin toggles bond energy on every emanating bond). Some untypical preferred configurations are shown in the Fig.~\ref{MaxFlip}.

\begin{figure}
\includegraphics[width=2.6in]{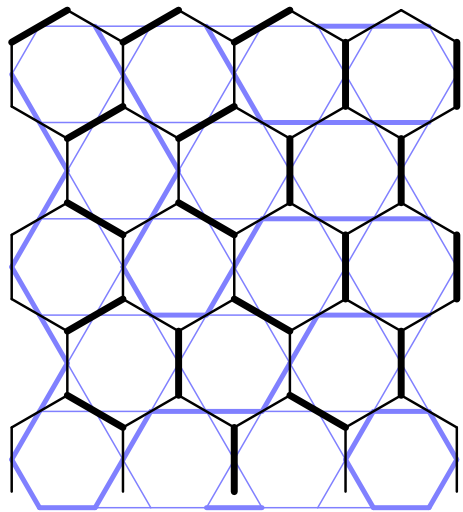}
\vskip -2mm
\caption{\label{MaxFlipHoneycomb}The maximally flippable states map to the honeycomb lattice hard-core dimer coverings. The Kagome triangle centers form a honeycomb lattice whose bonds go through the Kagome sites. For every Kagome bowtie with a C element from the Table \ref{SPBowties} put a dimer on the honeycomb lattice.}
\end{figure}

The total number of preferred configurations is macroscopically large. This can be demonstrated by observing that they map to the hard-core dimer coverings of the \emph{honeycomb} lattice. The preferred configurations have only A and C type sites. Two C-sites cannot be neighbors, but their number should be maximized, so that every C-site can be represented by a dimer on the corresponding \emph{honeycomb} lattice bond, as depicted in the Fig.~\ref{MaxFlipHoneycomb}. A transition graph can be found by overlapping any two \emph{honeycomb} lattice dimer coverings, and it consists of isolated loops, the smallest having six \emph{honeycomb} bonds. Therefore, the preferred configurations on the Kagome lattice are only locally different from one another, and may be transformed into one another by flipping six or more flippable spins (one at a time).

\begin{table}
\begin{displaymath}
\begin{array}{|c||c|c|c|c|}
\hline
& \textrm{A}_1 \includegraphics[height=0.6cm]{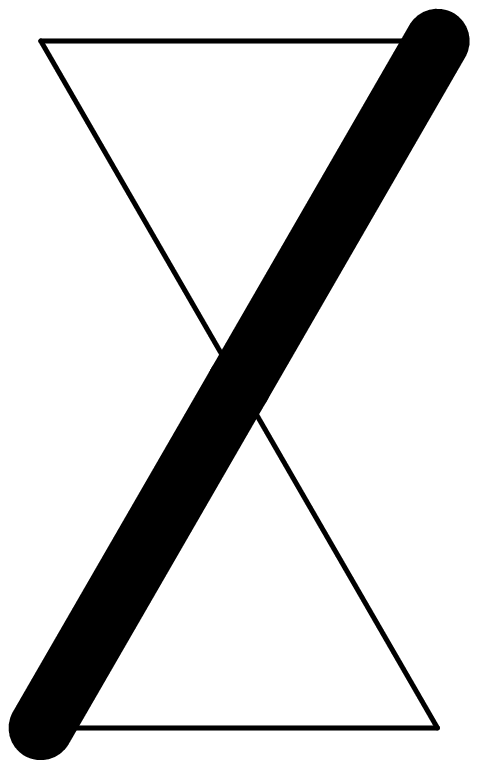}
& \textrm{A}_2 \includegraphics[height=0.6cm]{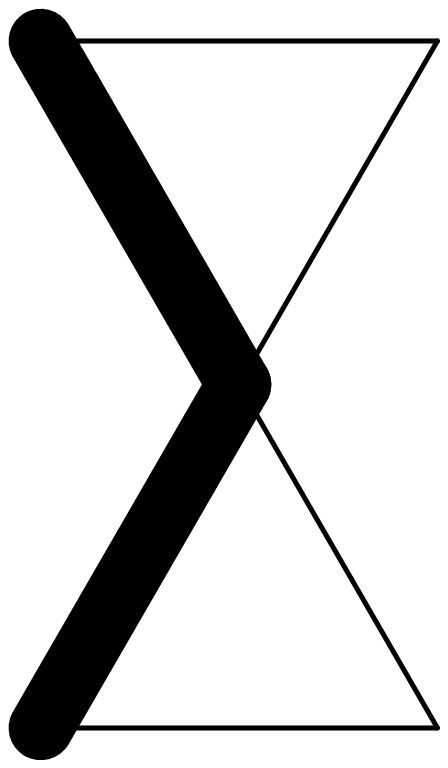}
& \textrm{B} \includegraphics[height=0.6cm]{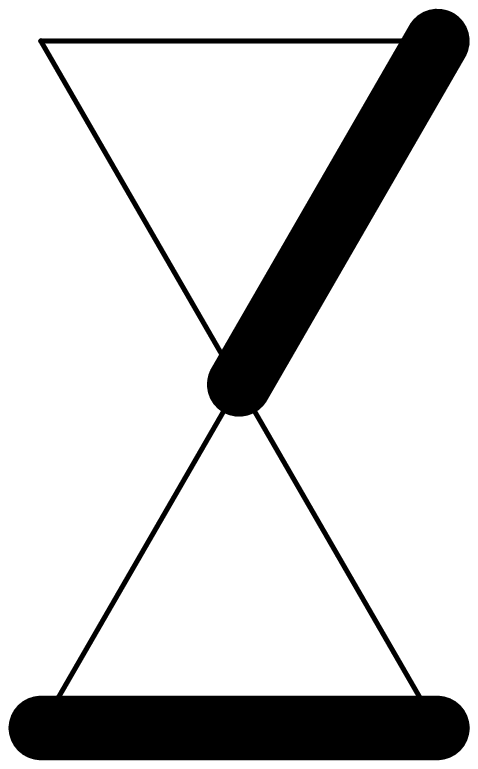}
& \textrm{C} \includegraphics[height=0.6cm]{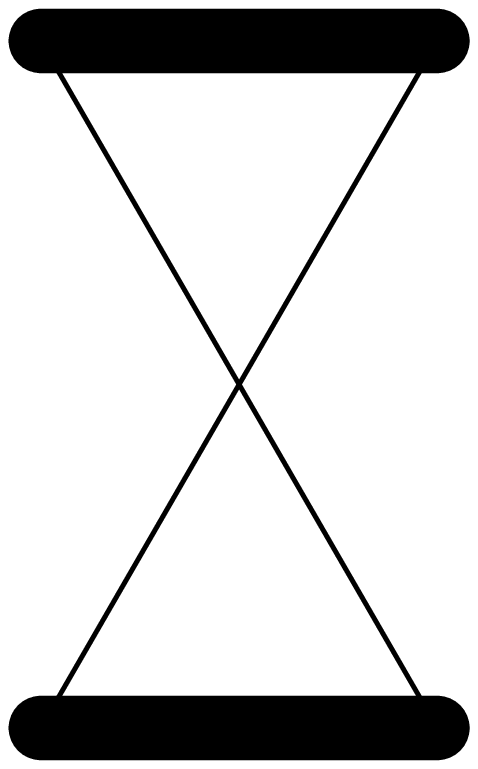}
\\ \hline \hline
\sum\limits_{j \in i}\xi_{ij}                        & \pm 2 &   0   & \pm 1 &   0   \\
\hline
\sum\limits_{j \in i}\varepsilon_{\bond{ij}}\xi_{ij} &   0   & \pm 2 & \pm 1 &   0   \\
\hline
\end{array}
\end{displaymath}
\caption{\label{SPBowties}All possible local configurations of dimers (frustrated bonds) at the saddle-points, and the corresponding values of the saddle-point vector components. The site $i$ sits at the center of the bowtie, and may be of type A (two varieties), B, or C. The description of the saddle-points in terms of dimers is given in the section ~\ref{TFIMprop}.}
\end{table}

\subsection{``Disorder-by-Disorder''}\label{TFIMdbd}

The conclusion so far is that the maximally flippable states may be entropicaly selected by fluctuations, to smaller or greater extent depending on the value of $g$. If we naively traced back connections between $g$ and the parameters of the original spin model, we would find that, generally speaking, smaller $g$ describes stronger further-neighbor and multiple-spin exchange processes.

In principle, the maximally flippable states will be mixed together in the ground state, since it takes only local fluctuations to change between them. Our findings so far have included only the effects of ``small'' fluctuations about various saddle-points. Even though a smaller value of $m^2$ can always be chosen to expand the scope of included fluctuations, too small $m^2$ may invalidate the approximations that made the calculations possible. Hence, certain ``large'' fluctuations are beyond reach of this formalism. Exactly these fluctuations decide whether the maximally flippable states are evenly mixed with all other states or not. If they are, the ground state is chaotic and disordered. The only chance for a valence bond order is if something suppressed the fluctuations into all but the maximally flippable states. Then, the effective degrees of freedom would be only the honeycomb lattice dimer coverings, and their local dynamics would yield a plaquette dimer long-range order typical of dimer models on bipartite lattices. It is our goal here to determine whether such long-range order might be stable.

There are several arguments that can be made in favor of the disordered phase. The first thing to observe is that the minimum of the free energy is extremely widely and evenly distributed over a large number of disordered states (mappable to the honeycomb lattice dimer coverings). This means that the system does not spend much time fluctuating near any particular one of them, likely ruling out a static order parameter, and thus lattice symmetry breaking. Secondly, due to the degeneracy in the action, it is possible to make local field changes that correspond to flipping a single spin without paying energy on the spatial links. Since such small fluctuations of the flippable spins are energetically controlled only along one dimension (imaginary time), there is nothing to stop them from proliferating. This would be true even if the ``smooth'' phase were obtained in the ``height'' action: the macroscopic degeneracy due to geometric frustration allows many spatially different patterns that break the ``height'' symmetry. If fact, every least frustrated state has a macroscopic number of flippable spins (see from ~(\ref{N3}) that $n_a \geqslant \frac{N}{3}$). Therefore, there is no mechanism to suppress fluctuations into any possible state. Note that for all but the maximally flippable states to be suppressed, the only favorable flipping processes would have to simultaneously involve at least six spins (dimer flip on a \emph{honeycomb} lattice hexagon). 

Another consequence of abundant single-spin fluctuations is absence of magnetic order in the ground state. In our problem this also contradicts possibility of the valence bond order, since it would be accompanied by a net Ising moment. All maximally flippable states have macroscopic magnetization $M = \pm \frac{N}{3}$. To see this, note in the Fig.~\ref{MaxFlip} that all flippable spins (A-type sites) must be aligned, since they are connected to each other either through one frustrated bond (dimer), or through two unfrustrated bonds.

It is apparent by now that all minimally frustrated spin configurations are mixed into the disordered and featureless ground state. The correlations are short-ranged since there is a macroscopic number of flippable spins in every least frustrated state, making the spins virtually independent. A property that distinguishes the Kagome from the other lattices is the formal lack of dispersion in the lattice field theory. We interpret this as a signal that the excitations are very heavy or perhaps even localized (exactly true in the small-$g$ limit). Indeed, strictly short-ranged spin-spin correlations, consistent with very undispersive modes, have been observed in the Monte Carlo simulations. \cite{FrIsing1, FrIsing2}

\subsection{Quantum-Mechanical Interpretation}\label{TFIMquant}

The analysis of the lattice field theory has yielded two essential results that can help us sketch the ground and excited states of the Hamiltonian ~(\ref{DiceDimerModel}). They are: a) no symmetry is spontaneously broken, b) the excitations have localized character (very large effective mass). We use the free energy $F(\bx)$ from the previous section as a rough indication of the probability amplitudes that different dimer configurations have in the ground state. The ground state is a smooth superposition of all possible configurations $|\psi\ra$ of frustrated bonds:
\begin{equation}\label{TFIMGnd}
|0\ra = \sum_{\psi} a_{\psi} |\psi\ra \ .
\end{equation}
The amplitudes of the similar states are roughly equal in magnitude. This is required in order for two states different by a single spin flip to give a large matrix element $\la\psi_1| (-\Gamma S^x) |\psi_2\ra$ and yield a significant energy gain. However, the amplitudes depend on flippability of the states $|\psi\ra$. The state with a larger number of flippable spins will have a larger probability $|a_{\psi}|^2$. 

Due to a very localized nature of excitations, we can say that the physics of this model is very similar to the physics of completely disconnected quantum spins in transverse field, for which all eigenstates are known. Although the actual flippable spins interact, their interaction seems to be largely inconsequential. This suggests that many good variational wavefunctions (for ground and excited states at $\Gamma \ll J_z$) can be obtained by a simple Gutzwiller's projection: take the states of the non-interacting Kagome spins in a transverse field and project them to the manifold of least frustrated states. All excitations are gapped, and the gap is $\sim \Gamma$.

Finally, we recall that this disordered quantum phase is not topological in the original spin model. Clearly, it is stable against small higher order perturbations in $\Gamma / J_z$. In fact, it obtains for all values of the transverse field $\Gamma$, without any intermediate phase transitions. \cite{KagIsing, FrIsing1, FrIsing2}


\section{Heisenberg Model With Easy Axis}\label{eahm}

In this section we analyze the XXZ model on the Kagome lattice, and its extensions:
\begin{equation}\label{EAHM2}
H = J_z \sum_{\bond{ij}} S^z_i S^z_j + J_{\bot} 
  \sum_{\bond{ij}} \bigl( S^x_i S^x_j + S^y_i S^y_j \bigr) \ ,
\end{equation}
with $J_z, J_{\bot} > 0$, and $J_{\bot} \ll J_z$. The analysis will closely follow that of the transverse field Ising model in the section ~\ref{tfim}, and rely on the notation and conventions defined there. Many similarities will be encountered, except that the formalism will be of greater complexity. One apparent difference, however, is that the total magnetization in $z$ direction is a good quantum number in this model.

\begin{figure}[b]
\includegraphics[width=1in]{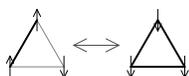}
\vskip -2mm
\caption{\label{ThreeFrust}Flipping a pair of aligned spins creates extra frustration on the triangle that contains them. Dimers denote the frustrated bonds.}
\end{figure}

As before, we begin by considering an effective theory that describes the physics at the energy scales well below $J_z$:
\begin{equation}\label{EAHMeff}
H_{\textrm{eff}} = \frac{J_{\bot}}{2} \sum_{\bond{ij}} \mathcal{P}_0
  \bigl( S^+_i S^-_j + S^-_i S^+_j \bigr) \mathcal{P}_0 + 
  \mathcal{O} \Bigl( \frac{J^2_{\bot}}{J_z} \Bigr) \ .
\end{equation}
This theory lives in the Hilbert space spanned by the least frustrated states of the pure Ising model, and $\mathcal{P}_0$ is the projection operator to this space. It describes dynamics of the flippable spin-pairs on the Kagome lattice bonds. In order for a pair of spins to be flippable, the spin configuration must be minimally frustrated before and after the pair is flipped. Note that this automatically requires that the two spins be antialigned (Fig.~\ref{ThreeFrust}). The effective theory can again be expressed as a soft-core dimer model in the same Hilbert space as the one that described the TFIM model, but with more complicated dynamics. We will reformulate it as a U(1) gauge theory, derive a dual lattice field theory for it, and explore the possible phases. This time we will find a valence bond crystal and a spin liquid.

\subsection{U(1) gauge theory}

\begin{figure}
\subfigure[{}]{\includegraphics[width=2.3in]{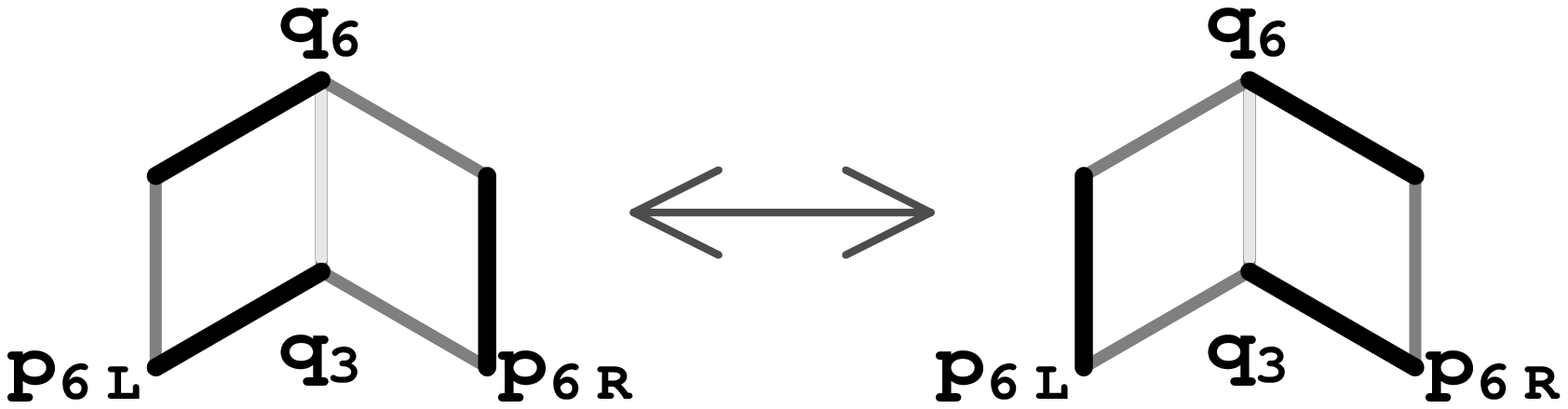}}
\vskip -3mm
\subfigure[{}]{\includegraphics[width=2.3in]{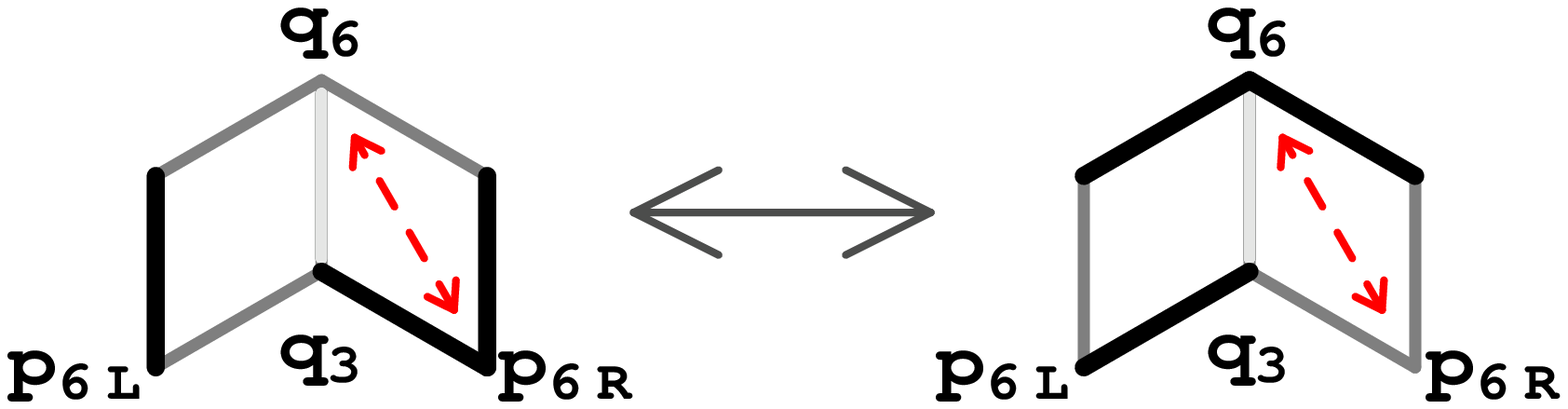}}
\vskip -3mm
\subfigure[{}]{\includegraphics[width=2.3in]{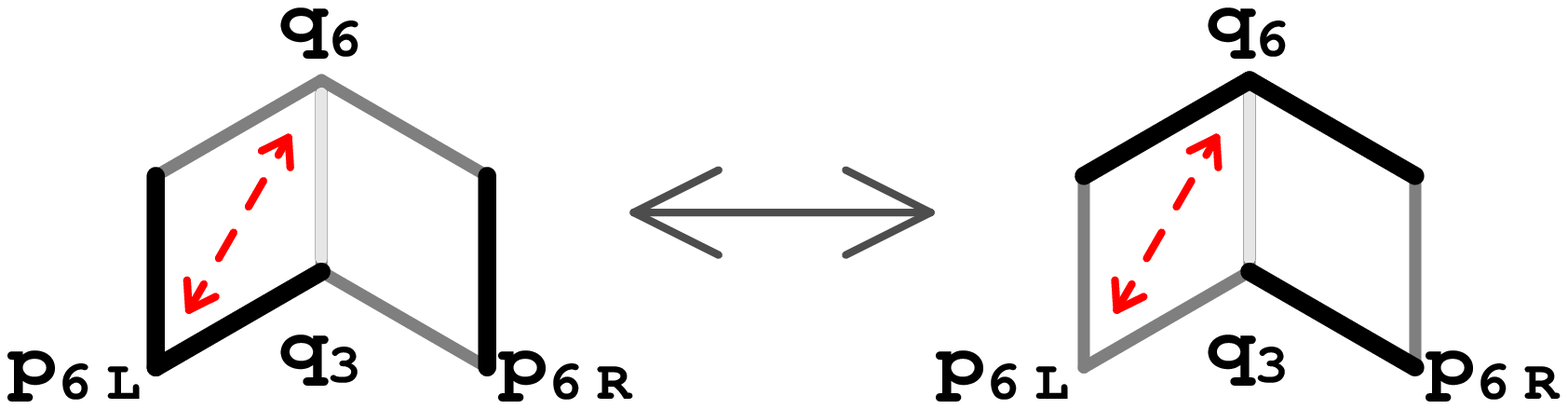}}
\vskip -3mm
\subfigure[{}]{\includegraphics[width=2.3in]{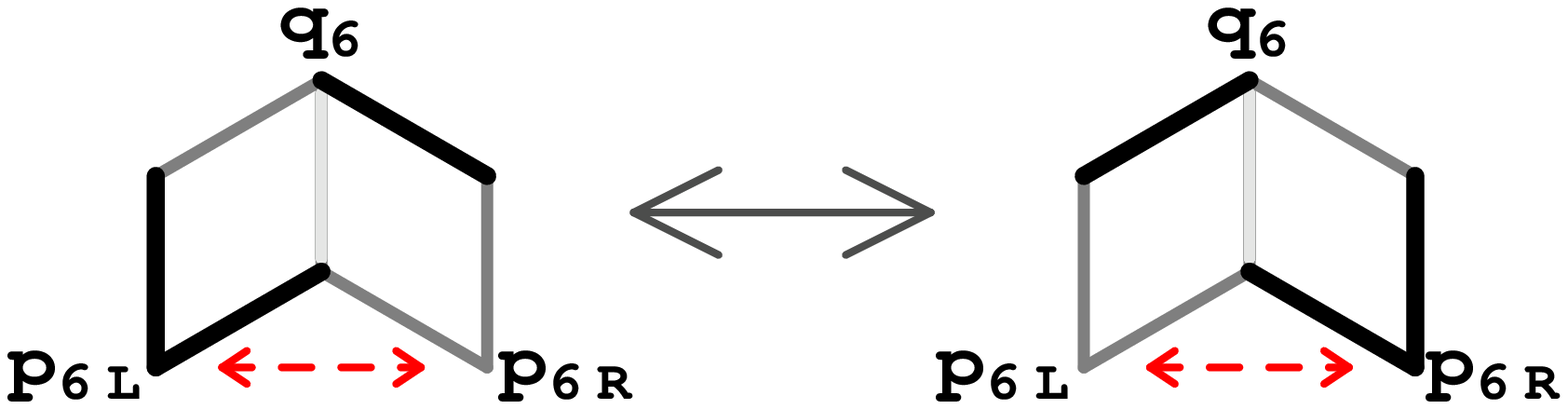}}
\vskip -3mm
\caption{\label{TwoDiceProc}Four possible low energy processes that keep frustration at the minimum. Recall that the minimum of frustration is achieved if there is one dimer emanating from every 3-coordinated dice site, and an even number of dimers emanating from every 6-coordinated site. The dashed arrows show between which two 6-coordinated sites a pair of dimers is exchanged (charge-2 boson hopping in the U(1) gauge theory). These processes preserve the global Ising magnetization: the bond between the two plaquettes is always unfrustrated.}
\end{figure}

The only difference between the XXZ effective dimer model and that of the TFIM model ~(\ref{DiceDimerModel}) is that now the elementary loops on which the dimers can be flipped enclose two dice lattice plaquettes, instead of one (since two spins are flipped at a time). There are four such processes, and they are shown in the Fig.~\ref{TwoDiceProc}.

\begin{table*}
\begin{displaymath}
\begin{array}{l @{\qquad , \qquad} l}
H^{(a)}_{\textrm{kin}} = J_{\bot} \sum\limits_{\DDPL^{}} 
   \cos \bigl( h^{(a)}_{\bond{q_6q_3}} \bigr) & h^{(a)}_{\bond{q_6q_3}} = 
   \varepsilon_{\bond{q_6q_3}} \sum\limits_{pq}^{\DDPLA^{}} \mathcal{A}_{pq} + 
   \varepsilon_{\bond{q_6q_3}} \sum\limits_{pq}^{\DDPLB^{}} \mathcal{A}_{pq} 
   \nonumber \\
H^{(b)}_{\textrm{kin}} = J_{\bot} \sum\limits_{\DDPL^{}}
   \cos \bigl( h^{(b)}_{\bond{q_6q_3}} \bigr) & h^{(b)}_{\bond{q_6q_3}} = 
   \varepsilon_{\bond{q_6q_3}} \sum\limits_{pq}^{\DDPLA^{}} \mathcal{A}_{pq} +
   \biggl( \bigl( \varphi_{q_6} - \varphi_{p_{6R}}  \bigr) \eta_{p_{6R}q_6} + 
      \sum\limits_{pq}^{\DDPLB^{}} \varepsilon_{\bond{pq}} \mathcal{A}_{pq} \biggr)
   \nonumber \\
H^{(c)}_{\textrm{kin}} = J_{\bot} \sum\limits_{\DDPL^{}}
   \cos \bigl( h^{(c)}_{\bond{q_6q_3}} \bigr) & h^{(c)}_{\bond{q_6q_3}} = 
   \biggl( \bigl( \varphi_{p_{6L}} - \varphi_{q_6}  \bigr) \eta_{q_6p_{6L}} + 
      \sum\limits_{pq}^{\DDPLA^{}} \varepsilon_{\bond{pq}} \mathcal{A}_{pq} \biggr) +
   \varepsilon_{\bond{q_6q_3}} \sum\limits_{pq}^{\DDPLB^{}} \mathcal{A}_{pq} 
   \nonumber \\
H^{(d)}_{\textrm{kin}} = J_{\bot} \sum\limits_{\DDPL^{}}
   \cos \bigl( h^{(d)}_{\bond{q_6q_3}} \bigr) & h^{(d)}_{\bond{q_6q_3}} = 
   \biggl( \bigl( \varphi_{p_{6L}} - \varphi_{q_6}  \bigr) \eta_{q_6p_{6L}} + 
      \sum\limits_{pq}^{\DDPLA^{}} \varepsilon_{\bond{pq}} \mathcal{A}_{pq}  \biggr) +
   \biggl( \bigl( \varphi_{q_6} - \varphi_{p_{6R}}  \bigr) \eta_{p_{6R}q_6} + 
      \sum\limits_{pq}^{\DDPLB^{}} \varepsilon_{\bond{pq}} \mathcal{A}_{pq} \biggr)
   \nonumber
\end{array}
\end{displaymath}
\caption{\label{DimerFlips2}Kinetic energy operators corresponding to the processes in the Fig. ~\ref{TwoDiceProc}. Notation for the sites is defined in the figure: $\bond{q_6q_3}$ is the bond shared between the two dice plaquettes, while $p_{6L}$ and $p_{6R}$ are the bottom 6-coordinated sites on the left and right plaquette respectively. Only the bonds with arrows are included in the sums.}
\end{table*}

The U(1) gauge theory is built the same way as in the section ~\ref{TFIMU1}. The electric field and the charge-2 bosons, whose fluctuations are controlled by the Gauss' Law ~(\ref{U1Gauss}) and the potential energy ~(\ref{HU}), represent the low energy degrees of freedom. The new form of the kinetic energy can be easily obtained by comparing the two-plaquette processes in the Fig.~\ref{TwoDiceProc} with the single-plaquette processes in the Fig.~\ref{DimerFlip1}. The two single-plaquette processes consistent with the low energy physics can be combined in four different ways to give the allowed two-plaquette processes. In combining them, the middle bond $\bond{q_6q_3}$ is flipped twice, so that there is no net change on it. This gives us the operators in the Table ~\ref{DimerFlips2}. The argument of each cosine is the sum of two corresponding single-plaquette circulations and boson hopping(s) from the expressions ~(\ref{DimerFlipA}) and ~(\ref{DimerFlipB}), but multiplied by the factors of $\eta_{\bond{pq}}$ and $\varepsilon_{\bond{pq}}$ in such a way that the contribution of the central bond $\bond{q_6q_3}$ is properly canceled out. Then, if we label the four processes by $\alpha=a,b,c,d$, the U(1) effective Hamiltonian is:
\begin{equation}\label{XXZU1}
H = U \sum_{\bond{pq}} E_{pq}^2
  + \sum_{\alpha} H^{(\alpha)}_{\textrm{kin}} \ .
\end{equation}
The $U \to \infty$ limit is an exact rewriting of the effective dimer model, while a finite $U$ introduces various new dynamical processes, defined on larger dimer loops (spin clusters), but consistent with the global spin-flip symmetry of the XXZ model.

\subsection{Lattice Field Theory}\label{EAHMft}

\begin{figure}
\centering
\subfigure[{}]{\includegraphics[width=0.8in]{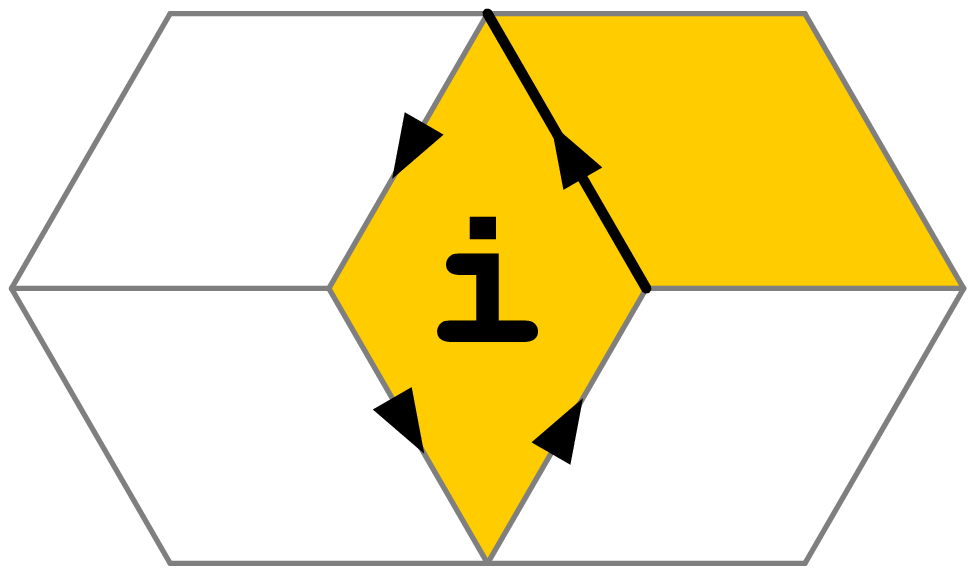}}
\subfigure[{}]{\includegraphics[width=0.8in]{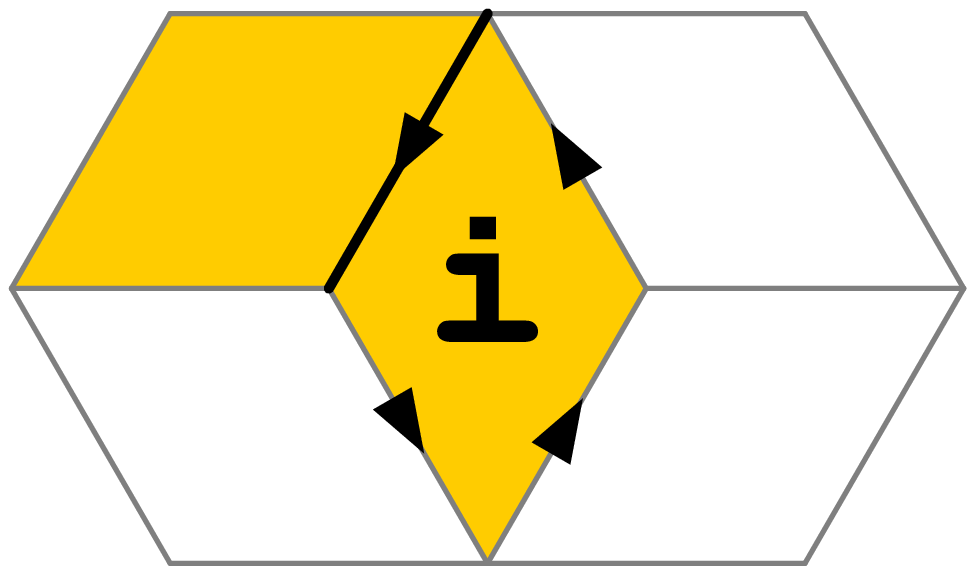}}
\subfigure[{}]{\includegraphics[width=0.8in]{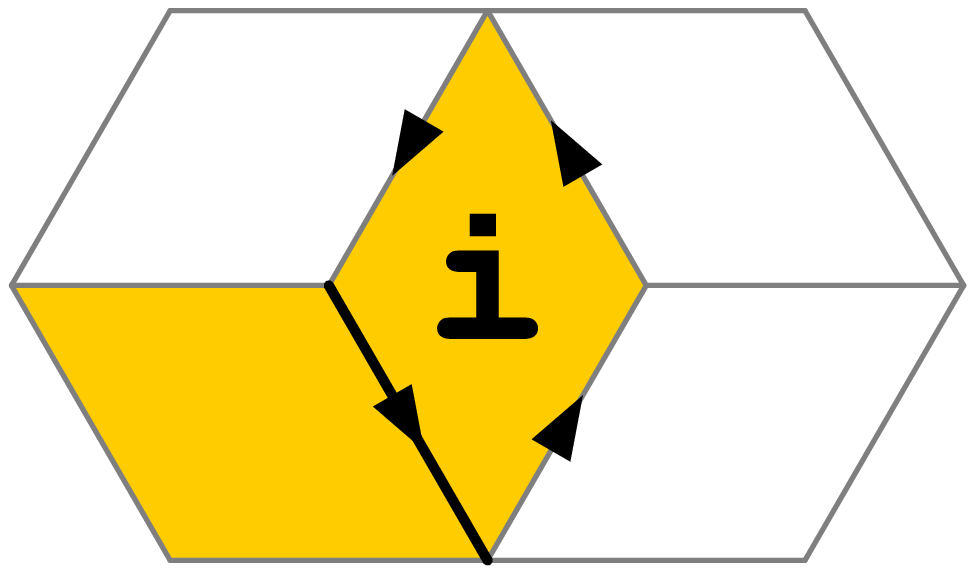}}
\subfigure[{}]{\includegraphics[width=0.8in]{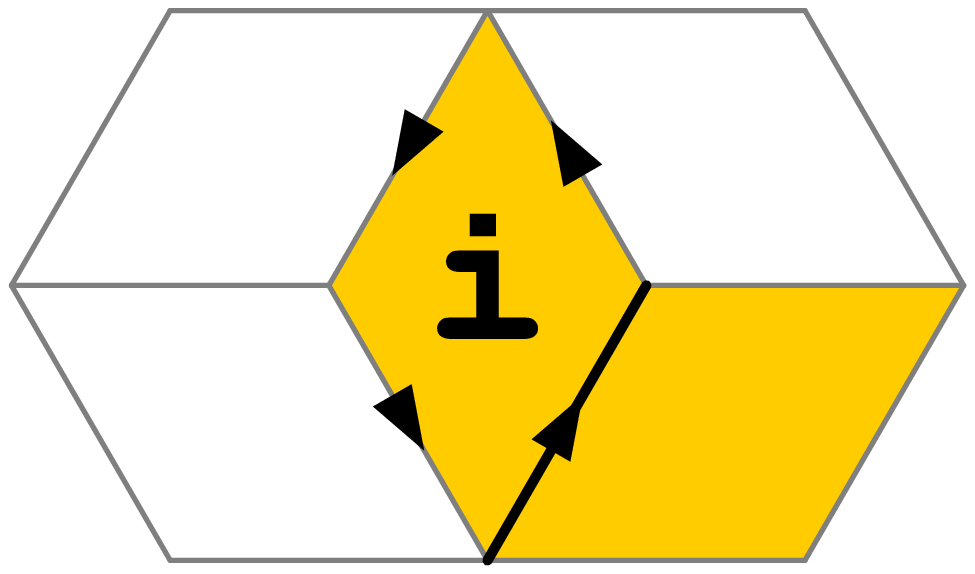}}
\vskip -3mm
\caption{\label{TwoDiceBJ}Explanation for the derivation of equations ~(\ref{EAHMBJ}): The goal is to associate $B_i$ to the curl of $\mathcal{A}_{pq}$, and $j_i$ to the curl of $\varepsilon_{\bond{pq}} \mathcal{A}_{pq}$. For each of the four plaquette pairs in the figures (a) through (d), a counter-clockwise circulation (of $\mathcal{A}_{pq}$ or $\varepsilon_{\bond{pq}} \mathcal{A}_{pq}$) is taken on the central plaquette, labeled by its dual Kagome site $i$. The fields $K^{(\alpha)}_{\bond{pq}}$ on the emphasized bonds will be coupled to this circulation, and we simply collect those coupled to the curl of $\mathcal{A}_{pq}$ into $B_i$, and those coupled to the curl of $\varepsilon_{\bond{pq}} \mathcal{A}_{pq}$ into $j_i$. However, there is a small complication. Let us call a plaquette ``left'' or ``right'' according to its position when the plaquette pair is rotated to point ``upward'' like in the Fig. ~\ref{TwoDiceProc}. Then, for the pairs (a) and (c), the plaquette $i$ is ``left'', while for the pairs (b) and (d), $i$ is the ``right'' plaquette. The fields $K^{(b)}_{\bond{pq}}$ and $K^{(c)}_{\bond{pq}}$ couple to circulations of the different quantities on the ``left'' and ``right'' plaquettes (see Table ~\ref{DimerFlips2}). Therefore, the ``left'' and ``right'' must be distinguished. Notice that the oriented emphasized bond in the figures (a) and (c) points from the 3-coordinated site toward the 6-coordinated site ($\eta_{pq}=-1$), and the opposite in the figures (b) and (d) ($\eta_{pq}=+1$). This can be used to determine when the circulation is made on the ``left'' or on the ``right'' plaquette, and this is the origin of the $\eta_{pq}$ terms in ~(\ref{EAHMBJ}).}
\end{figure}

We proceed by writing the path-integral for the U(1) Hamiltonian, respecting the constraints given by the Gauss' Law ~(\ref{U1Gauss}). The action will contain the Berry's phase ~(\ref{U1Berry}) and the potential energy term ~(\ref{U1pot}) as before. However, this time the Villain's approximation will give rise to four integer-valued massive fields $K^{(\alpha)}_{\bond{pq}}$ ($\alpha=a,b,c,d$) that live on the dice lattice bonds, and couple to the arguments of cosines $h^{(\alpha)}_{\bond{pq}}$ from the Table ~\ref{DimerFlips2}:
\begin{equation}\label{EAHMActKin1}
S_{\textrm{kin}} = \sum_{\tau} \sum_{\bond{pq}} \sum_{\alpha} \Biggl\lbrack
     g \Bigl(K^{(\alpha)}_{\bond{pq}}\Bigr)^2 
   + i K^{(\alpha)}_{\bond{pq}} \Bigl( h^{(\alpha)}_{\bond{pq}} + \pi \Bigr)
     \Biggr\rbrack \ . \nonumber
\end{equation}
An additional Berry's phase $\sim i\pi K^{(\alpha)}_{\bond{pq}}$ appears because the coupling $J_{\bot}$ is positive. It is again possible to define the magnetic field $B_i$ and the particle current $j_{p_6q_6}$. The magnetic field couples to the plain plaquette curl of the vector potential, while the current couples to the boson hopping:
\begin{eqnarray}\label{EAHMActKin}
S_{\textrm{kin}} & = & \sum_{\tau} \sum_{\bond{pq}} \sum_{\alpha} \Biggl\lbrack
     g \Bigl(K^{(\alpha)}_{\bond{pq}}\Bigr)^2 + i \pi K^{(\alpha)}_{\bond{pq}} 
     \Biggr\rbrack \nonumber \\
   & + & \sum_{\tau} \Biggl\lbrack i \sum_i B_i \sum_{pq}^{\CDPL_i} \mathcal{A}_{pq} \\
   & + & i \sum_{\bond{p_6q_6}}
      j_{p_6q_6} \Biggl( \varphi_{q_6} - \varphi_{p_6} +
      \eta_{p_6q_6} \sum_{pq}^{\CDPL_{}} 
      \varepsilon_{\bond{pq}} \mathcal{A}_{pq} \Biggr) \Biggr\rbrack \ . \nonumber
\end{eqnarray}
In terms of the fields $K^{(\alpha)}_{\bond{pq}}$, the magnetic field and the particle current (labeled by the Kagome sites) are:
\begin{eqnarray}\label{EAHMBJ}
B_i & = & \sum^{\CDPL_i}_{pq} \varepsilon_{\bond{pq}} \Biggl( K^{(a)}_{\bond{pq}} +
    \frac{1-\eta_{pq}}{2} K^{(b)}_{\bond{pq}} +
    \frac{1+\eta_{pq}}{2} K^{(c)}_{\bond{pq}}
    \Biggr) \ , \nonumber \\
j_i & = & \sum^{\CDPL_i}_{pq} \Biggl( K^{(d)}_{\bond{pq}} +
    \frac{1-\eta_{pq}}{2} K^{(c)}_{\bond{pq}} +
    \frac{1+\eta_{pq}}{2} K^{(b)}_{\bond{pq}} 
    \Biggr) \ .
\end{eqnarray}
We sketch the derivation of these equations in the Fig. ~\ref{TwoDiceBJ} and its caption. The magnetic field and the particle current are integer-valued fields, and the action is seemingly reduced to the form that pertains to the TFIM model. When the vector potential $\mathcal{A}_{pq}$ and the particle phase $\varphi_{p_6}$ are integrated-out, the same current conservation ~(\ref{U1CurrCon}) and Maxwell's ~(\ref{U1Maxwell}) equations are recovered.

We can again introduce the height fields $\chi_i$ and $\lambda_i$ as in ~(\ref{ChiLambda}). For this purpose, it would be convenient to express the $K^{(\alpha)}_{\bond{pq}}$ fields as time derivatives. Since the path-integral has a closed boundary condition in imaginary time, and $\sum_{\tau} K^{(\alpha)}_{\bond{pq}, \tau}$ need not be zero, we can generally write:
\begin{equation}\label{kappa}
K^{(\alpha)}_{\bond{pq},\tau} = \Delta_{\tau} \kappa^{(\alpha)}_{\bond{pq},\tau} + 
  \delta_{\tau, \beta} \widetilde{K}^{(\alpha)}_{\bond{pq}} \ ,
\end{equation}
where the time dependencies have been explicitly shown ($\beta \to \infty$ is the ``last'' moment of time). Ignoring for a moment the magnetic field and current terms from ~(\ref{EAHMActKin1}), the kinetic part of the action becomes:
\begin{eqnarray}\label{EAHMActKin2}
S_{\textrm{kin}} & \sim & \sum_{\bond{pq},\alpha} \Biggl\lbrack \sum_{\tau=0}^{\beta-1} 
     g \Bigl( \Delta_{\tau} \kappa^{(\alpha)}_{\bond{pq}} \Bigr)^2 \\
   & & + g \Bigl( \kappa^{(\alpha)}_{\bond{pq}, 0} - \kappa^{(\alpha)}_{\bond{pq}, \beta} +
                        \widetilde{K}^{(\alpha)}_{\bond{pq}} \Bigr)^2
       + i\pi \widetilde{K}^{(\alpha)}_{\bond{pq}} \Biggr\rbrack \nonumber \\
   & \rightarrow & \sum_{\bond{pq},\alpha} \Biggl\lbrack
       \sum_{\tau=0}^{\beta-1} g\Bigl( \Delta_{\tau} \kappa^{(\alpha)}_{\bond{pq}} \Bigr)^2
       - i\pi \Bigl( \kappa^{(\alpha)}_{\bond{pq}, 0} - \kappa^{(\alpha)}_{\bond{pq}, \beta} \Bigr)
       \Biggr\rbrack \nonumber
\end{eqnarray}
The field $\widetilde{K}^{(\alpha)}_{\bond{pq}}$ was integrated out in the last line, and an emerging additive constant was discarded. The boundary conditions in imaginary time now appear open, and the extra Berry's phase, due to the positive $J_{\bot}$, appears only at the boundary. In fact, the Berry's phase is sensitive only to the parity of the integer-valued fields at the boundary. Whenever the fluctuations render this parity short-range correlated along time, one may expect that the Berry's phase will not affect the macroscopic properties of the theory. This will certainly happen in any disordered phase. However, it can also happen in a ``smooth'' phase that describes a plaquette valence-bond order: the macroscopic degeneracy created by geometric frustration allows many locally different ``smooth'' states, and the small fluctuations between them are extremely abundant, especially on the corner-sharing lattices such as the Kagome.

Therefore, we will assume in the following that the Berry's phase shapes only certain local properties of fluctuations, and neglect it for the purposes of discussing the possible phases of the theory. The affected local properties can be revealed from a microscopic point of view. The positive value of the coupling $J_{\bot}$ in the XXZ model ~(\ref{EAHM2}) prefers the spin singlet formation on the Kagome bonds: $|\uparrow\downarrow\ra - |\downarrow\uparrow\ra$. If $J_{\bot}$ were negative, the symmetric triplets would be favored instead: $|\uparrow\downarrow\ra + |\downarrow\uparrow\ra$. Note that at least due to the strong Ising antiferromagnetic interaction, the ``true ferromagnetic'' nature of the negative $J_{\bot}$ would be suppressed. The macroscopic physics of such triplet bonds must be very similar to that of the singlet bonds, because in both cases every spin can be paired with only one of its neighbors. Even the same higher order processes, such as the valence-bond movements on the closed loops, would be preferred by either sign of $J_{\bot}$.

Passing completely to the Kagome lattice notation, and neglecting the Berry's phase, the action of the final field theory becomes:
\begin{eqnarray}\label{EDHAction}
S & = & g \sum_{\tau} \sum_{\bond{ij}} \Biggl\lbrack
      \sum_{\alpha} \Bigl(\Delta_{\tau}\kappa^{(\alpha)}_{\bond{ij}}\Bigr)^2 \\
  & & + \Bigl( \chi_i - \chi_j +
        \varepsilon_{\bond{ij}} ( \lambda_i - \lambda_j ) +
        \zeta_{ij} \Bigr)^2 \Biggr\rbrack
  \ , \nonumber      
\end{eqnarray}
where the fluctuations of the height fields are constrained by the equations ~(\ref{EAHMBJ}) in the dual form:
\begin{eqnarray}\label{EDHconstr}
\chi_i 
    & = & \sum_{j \in i} \varepsilon_{\bond{ij}} \Bigl( \kappa^{(a)}_{\bond{ij}} +
    \frac{1-\eta_{ij}}{2} \kappa^{(b)}_{\bond{ij}} +
    \frac{1+\eta_{ij}}{2} \kappa^{(c)}_{\bond{ij}}
    \Bigr) \ , \nonumber \\
\lambda_i
    & = & \sum_{j \in i} \Bigl( \kappa^{(d)}_{\bond{ij}} +
    \frac{1-\eta_{ij}}{2} \kappa^{(c)}_{\bond{ij}} +
    \frac{1+\eta_{ij}}{2} \kappa^{(b)}_{\bond{ij}} 
    \Bigr) \ .
\end{eqnarray}

\subsection{Important Properties}\label{EAHMprop}

The action ~(\ref{EDHAction}) resembles very much that of the TFIM model ~(\ref{DHAction}). However, it could give rise to a very different physics. Certain kind of fluctuations are forbidden by the action ~(\ref{EDHAction}), and the remaining ones might be able to entropicaly select some ordered state. The forbidden fluctuations are those that change the total magnetization of the Kagome Ising antiferromagnet. The mechanism for this is provided by the constraints on the spatial configurations of $\chi_i$ and $\lambda_i$, which emerge from ~(\ref{EDHconstr}). The natural degrees of freedom in this field theory are $\kappa^{(\alpha)}_{\bond{ij}}$, and they live on the Kagome bonds, reflecting the nature of the XXZ perturbation.

\begin{table*}
\begin{displaymath}
\begin{array}{l @{\qquad} l}
\bigl(\bC_{\textrm{pot}}\bkp\bigr)^{(a)}_{\bond{ij}} = \varepsilon_{\bond{ij}} \Bigl(
      4\chi_i - \sum\limits_{k \in i} (\chi_k + \varepsilon_{\bond{ik}}\lambda_k)
      \Bigr) + \varepsilon_{\bond{ij}} \Bigl(
      4\chi_j - \sum\limits_{k \in j} (\chi_k + \varepsilon_{\bond{jk}}\lambda_k) 
      \Bigr)
&
\bx^{(a)}_{\bond{ij}} =
      \varepsilon_{\bond{ij}} \sum\limits_{k \in i} \xi_{ik} +
      \varepsilon_{\bond{ij}} \sum\limits_{k \in j} \xi_{jk}
      \nonumber \\
\bigl(\bC_{\textrm{pot}}\bkp\bigr)^{(b)}_{\bond{ij}} = \Bigl\lbrack \Bigl(
      4\lambda_i - \sum\limits_{k \in i} (\lambda_k + \varepsilon_{\bond{ik}}\chi_k)
      \Bigr) + \varepsilon_{\bond{ij}} \Bigl( 
         4\chi_j - \sum\limits_{k \in j} (\chi_k + \varepsilon_{\bond{jk}}\lambda_k)
      \Bigr) \Bigr\rbrack_{i \to j}
&
\bx^{(b)}_{\bond{ij}} = \Bigl\lbrack
      \sum\limits_{k \in i} \varepsilon_{\bond{ik}} \xi_{ik} + 
      \varepsilon_{\bond{ij}} \sum\limits_{k \in j} \xi_{jk}
      \Bigr\rbrack_{i \to j} \nonumber \\
\bigl(\bC_{\textrm{pot}}\bkp\bigr)^{(c)}_{\bond{ij}} = \Bigl\lbrack 
      \varepsilon_{\bond{ij}} \Bigl( 
      4\chi_i - \sum\limits_{k \in i} (\chi_k + \varepsilon_{\bond{ik}}\lambda_k)
      \Bigr) + \Bigl(
      4\lambda_j - \sum\limits_{k \in j} (\lambda_k + \varepsilon_{\bond{jk}}\chi_k) 
      \Bigr) \Bigr\rbrack_{i \to j}
&
\bx^{(c)}_{\bond{ij}} = \Bigl\lbrack
      \varepsilon_{\bond{ij}} \sum\limits_{k \in i} \xi_{ik} +
      \sum\limits_{k \in j} \varepsilon_{\bond{jk}} \xi_{jk}
      \Bigr\rbrack_{i \to j} \nonumber \\
\bigl(\bC_{\textrm{pot}}\bkp\bigr)^{(d)}_{\bond{ij}} = \Bigl(
      4\lambda_i - \sum\limits_{k \in i} (\lambda_k + \varepsilon_{\bond{ik}}\chi_k)
      \Bigr) + \Bigl(
      4\lambda_j - \sum\limits_{k \in j} (\lambda_k + \varepsilon_{\bond{jk}}\chi_k) 
      \Bigr)
&
\bx^{(d)}_{\bond{ij}} = 
      \sum\limits_{k \in i} \varepsilon_{\bond{ik}} \xi_{ik} + 
      \sum\limits_{k \in j} \varepsilon_{\bond{jk}} \xi_{jk}
      \nonumber \\
\end{array}
\end{displaymath}
\caption{\label{EDHMatrAct2}Action of the spatial (potential) part of the coupling matrix on the field vectors (left), and the components of the saddle-point vectors (right). For the components of type (b) and (c), the sites $i$ and $j$ must be chosen in such a way that the bond orientation is from $i$ toward $j$.}
\end{table*}

Therefore, we will adapt the analysis of the TFIM model to these new degrees of freedom, and write the action ~(\ref{EDHAction}) in the matrix form. First, we note that the action is minimized by the same height field configurations $\chi_i$ and $\lambda_i$ as before. By shifting variables and expanding about a particular saddle-point, the potential part of the action becomes:
\begin{equation}\label{EDHAction2}
S_{\textrm{pot}} = g \sum_{\tau} \sum_{\bond{ij}}
  \Bigl( \chi_i - \chi_j +
  \varepsilon_{\bond{ij}} ( \lambda_i - \lambda_j ) +
  \xi_{ij} \Bigr)^2 \ ,      
\end{equation}
where $\xi_{ij}$ has been defined in the section ~\ref{TFIMprop}. Then, we apply the resummation formula:
\begin{equation}
\sum_i a_i \sum_{j \in i} b_{(ij)} = \sum_{\bond{ij}}
  \bigl( b_{(ij)}a_i + b_{(ji)}a_j \bigr)
\end{equation}
to the expressions ~(\ref{EDHconstr}), and substitute the result in ~(\ref{DHMatrAct}). This gives us the matrix form of the action ~(\ref{EDHAction}):
\begin{equation}\label{EDHMatrAct}
\frac{S}{g} = \bkp^T \bC \bkp +
  ( \bkp^T \bx + \bx^T \bkp ) \ .
\end{equation}
Components of the vector $\bkp$ are the $\kappa^{(\alpha)}_{\bond{ij}}$ fields. The structure of the coupling matrix in terms of the natural degrees of freedom, and the saddle-point vectors are given in the Table ~\ref{EDHMatrAct2}.

Now, we repeat the analysis from the sections ~\ref{TFIMprop} and ~\ref{TFIMfluct} in order to find the effect of fluctuations. The crucial pieces of information are how the saddle-point vectors are normalized, and how the coupling matrix acts on them:
\begin{itemize}
\item all saddle-point vectors $\bx$ have the same norm:
\begin{equation}\label{VectNorm2}
\bx^T\bx = \textrm{const.} \ ;
\end{equation}
\item all saddle-point vectors $\bx$ are degenerate eigenvectors of the coupling matrix $\bC$:
\begin{equation}\label{EigenVect2}
\bC\bx = 36\bx \ .
\end{equation}
\end{itemize}
In full analogy to the TFIM case, the coupling matrix $\bC$ is completely dispersionless. There are 24 bare modes per unit-cell of the Kagome lattice (unit-cell has six bonds), and all of them are localized. Only four of them have a non-zero eigenvalue at zero frequency (equal to 36), while the other 20 are ``gapless'' and unphysical fluctuations (due to the redundancy of representation).

\subsection{Effect of fluctuations}\label{EAHMfluct}

In the quest for potential order-by-disorder, we proceed in exactly the same fashion as before. The XXZ model only brings a new complication: the total Ising magnetization is conserved. Every value of total magnetization defines a separate sector of states, and the fluctuations in the lattice field theory ~(\ref{EDHAction}) cannot mix the states from different sectors. In principle, the entropical effects of fluctuations should be investigated for every sector separately. However, we only need to focus to the sector of zero Ising magnetization, since the XXZ coupling clearly favors it.

We introduce the free energy $F(\bx)$ of fluctuations about a saddle-point $\bx$, and search for the saddle-points that minimize it. For small $g$ the same situation occurs as in the TFIM case (see equation ~(\ref{SmallG})): there is no entropical selection of the saddle-points. Thus, a disordered ground state is obtained, which in the XXZ case in fact has non-trivial topology, as will be argued at the end of the section ~\ref{EAHMvar}. New interesting things happen for large $g$. The free energy $F(\bx)$ in the large $g$ limit, after the Poisson resummation, is given by the approximate expression analogous to ~(\ref{TFIMlargeG}):
\begin{eqnarray}\label{EAHMlargeG}
e^{-F(\bx)} & \propto & \sum_{\boldsymbol{\mu}} \exp \Bigl\lbrack -\frac{(36\pi)^2}{g} 
  \boldsymbol{\mathcal{M}}^T(\bC+m^2)^{-1}\boldsymbol{\mathcal{M}} \nonumber \\
  & & +\frac{i\pi}{36+m^2}(\boldsymbol{\delta\mu}^T\bx+\bx^T\boldsymbol{\delta\mu})
  \Bigr\rbrack \ ,
\end{eqnarray}
where the integer Poisson fields $\mu^{(\alpha)}_{\bond{ij}}$ ($\alpha=a \dots d$), forming the vector $\boldsymbol{\mu}$, have been decomposed as:
\begin{equation}
\mu^{(\alpha)}_{\bond{ij}} = 36\mathcal{M}^{(\alpha)}_{\bond{ij}} +
   \delta\mu^{(\alpha)}_{\bond{ij}} \quad , \quad 
   \delta\mu^{(\alpha)}_{\bond{ij}} \in \lbrace 0 \dots 35 \rbrace \ .
\end{equation}
The free energy is minimized when the saddle-point vector $\bx$, given in the Table ~\ref{EDHMatrAct2}, has the maximum number of zero components. All other possible values of the components $\bx$ are integers and factors of 36, so that only the zero components avoid destructive interference in ~(\ref{EAHMlargeG}).

In order to discover which saddle-point dimer coverings are preferred and minimize the free energy, we need to characterize them in terms of the local dimer configurations at the Kagome sites, bonds and triangles. The Kagome sites have already been characterized by the number of dimers emanating from them, in the Table ~\ref{SPBowties}. All non-equivalent dimer arrangements in the neighborhood of a Kagome bond are systematically shown in the Table ~\ref{SPBondSituations}, together with the corresponding numbers of the zero components of the saddle-point vector. Finally, the triangles can be characterized by the types of sites at their corners, and all possibilities are given in the Table ~\ref{SPTriangles}. For every allowed type of triangle one can find three situations in the Table ~\ref{SPBondSituations}, corresponding to the three bonds on the triangle (one of which is frustrated), and collect the total number of the saddle-point zero components that such a triangle would contribute. Adding contributions of all triangles, that is all bonds, we obtain the following ``scoring'' number that should be maximized:
\begin{equation}
n'_p = 8n_{bbb} + 2n_{abb} + 2n_{aab} + 6n_{aac} + 2n_{abc} + 4n_{bbc} \ .
\end{equation}
The quantities $n_{bbb} \dots n_{bbc}$ denote the total numbers of various kinds of triangles in a given saddle-point dimer configuration. At this stage, we have to investigate possible relationships between these numbers. The first thing to note is that the total number of Kagome triangles is:
\begin{equation}\label{TriNum}
n_{bbb} + n_{abb} + n_{aab} + n_{aac} + n_{abc} + n_{bbc} = \frac{2N}{3} \ .
\end{equation}
Then, using the Table ~\ref{SPTriangles}, we can count the total numbers of A, B, and C sites:
\begin{eqnarray}\label{NaNbNc}
n_a & = & \frac{1}{2} ( 2n_{aab} + 2n_{aac} + n_{abb} + n_{abc} ) \ , \\
n_b & = & \frac{1}{2} ( 3n_{bbb} + 2n_{abb} + 2n_{bbc} + n_{aab} + n_{abc} ) 
  \ , \nonumber \\
n_c & = & \frac{1}{2} ( n_{aac} + n_{abc} + n_{bbc} ) \ . \nonumber
\end{eqnarray}
By combining these equations with ~(\ref{N3}), one finds that $n_a$, $n_b$, $n_c$, $n_{abb}$, and $n_{abc}$ can be expressed in terms of the independent variables $n_{bbb}$, $n_{aac}$, $n_{bbc}$, and $n_{aab}$. The ``scoring'' number $n'_p$ can now be simplified using the identity ~(\ref{TriNum}). The quantity that has to be maximized is:
\begin{equation}\label{EAHMscore}
n_p = 3n_{bbb}+2n_{aac}+n_{bbc} \ ,
\end{equation}
and the variables appearing in it are independent, although subject to inequalities $0 \leqslant n_a \leqslant N$, \dots, $0 \leqslant n_{bbc} \leqslant 2N/3$.

\begin{table}
\begin{displaymath}
\begin{array}{|c|c|c|c|c|c|c|}
\hline
  \includegraphics[height=0.75cm]{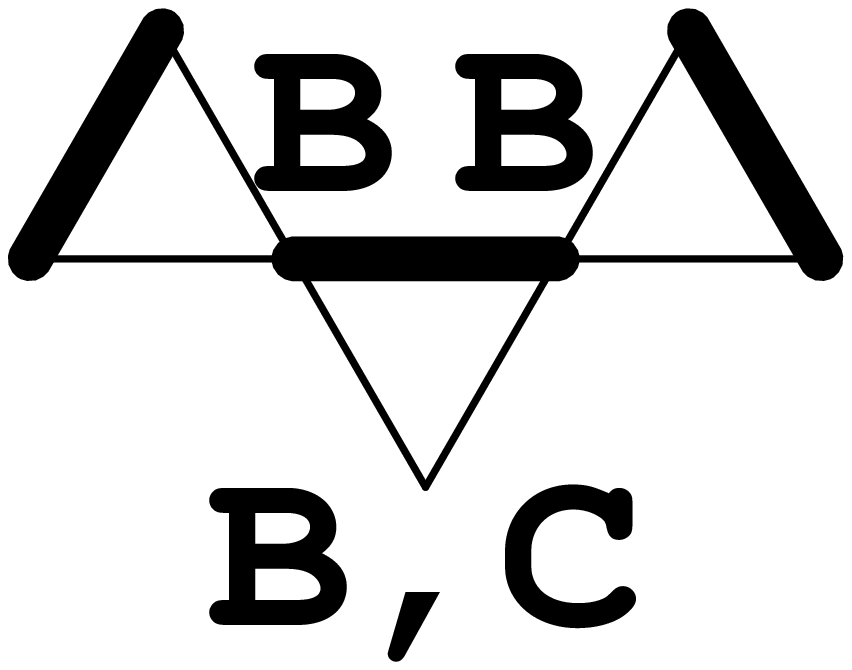}
& \includegraphics[height=0.75cm]{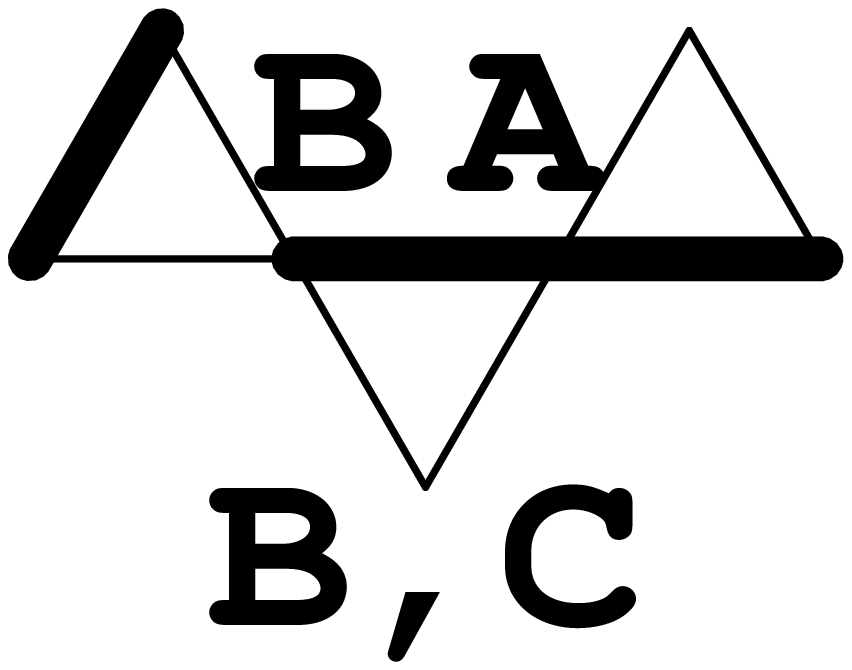}
& \includegraphics[height=0.75cm]{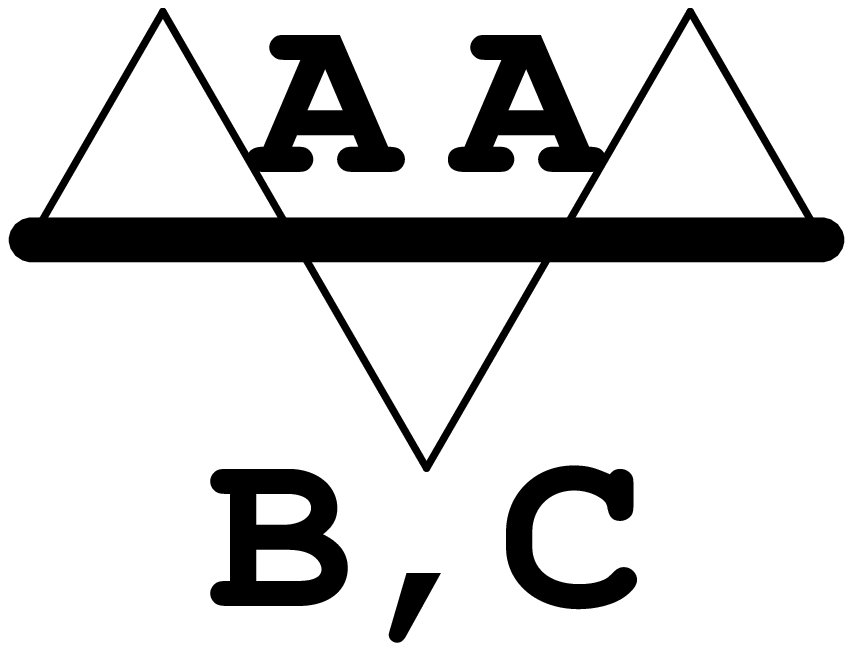}
& \includegraphics[height=0.75cm]{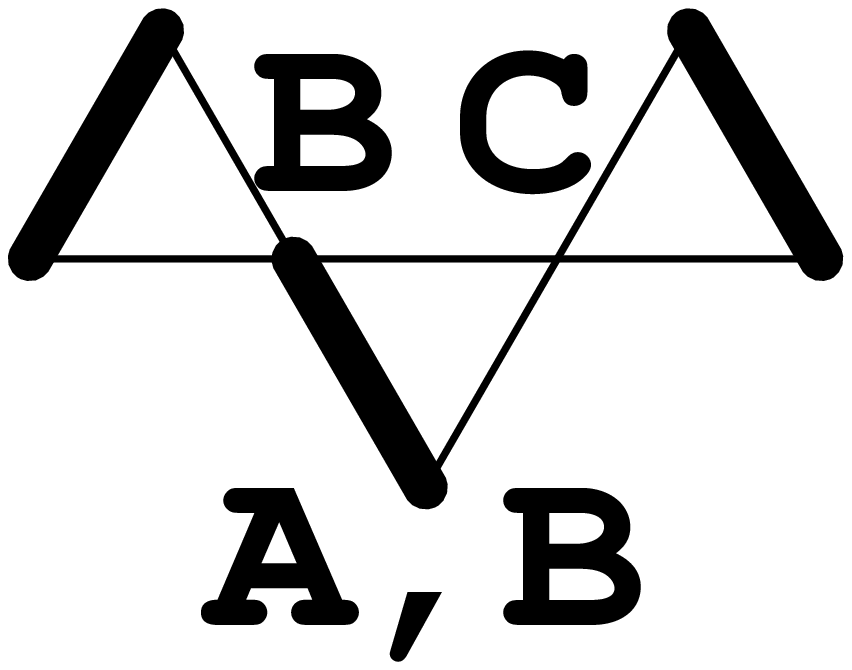}
& \includegraphics[height=0.75cm]{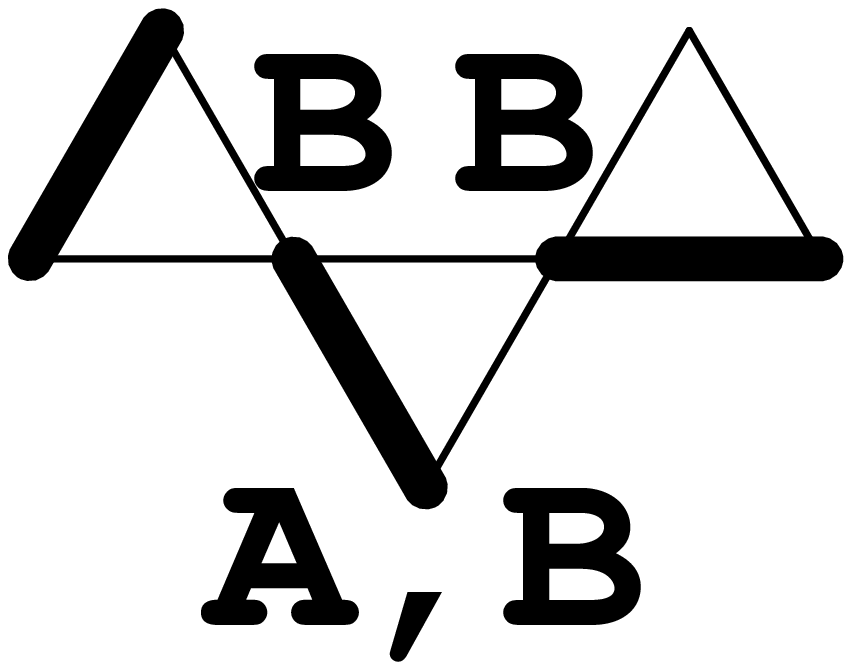}
& \includegraphics[height=0.75cm]{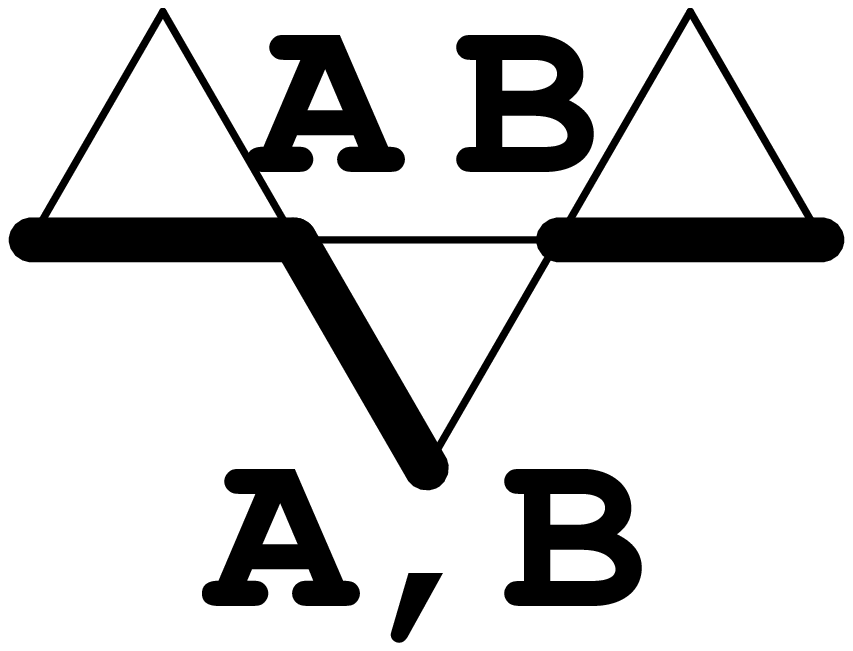}
& \includegraphics[height=0.75cm]{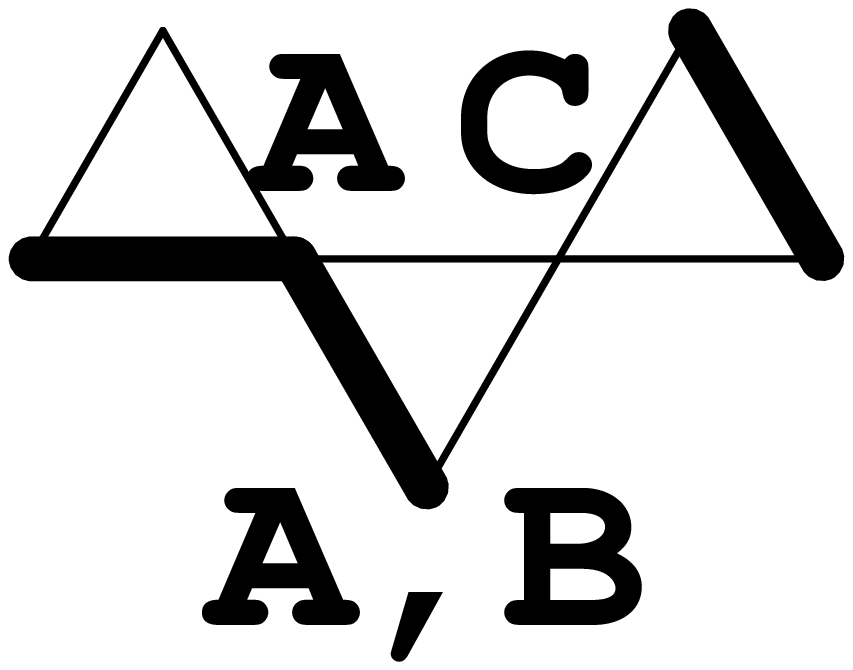}
\\ \hline
  4 & 0 & 2 & 0 & 2 & 0 & 2 \\
\hline
\end{array}
\end{displaymath}
\caption{\label{SPBondSituations}All possible non-equivalent configurations of frustrated bonds in the neighborhood of a Kagome bond. The types of sites are labeled according to the scheme from the Table ~\ref{SPBowties} (for the bottom site on the central triangle there are always two options). This table shows the number of zero-valued components $\xi^{(\alpha)}_{\bond{ij}}$, $\alpha=a \dots d$  of the saddle-point vector, for the horizontal bond $\bond{ij}$ on the central triangle.}
\end{table}

\begin{table}
\begin{displaymath}
\begin{array}{c c c c c c}
\includegraphics[height=1cm]{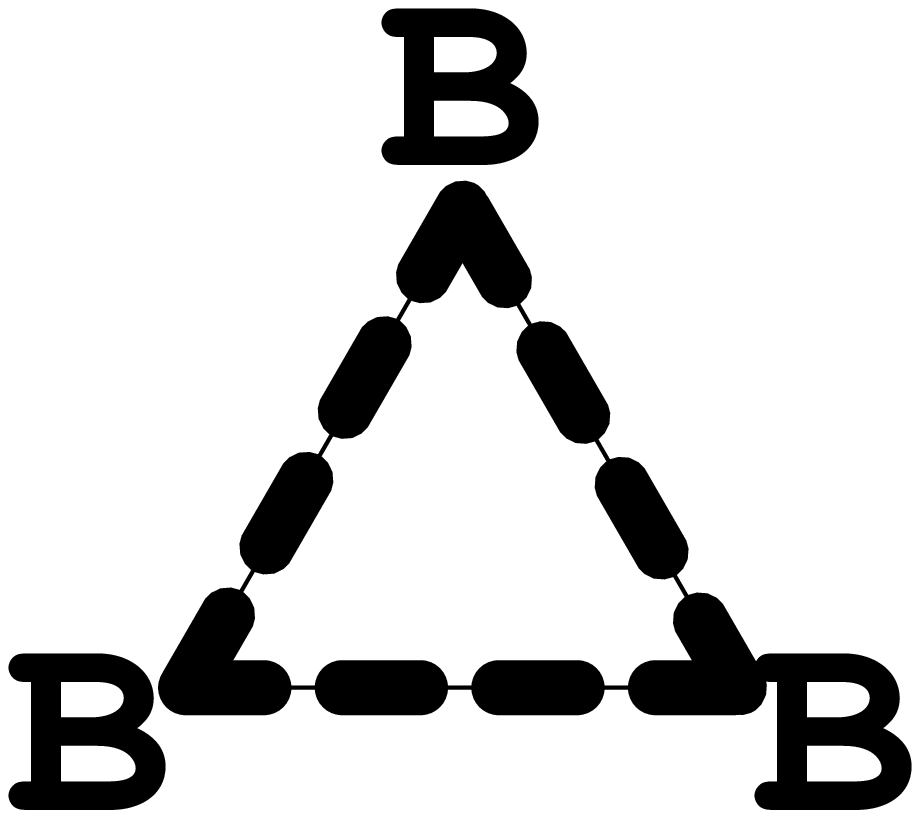} &
\includegraphics[height=1cm]{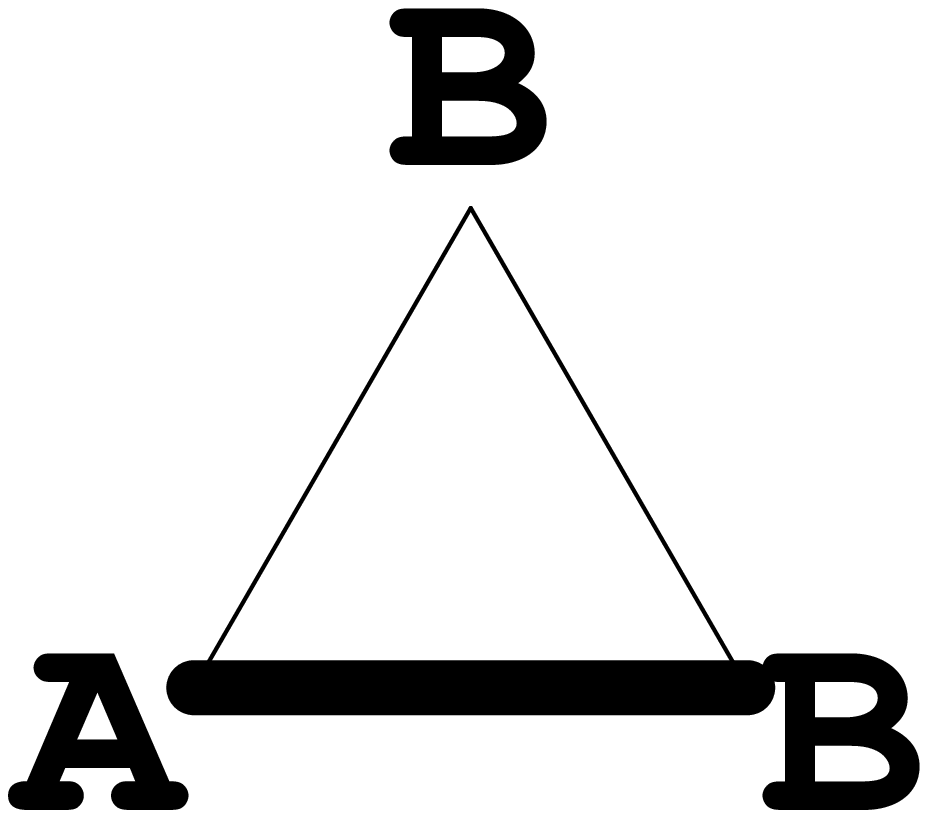} &
\includegraphics[height=1cm]{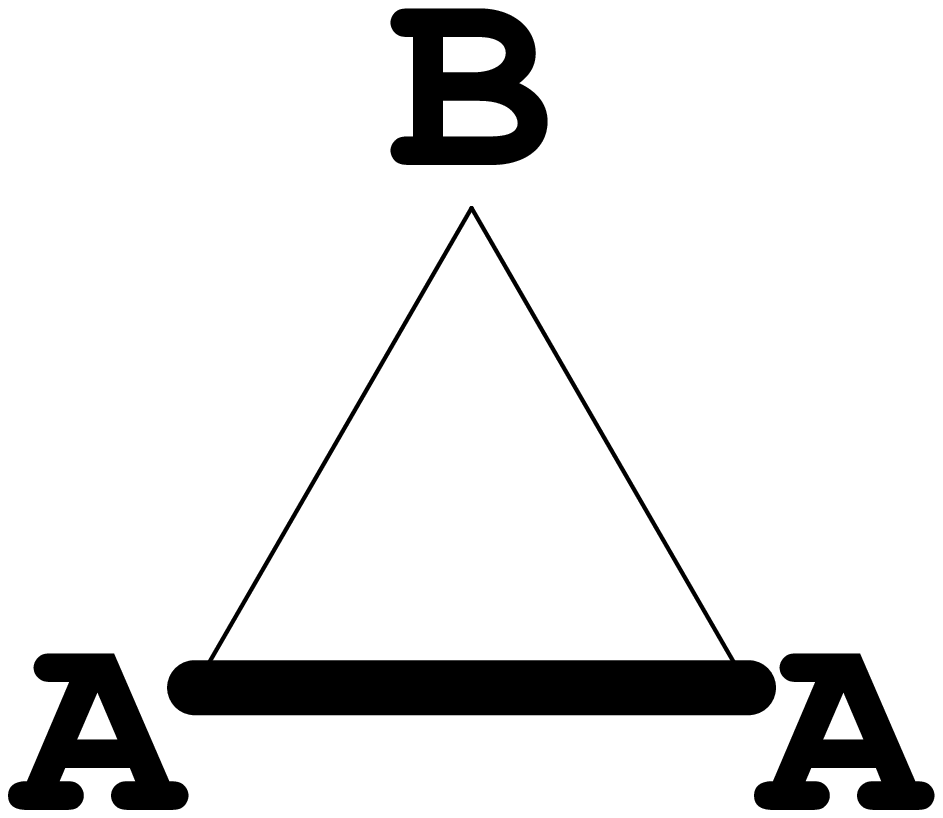} &
\includegraphics[height=1cm]{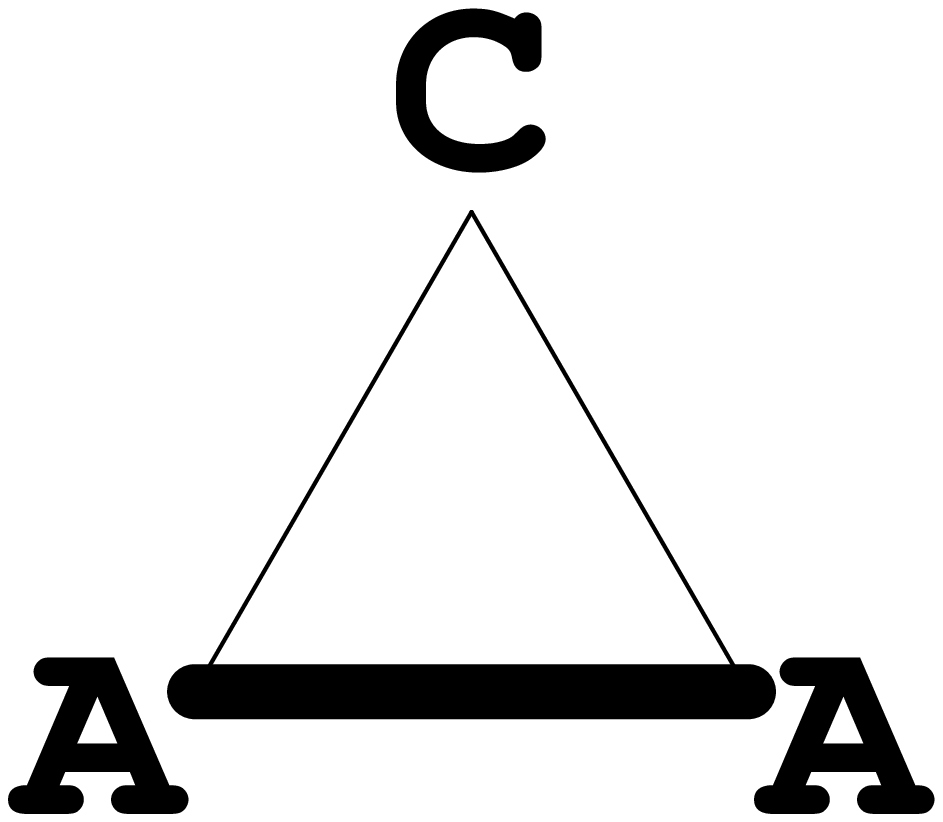} &
\includegraphics[height=1cm]{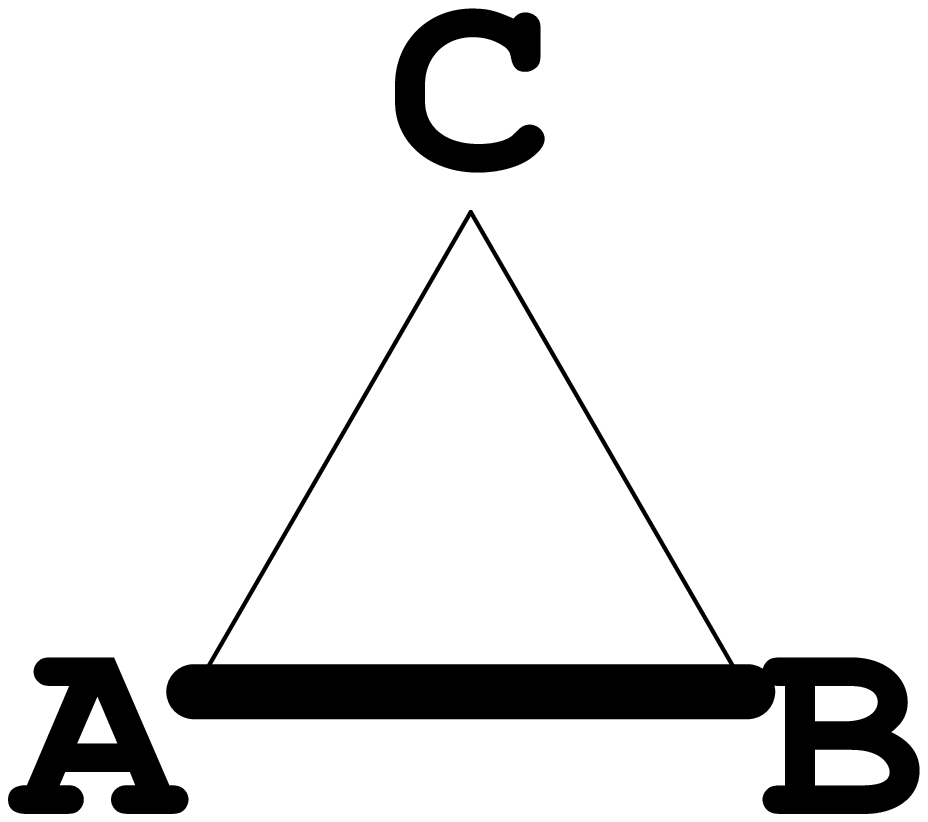} &
\includegraphics[height=1cm]{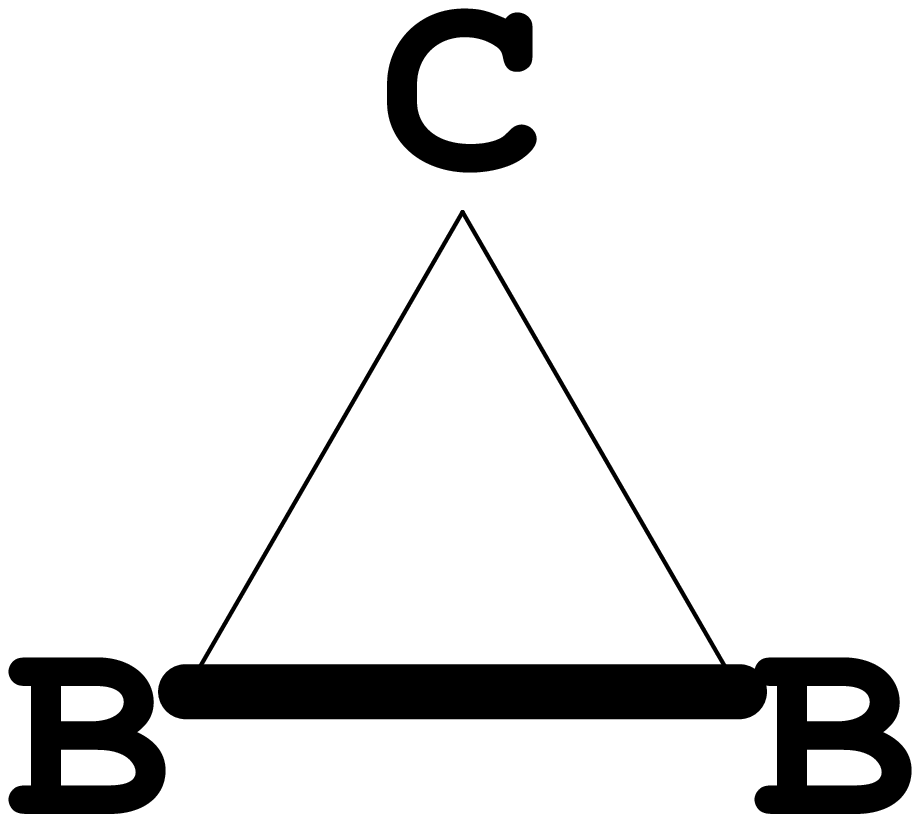} \\
\textrm{BBB} & \textrm{ABB} & \textrm{AAB} &
\textrm{AAC} & \textrm{ABC} & \textrm{BBC}
\end{array}
\end{displaymath}
\vskip -4mm
\caption{\label{SPTriangles}Characterization of all possible kinds of Kagome triangles. The site types A,B, and C are defined in the Table ~\ref{SPBowties}. Note that two C-sites can never be neighbors, and that three A-sites cannot sit on the same triangle. The dimers emphasize which bond must be frustrated in a given situation (there are multiple choices only for the case BBB).}
\end{table}

\begin{figure}
\includegraphics[width=2.6in]{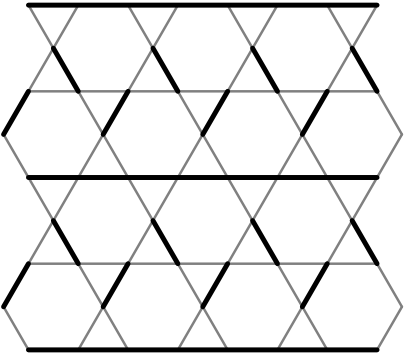}
\vskip -2mm
\caption{\label{EAHMstripeSP}The preferred saddle-point configuration with zero magnetization, selected by the XXZ fluctuations. The lattice symmetries are ``broken'' in the stripe-like fashion. There are only type A and B sites in this state, and the number of BBB triangles is large (they sit between the straight chains of dimers). The total magnetization is zero: every two spins connected by a vacant bond are antialigned, so that the straight chains alternate in magnetization, as well as the dimers in the middle along the chains.}
\end{figure}

Finding the absolute maximum of $n_p$ is a well posed problem of linear programming. It can be shown that it is obtained when the number of C-type sites is maximized, which gives rise to the very same states preferred by the fluctuations of the TFIM model (see Fig.~\ref{MaxFlip}). However, these are not the saddle-points that we are looking for: they have a macroscopic magnetization. Besides, instead of being maximally flippable, as in the case of the TFIM model, they are now minimally flippable; in fact, they have no flippable spin-pairs at all. If we want to find the configurations with zero magnetization that maximize $n_p$, we must explore a path different from having a large number of C sites. This is, in principle, a difficult problem, and an exact analytical solution is not available at this time. Instead, we guess that $n_{bbb}$ should be made as large as possible. This yields the configurations without any C sites, and with a large number of B sites (see ~(\ref{N3})). The best choice is: $n_a=N/3$, $n_b=2N/3$, $n_{bbb}=n_{aab}=N/3$. It is possible, though relatively complicated, to demonstrate that the configurations with these parameters break the lattice symmetries in a unique stripe-like fashion shown in the Fig.~\ref{EAHMstripeSP}. It turns out that such states are not magnetized at all, and that they have a large number of flippable spin-pairs. Their scoring number ($n_p=N$) is significantly larger than that of a typical unmagnetized state (a fraction of $N$). Therefore, they are excellent candidates for the preferred configurations. No better configurations were found when every least frustrated state with zero magnetization was explicitly examined using the computer (the sample had 24 sites and closed boundary conditions).

In conclusion, for large $g$ the preferred configurations of frustrated bonds that minimize the free energy break the lattice symmetries in the stripe-like way, as shown in the Fig.~\ref{EAHMstripeSP}. This indicates that a valence-bond ordered phase could be realized in the XXZ model when dynamics is dominated by the short-ranged spin-pair flips (the larger $g$, the weaker further-neighbor and multiple-spin exchange). It is now necessary to verify stability of such an ordered phase.

\subsection{Stability of the Valence-Bond Order}

A usual way to determine whether fluctuations ultimately destroy the long-range order involves renormalization group (RG). The lattice field theory ~(\ref{EDHAction}) of the XXZ model resembles an integer-valued height model, and one might naively hope that the RG arguments could be applicable to it. In a standard and simple integer-valued height model (on the square lattice, for example) one first softens the integer constraints for the height fields by writing a sine-Gordon theory. Then, one checks how the sine-Gordon coupling flows under RG, starting from various parameter values in the theory. If it flows toward zero, then the integer constraints are irrelevant at the macroscopic scales, and the height model may be found in the ``rough'' disordered phase. Alternatively, the flow can be toward infinity, in which case the ``smooth'' long-range ordered phase is realized. In the context of frustrated magnetism, the appropriate height model typically comes with a background ($\bx$ in our case), so that the ``smooth'' phase also breaks the lattice symmetries. \cite{DimerSG}

Therefore, let us write a sine-Gordon theory for the XXZ model, based on the action ~(\ref{EDHMatrAct}):
\begin{equation}\label{EAHMsg}
S_{\textrm{sg}} = g\bkp^T \bC \bkp + g( \bkp^T \bx + \bx^T \bkp ) 
  - \gamma \sum_{\tau, \bond{ij}, \alpha} \cos \bigl( 2\pi \kappa^{(\alpha)}_{\bond{ij}} 
    \bigr) \ . \nonumber
\end{equation}
The $\kappa^{(\alpha)}_{\bond{ij}}$ fields are now real-valued, and their deviation from integers is penalized by the sine-Gordon term, especially in the large $\gamma$ limit. It is convenient to shift the variables $\bkp$ by $\bC^{-1}\bx = \bx / 36$, and remove the linear terms:
\begin{equation}\label{EAHMsg2}
S_{\textrm{sg}} = g\bkp^T \bC \bkp 
  - \gamma \sum_{\tau, \bond{ij}, \alpha} \cos \Biggl( 2\pi \Bigl( 
    \kappa^{(\alpha)}_{\bond{ij}} - \frac{1}{36}\xi^{(\alpha)}_{\bond{ij}}
    \Bigr) \Biggr) \ . \nonumber
\end{equation}
Unfortunately, it is not possible to directly apply the RG treatment to this theory. The bare modes (modes of the coupling matrix $\bC$) are not only dispersionless, but some of them appear gapless as well. The ``gapless'' bare modes are redundancy of representation, but they still pose a technical difficulty. It is through the sine-Gordon coupling that they at least acquire dispersion. The sine-Gordon term mixes the bare modes when they describe non-integer fluctuations of the height fields. Let us relabel the fields $\kappa^{(\alpha)}_{\bond{ij}}$ as $\kappa_{n,\rv}$, where $\rv$ is a vector specifying a Kagome lattice unit-cell, and $n \in \lbrace 1,2\dots 24 \rbrace$ is an index specifying the bond $\bond{ij}$ (one of six) and the flavor $\alpha$ (one of four) within the unit-cell. Then, we can express the fields as the linear combinations of the 24 local bare modes $\Phi_{n,\rv}$:
\begin{equation}
\kappa_{n,\br} = \sum_{m=1}^{24} \sum_{\drv} W_{nm,\drv} \Phi_{m,\rv+\drv} \ .
\end{equation}
One can formally integrate out the four physical (massive) modes: $\Phi_{21,\rv}, \Phi_{22,\rv}, \Phi_{23,\rv}, \Phi_{24,\rv}$, and obtain the effective theory as a perturbative expansion in $\gamma$. The effective theory is a complicated expression involving cosine terms whose arguments are linear combinations of the remaining twenty modes:
\begin{eqnarray}\label{EffSG}
S_{\textrm{eff}} & = & g\sum_{n=1}^{20} \sum_{\rv} ( \Delta_{\tau} \Phi_{n,\rv} )^2 \\
   & - & \gamma \sum_{n,\rv} C_n \cos \Biggl( 
         2\pi \sum_{m=1}^{20} \sum_{\drv} W_{nm,\drv} \Phi_{m,\rv+\drv} \nonumber \\
   & & - 2\pi \frac{1}{36} \xi_{n,\rv} \Biggr) + \mathcal{O} ( \gamma^2 ) \ . \nonumber
\end{eqnarray}
The redundancy of representation survives in the effective theory through periodicity of the cosines. However, the redundancy is easily removed by treating the mode amplitudes in the effective theory as angles: the physical degrees of freedom (which have been integrated-out) enter the effective theory precisely trough the residual $\lbrack 0,2\pi)$ amplitudes of the remaining modes.

\begin{figure}
\subfigure[{}]{
\includegraphics[width=3.1in,viewport=90 477 500 710, clip]{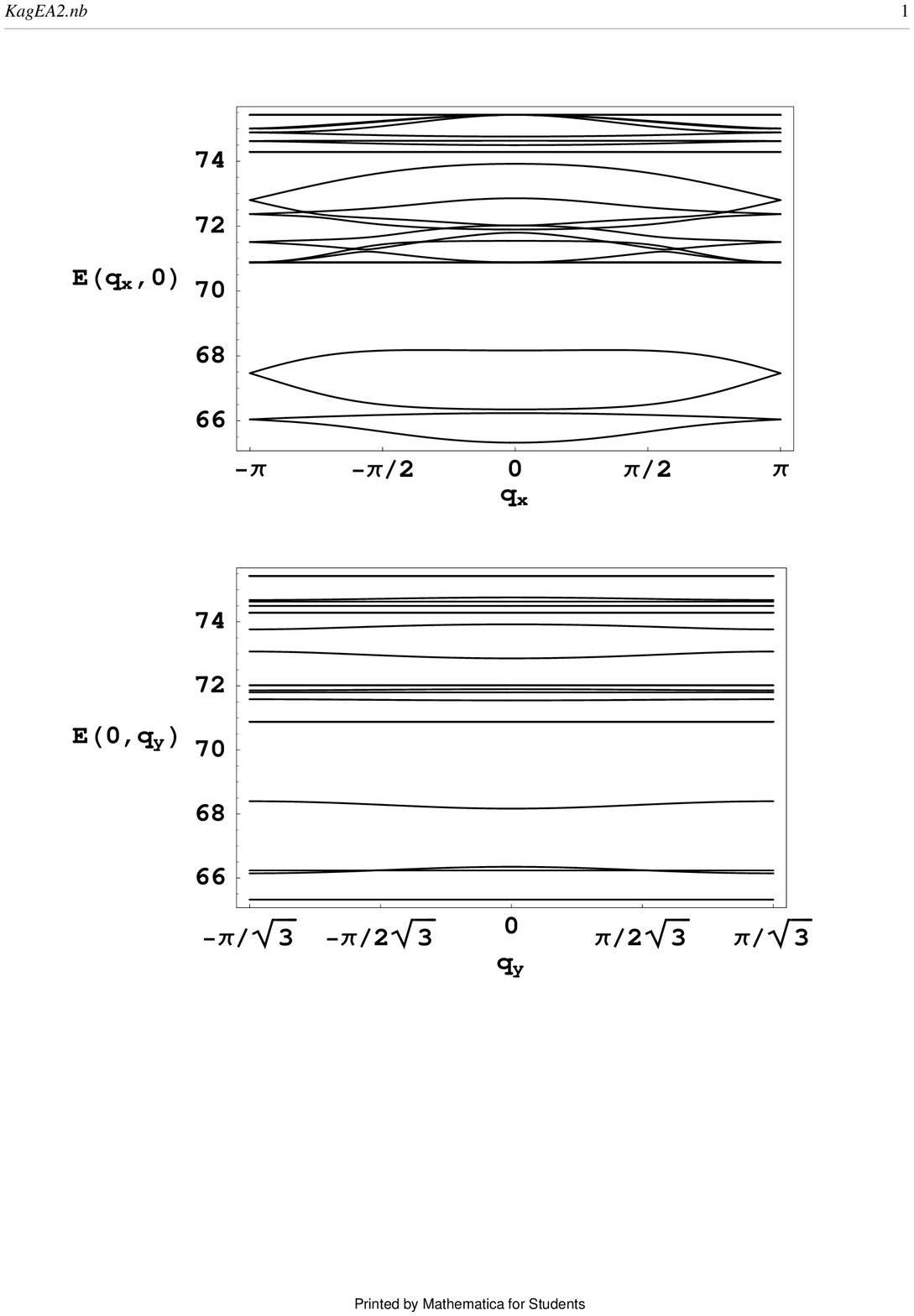}}
\subfigure[{}]{
\includegraphics[width=3.1in,viewport=90 217 500 450, clip]{eahm-dispersion.eps}}
\vskip -2mm
\caption{\label{ZeroDisp}Effective band structure of the sine-Gordon theory ~(\ref{EffSG}) for the candidate valence-bond order in Fig.~\ref{EAHMstripeSP} (zero frequency, quadratic approximation). The wavevector $\boldsymbol{q}$ is taken (a) along the stripes, (b) perpendicular to the stripes. The vertical scale is given in arbitrary units, but proportional to $\gamma$. The Kagome lattice unit-cell had to be doubled, so that 40 modes are shown; only the dispersionless branches are degenerate. Note that the lowest lying mode is dispersive in only one spatial direction, but also that some higher modes have dispersion in the other direction.}
\end{figure}

Let us for a moment expand the cosine terms to the quadratic order, and obtain a Gaussian theory for the twenty modes. In absence of the saddle-point background $\bx = 0$, such a theory would be trivial:
\begin{equation}
\xi_{n,\rv} = 0 \quad \Rightarrow \quad S_{\textrm{gauss}} = \sum_{n,\rv} 
  \Bigl\lbrack ( \Delta_{\tau} \Phi_{n,\rv} )^2 + m^2 \Phi_{n,\rv}^2 \Bigr\rbrack \ .
  \nonumber
\end{equation}
This means that the effective theory ~(\ref{EffSG}) retains the nature of the original sine-Gordon model: small fluctuations of the bare modes are gapped. However, we are concerned only with the vicinity of the candidate valence-bond ordered state (Fig.~\ref{EAHMstripeSP}). For the saddle-point vector $\bx$ that describes this stripe pattern the effective band structure in the Gaussian approximation is shown in the Fig.~\ref{ZeroDisp}. Apart from having frequency dependence, the lowest lying mode is dispersive in the direction along the stripes, but not in the perpendicular direction. Nevertheless, full spatial dispersion can be expected if one goes beyond the Gaussian approximation, because the lowest lying mode is actually coupled to some higher modes that are dispersive in the perpendicular direction. The results of this approximation are only good for arguing that dispersion emerges. There is no simple way of telling what changes beyond the quadratic approximation. Therefore, the correct way in which the lattice symmetries may be eventually broken may not be possible to guess from this information.

We can now write the effective theory in a form that manifestly separates the dispersive and sine-Gordon parts. If we partially expand the cosines from ~(\ref{EffSG}) in the following way (supressing the $n$ and $\rv$ indices of $\xi_{n,\rv}$ and $x_{n,\rv} = \sum\limits_{m,\drv} W_{nm,\drv} \Phi_{m,\rv+\drv}$):
\begin{eqnarray}
\sum_{n,\rv} C_n \cos \Bigl( 2\pi x - \frac{2\pi\xi}{36} \Bigr) & = &
      \sum_{n,\rv} C_n \Bigl\lbrace 
      2\pi x \sin \Bigl(\frac{2\pi\xi}{36} \Bigr) + \nonumber \\
  a\cos (2\pi x) + \Bigl\lbrack \cos \Bigl(\frac{2\pi\xi}{36} \Bigr) & - & a \Bigr\rbrack 
      \Bigl( 1 - \frac{(2\pi x)^2}{2} \Bigr) \Bigr\rbrace + \mathcal{O}(x^3)
      \ , \nonumber
\end{eqnarray}
then for a proper choice of the constant $a$ the effective theory becomes:
\begin{eqnarray}\label{EffSG2}
S_{\textrm{eff}} & \approx & 
         g \boldsymbol{\Phi}^T \bC_{\textrm{eff}}(\bx) \boldsymbol{\Phi} 
         -\gamma \Bigl( \boldsymbol{h}^T(\bx) \boldsymbol{\Phi} + 
                        \boldsymbol{\Phi}^T \boldsymbol{h}(\bx) \Bigr) \\
   & - & a\gamma \sum_{n,\rv} C_n \cos \Biggl( 
         2\pi \sum_{m=1}^{20} \sum_{\drv} W_{nm,\drv} \Phi_{m,\rv+\drv} \Biggr) 
         \ , \nonumber
\end{eqnarray}
where the coupling matrix $\bC_{\textrm{eff}}(\bx)$ collects all space-time dispersion (brought up by the non-trivial $\bx \neq 0$), and has no gap at $\boldsymbol{q} = 0$. The remaining cosines open up a gap for small fluctuations, thereby justifying the quadratic expansion that took place. This is an effective sine-Gordon theory. As a matter of principle, the RG treatment is now applicable. Even if the dispersion were ultimately created only in one spatial dimension (along the stripes), combined with the dispersion in time it would give a ``smooth'' phase for sufficiently large $g$ and $\gamma$. If the full ($2+1$)D dispersion were obtained, then only the ``smooth'' phase would exist, since the sine-Gordon coupling would always flow toward infinity. In any event, existence of the ``smooth'' phase means that the valence-bond long-range order for large $g$ is stable. For small $g$, however, the ``smooth'' phase is disordered since the fluctuations cannot select an ordered state from the degenerate manifold.

This concludes our discussion of the lattice field theory. In the following, we take a completely different point of view, and provide a more physical picture of the discovered XXZ phases.

\subsection{Nature of the XXZ Phases}\label{EAHMvar}

In this section we use some simple physical arguments and show that the short-range valence-bond picture applies extremely naturally to the XXZ models on the corner-sharing lattices. This will allow us to identify a physical ``order parameter'' for the valence-bond ordered phase of the Kagome XXZ model, and construct qualitatively good variational wavefunctions. Also, we will argue that the disordered phase of the Kagome (and any other) XXZ model has non-trivial topology.

\begin{figure}
\includegraphics[width=1in]{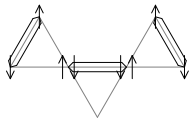}
\vskip -2mm
\caption{\label{FlipPair}A flippable pair of spins sits on the horizontal bond of the central triangle. The pairs of anticorrelated spins are emphasized.}
\end{figure}

Let us seek variational ground states of the Hamiltonian ~(\ref{EAHM}) that are described in terms of the singlet bonds. The energy minimum requirements shaped by the $J_z$ and $J_{\bot}$ terms can be met by following these criteria:
\begin{itemize}
\item number of frustrated bonds is minimized,
\item total Ising magnetization is zero, 
\item number of flippable spin-pairs is maximized.
\end{itemize}
First, we explore the circumstances in which a pair of spins on a Kagome bond is \emph{flippable}. The XXZ perturbation $\sim J_{\bot}$ in ~(\ref{EAHM}) can flip a pair of antialigned spins, but one must make sure that both the initial and final states are minimally frustrated. The Fig.~\ref{FlipPair} shows a flippable pair of spins. Regardless of whether the two antialigned spins on the central horizontal bond are in one or the other state, every triangle has exactly one frustrated bond, which is a condition for minimum frustration. This requires that the two opposite bonds on the neighboring triangles hold a pair of antialigned spins each. Clearly, energy will be gained by allowing the flippable pair of spins to resonate between the two possible states and form a singlet bond. If the other two pairs of antialigned spins were also flippable, more energy could be gained by turning them into the singlet bonds too. The attempt to create as many flippable pairs as possible leads to the hard-core dimer coverings of the Kagome lattice, where every dimer represents a singlet bond (in contrast to the earlier representation, when the dimer was a \emph{frustrated} bond). The corner-sharing structure of the Kagome lattice makes this picture extremely natural by allowing the singlet pairs to be close-packed. The easy-axis anisotropy clearly shapes them as good degrees of freedom.

\begin{figure}
\subfigure[{}]{\includegraphics[width=1in]{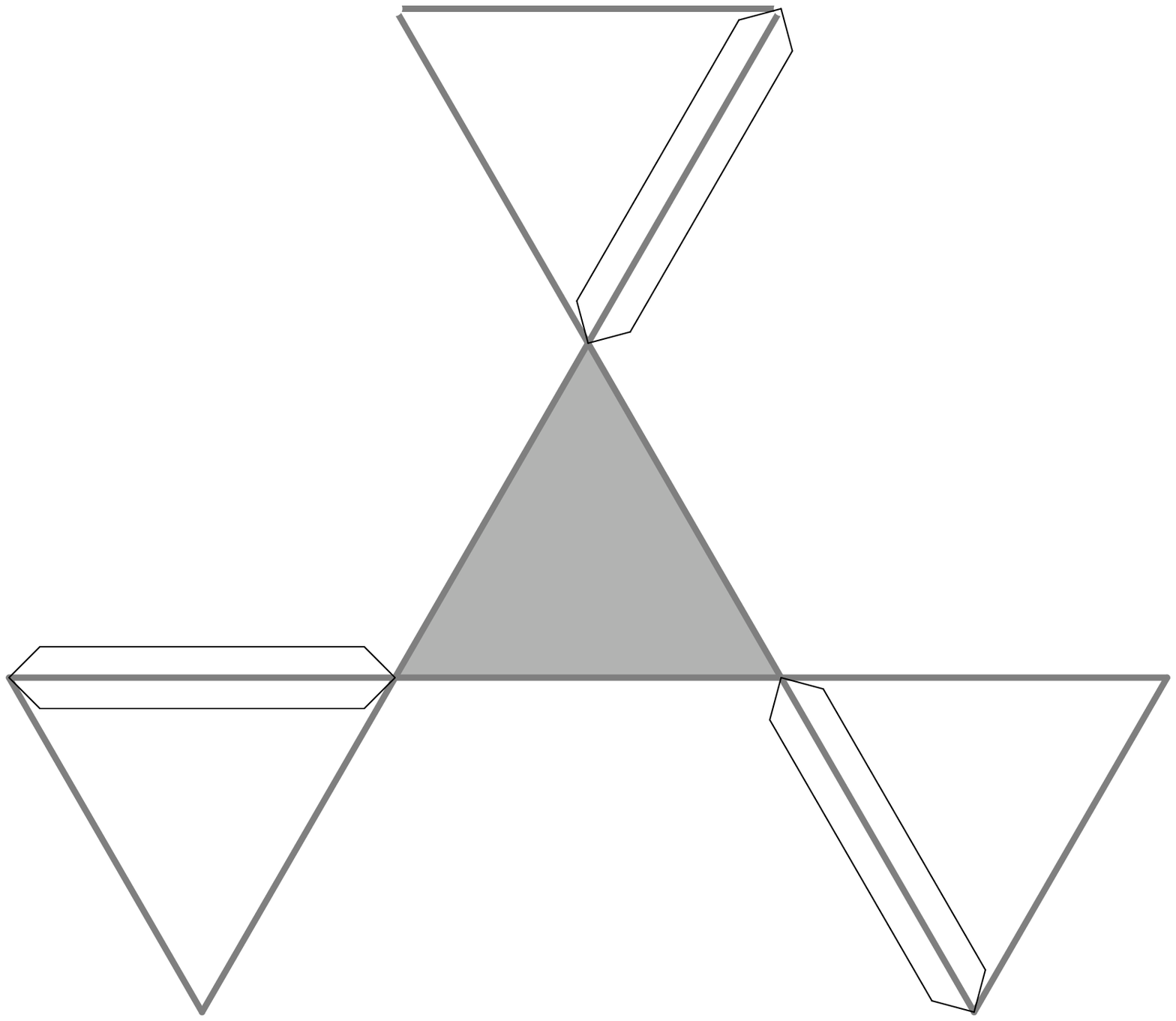}}
\subfigure[{}]{\includegraphics[width=1in]{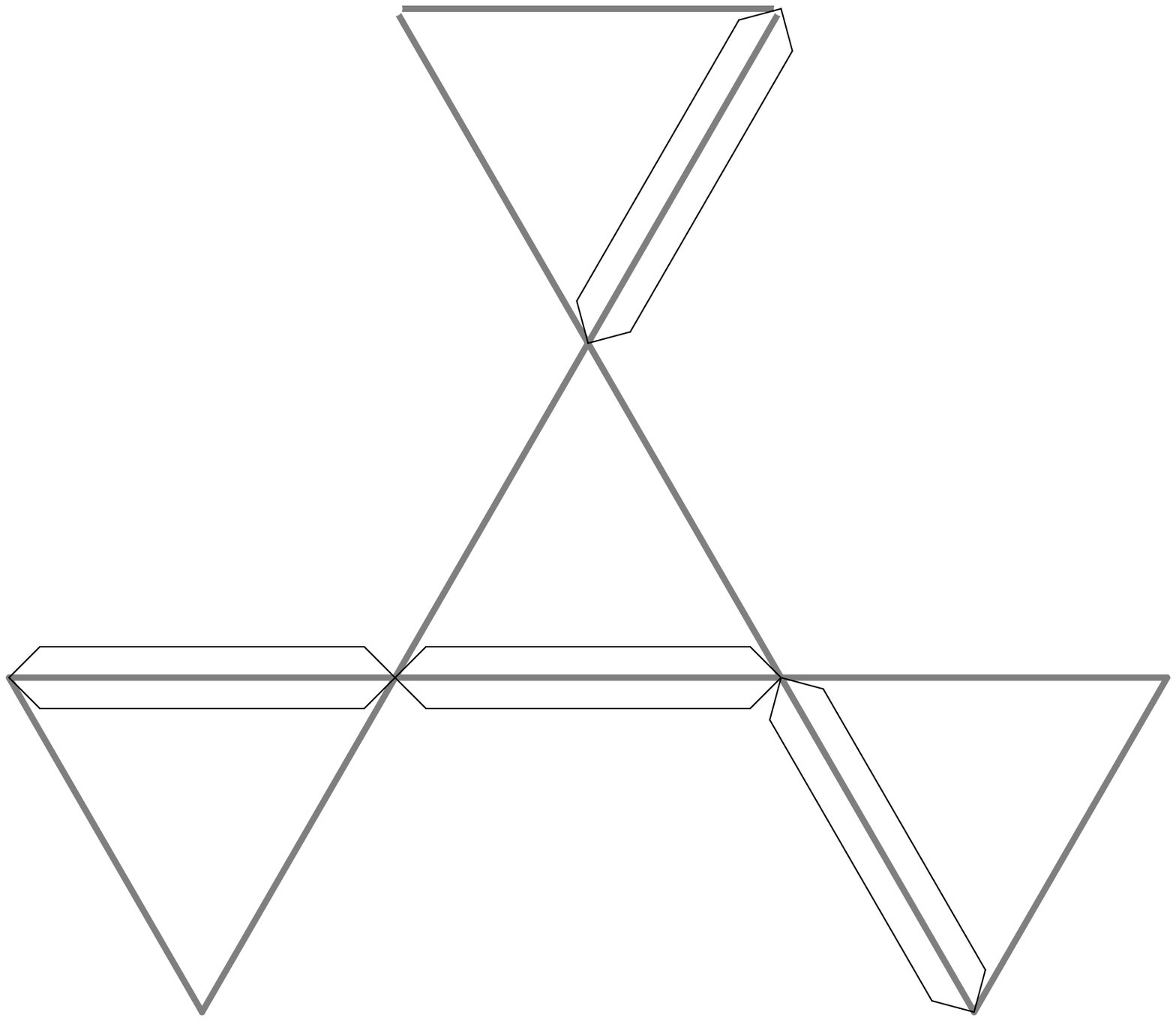}}
\vskip -2mm
\caption{\label{Defect}(a) A defect triangle (shaded) holds no dimers on its bonds. As the neighboring singlet bonds fluctuate independently, all three spins on it are occasionally aligned, making all three bonds frustrated. (b) A dimer is placed on the defect triangle, relaxing its frustration at all times, but simultaneously correlating antiferromagnetically a group of four spins.}
\end{figure}

However, the hard-core dimer coverings of the Kagome lattice are not quite acceptable. It is known that they unavoidably have a fixed number of so called \emph{defect} triangles that hold no dimers on their bonds. \cite{KagSUN, KagHe} This means that there would be macroscopically many triangles with all three bonds occasionally frustrated, and the first criterion would be violated. An example of a defect triangle is shown shaded in the Fig.~\ref{Defect}(a). In order to make sure that two of its bonds are always unfrustrated, two of its spins must be always anticorrelated. One variational mechanism that achieves this is the following. Let us fix the relative antialigned states of two spins on the defect triangle, and denote those two spins by a dimer, like in the Fig.~\ref{Defect}(b). This allows at least the dimer on the top triangle to resonate as a singlet bond, but there are now four spins connected by the dimers, thus antiferromagnetically correlated. Certainly, those four spins could also resonate together while keeping their correlations, but the energy gain would be much smaller than that brought by a singlet bond (this is a higher order process). Generally, two singlet bonds are lost for every defect triangle. The dimers no longer represent only the singlet bonds, but any pair of anticorrelated spins.

\begin{figure}
\subfigure[{}]{\includegraphics[width=1.4in]{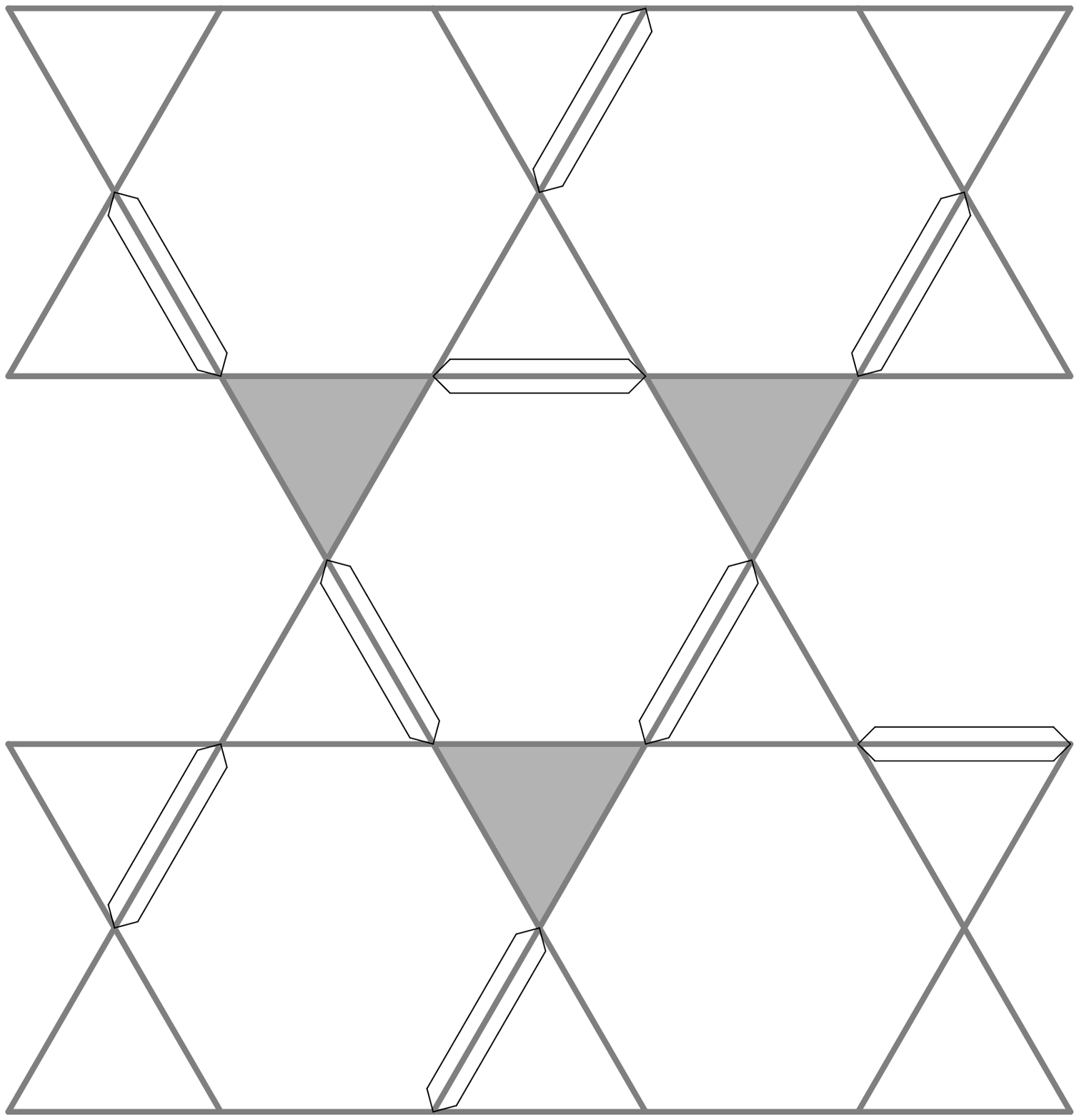}}
\subfigure[{}]{\includegraphics[width=1.4in]{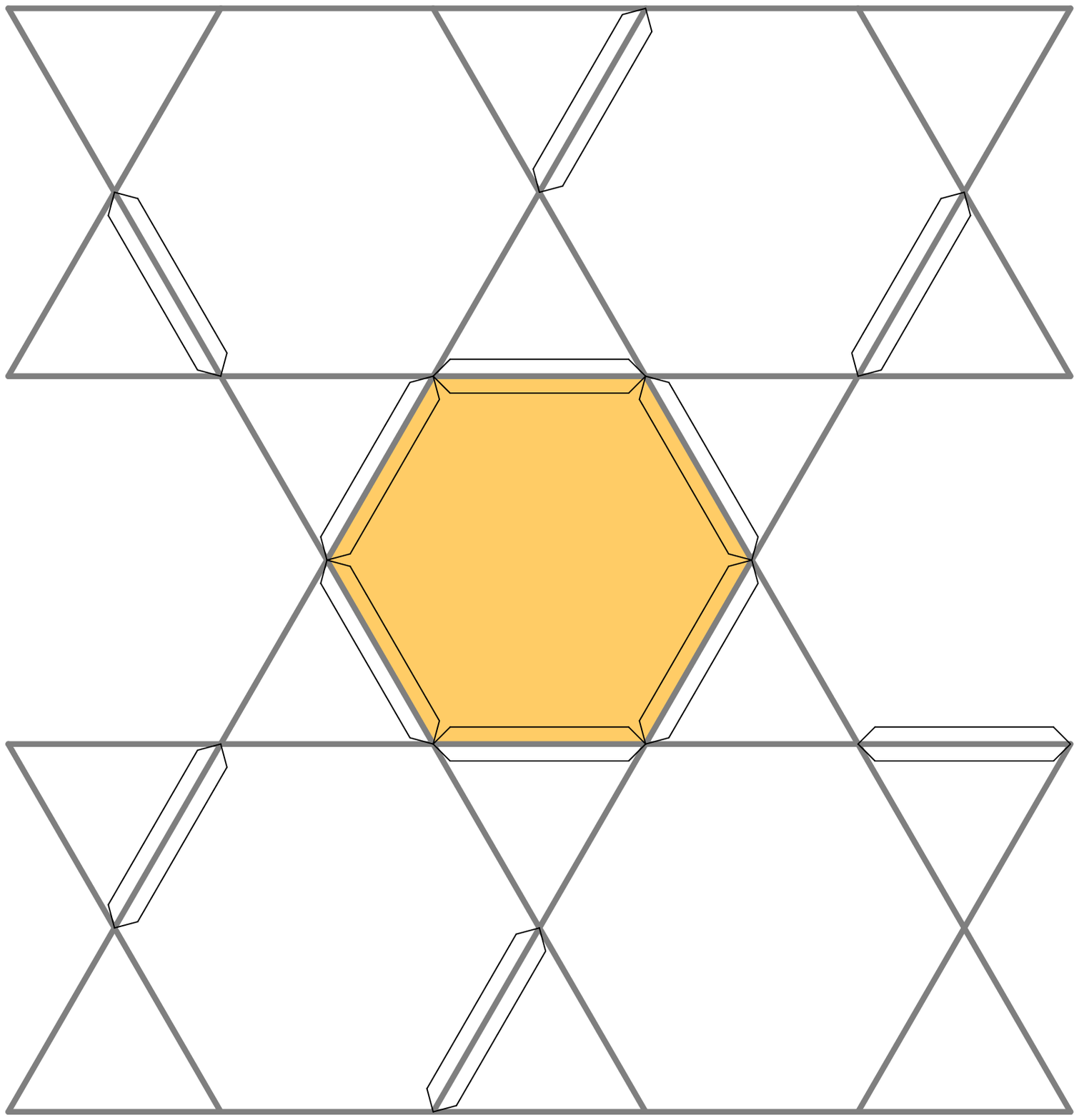}}
\vskip -2mm
\caption{\label{PerfectHex}(a) A perfect hexagon holds three dimers, and has three defect triangles (shaded) around it in a hard-core dimer covering. (b) Covering all bonds of the perfect hexagon by dimers removes the defect triangles, and correlates antiferromagnetically all spins on the hexagon. Only three singlet bonds are lost per three defect triangles.}
\end{figure}

\begin{figure}
\subfigure[{}]{\includegraphics[width=1.6in]{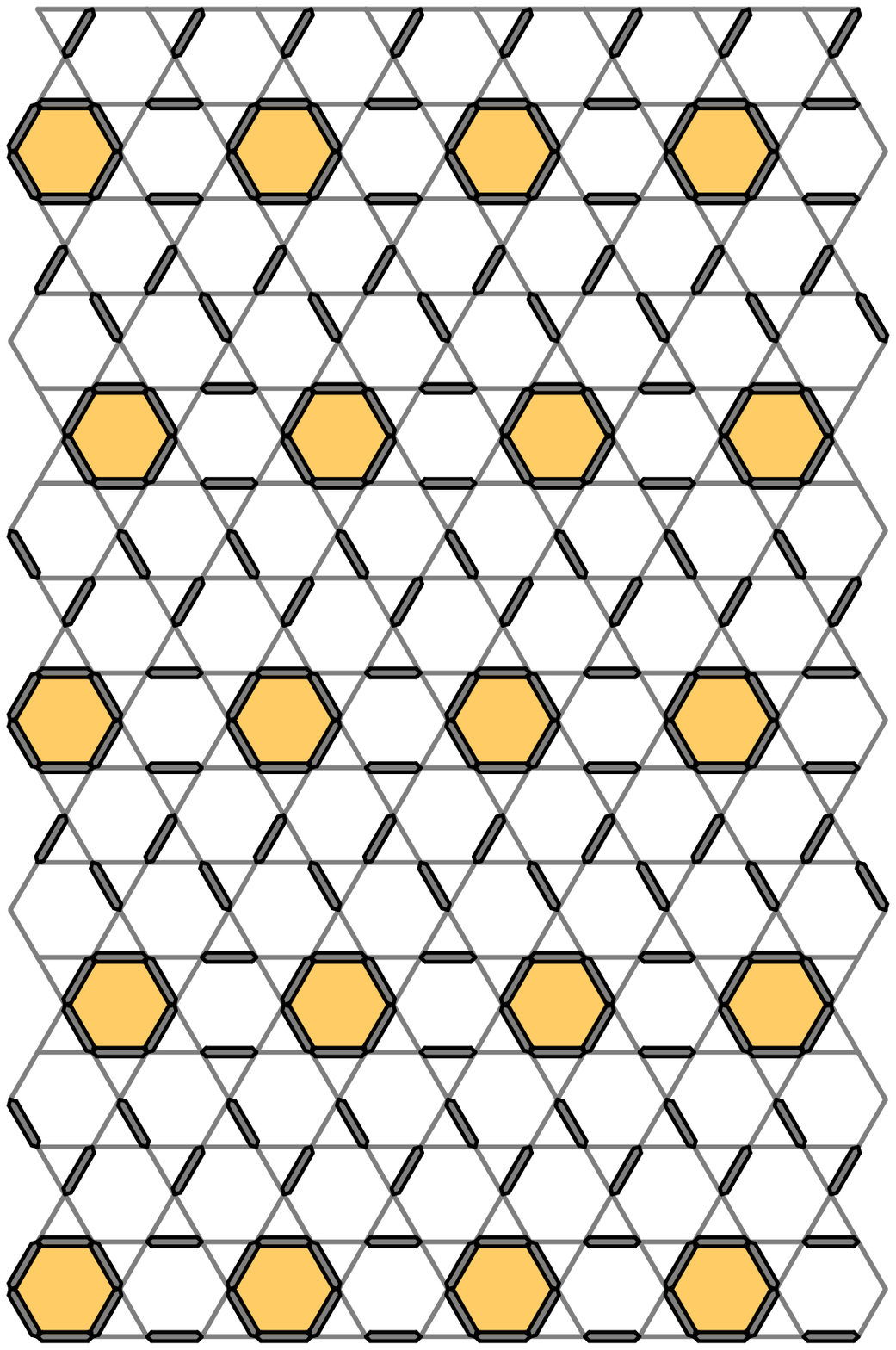}}
\subfigure[{}]{\includegraphics[width=1.6in]{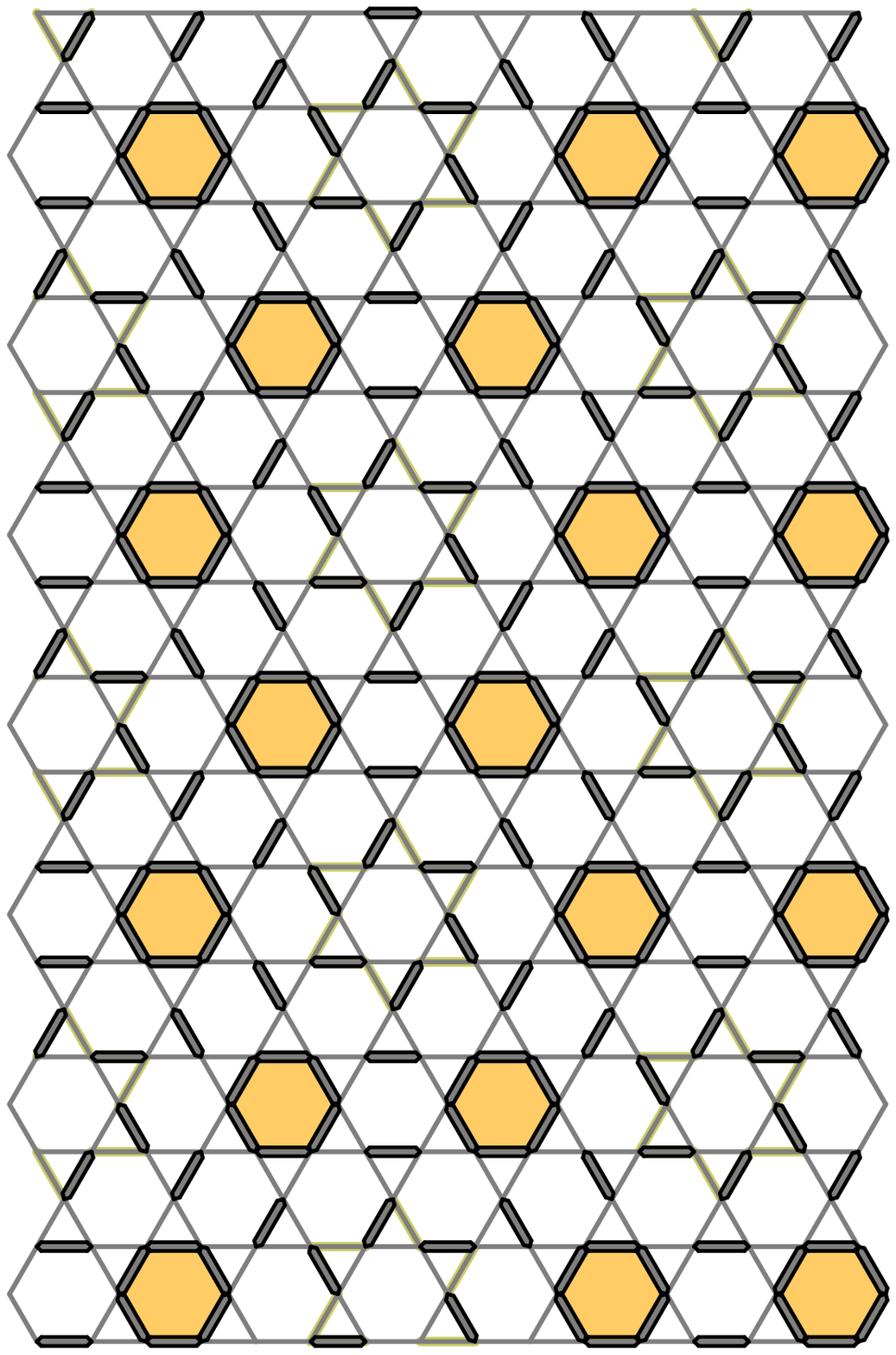}}
\vskip -2mm
\caption{\label{VarStates}Characteristic variational ground-states: (a) stripe pattern, (b) honeycomb pattern. The number of emphasized perfect hexagons is maximized ($1/6$ of all hexagons). In the \emph{honeycomb} pattern there are star-shaped resonant loops of singlet bonds, one sitting inside every honeycomb super-cell. There are two possible singlet bond arrangements on every loop, and more energy can be gained by resonant fluctuations between them.}
\end{figure}

This situation can be improved. It is possible to arrange the defect triangles close to each other in such a way that they share the singlet bonds that are going to be lost. Consider a so called \emph{perfect} hexagon in the Fig.~\ref{PerfectHex}(a). It holds three dimers on its bonds, and therefore has three defect triangles around it in a hard-core dimer covering. By putting three extra dimers on the hexagon, the defect triangles are removed, and the six spins on the hexagon are forced to be antiferromagnetically correlated. Only three singlet bonds are lost per three defect triangles. This is clearly energetically favorable, and consequently the singlet bonds will arrange in a way that maximizes the number of perfect hexagons. Every perfect hexagon can then gain additional energy by correlated fluctuations of its six spins. Furthermore, the groups of singlet bonds may be able to collectively resonate on the closed resonant loops. There are many variational states that maximize the number of perfect hexagons, and we show some characteristic examples in the Fig.~\ref{VarStates}.

\begin{figure}
\subfigure[{}]{\includegraphics[angle=90,height=1.6in]{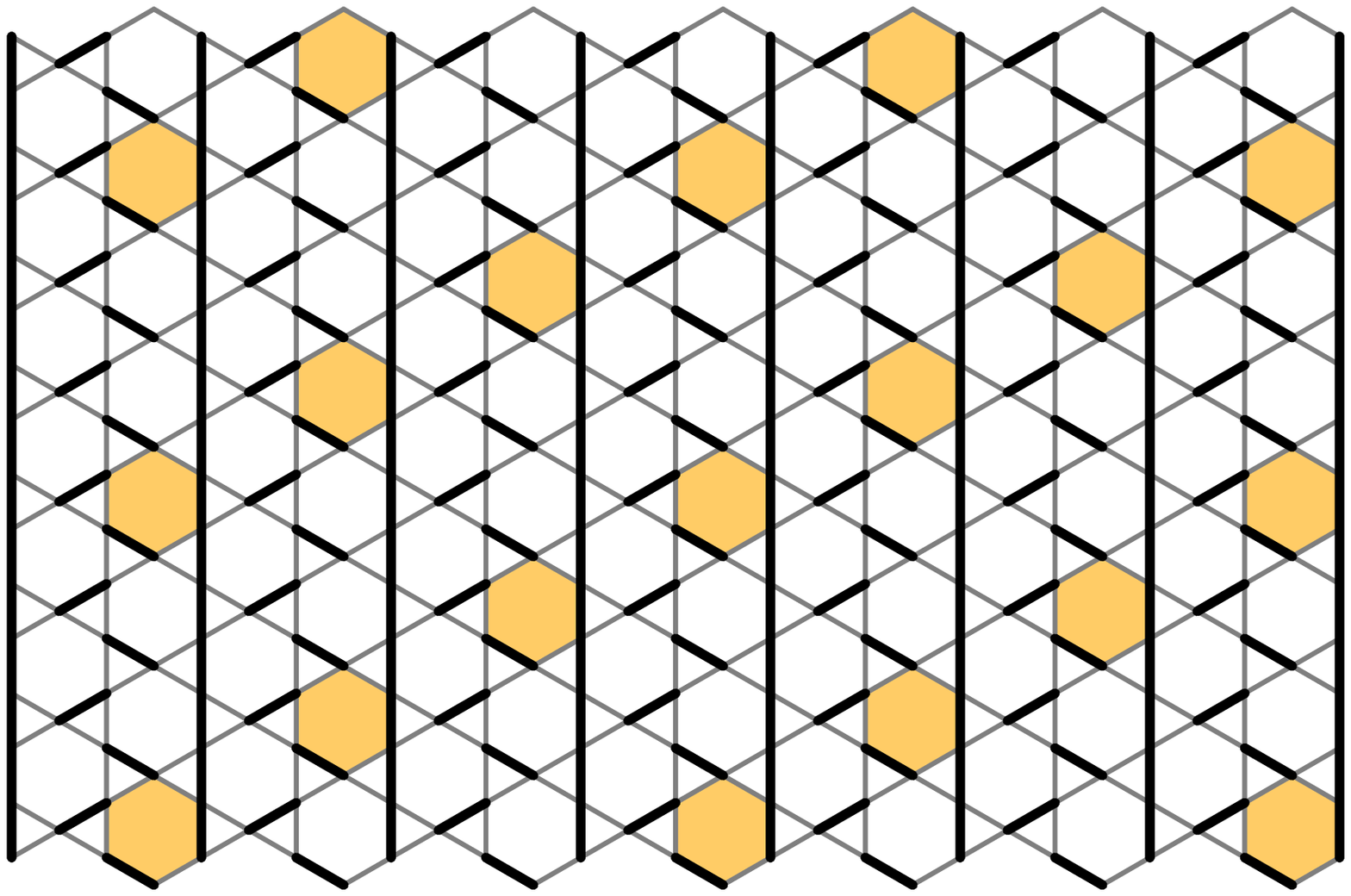}}
\subfigure[{}]{\includegraphics[angle=90,height=1.6in]{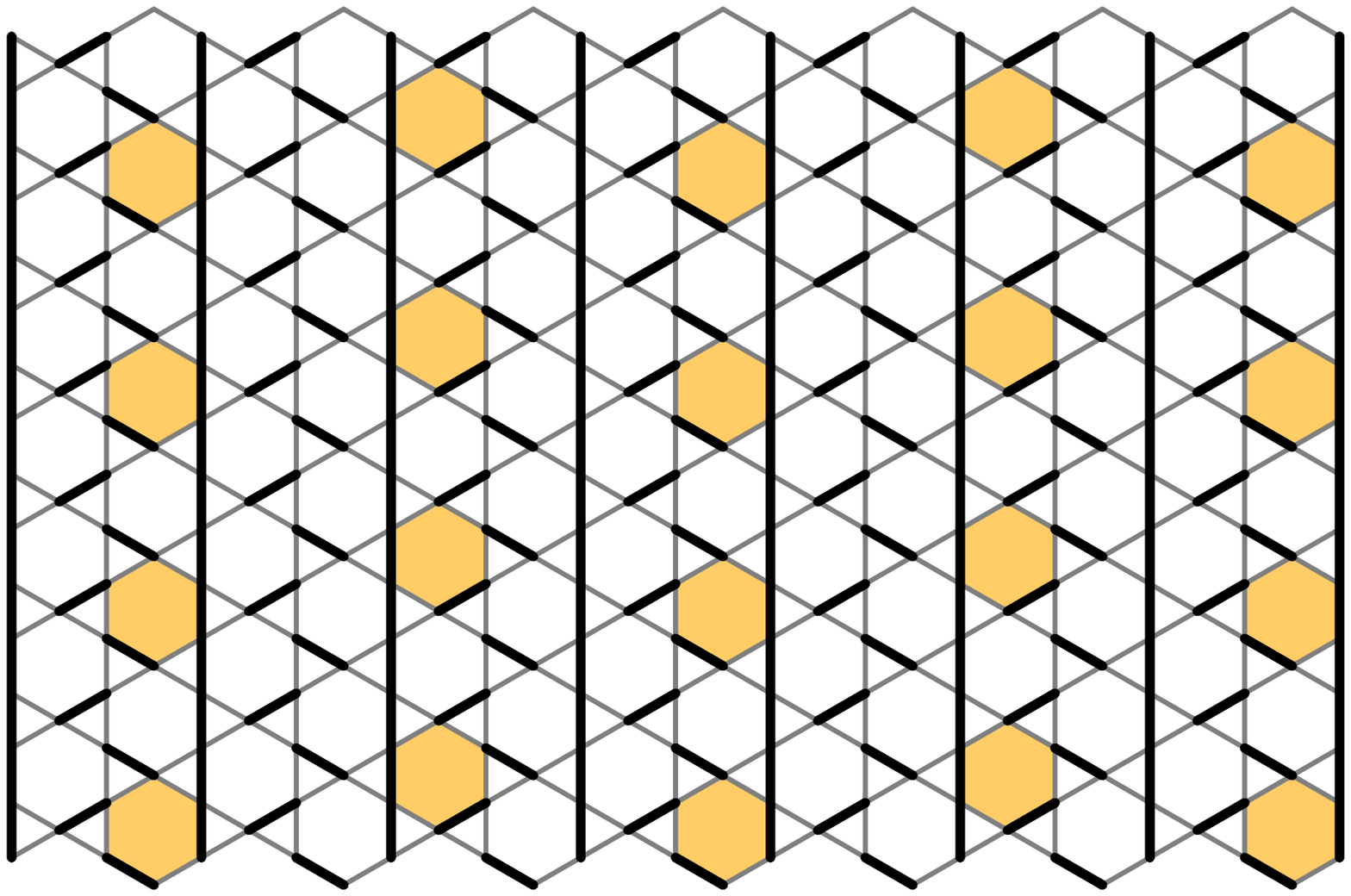}}
\subfigure[{}]{\includegraphics[angle=90,height=1.6in]{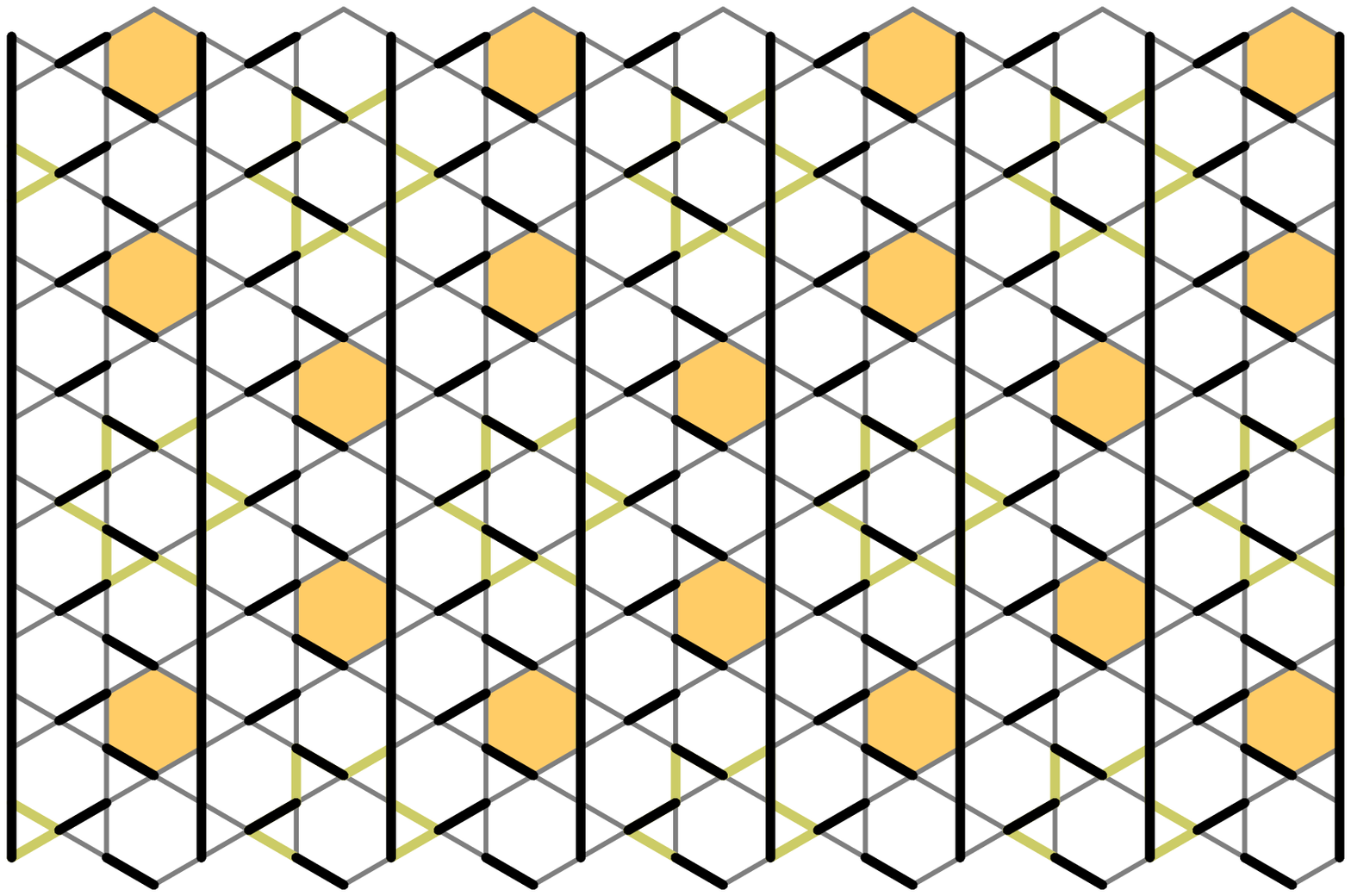}}
\vskip -2mm
\caption{\label{StripeComp}Overlap between the preferred configuration of frustrated bonds and the variational states. Dimers represent the frustrated bonds, and the emphasized hexagons are perfect. (a) and (b) demonstrate two possible ways of overlapping the stripe variational states, and (c) demonstrates the overlap with the honeycomb variational state. The repeating unit-cell of the overlap has 18 sites in (a), and 36 sites in (b) and (c).}
\end{figure}

This variational picture seems to apply very well whenever the physics of the Kagome spin models is describable by the short-range valence bonds. We now apply it to the ordered phase of the XXZ model. The lattice field theory in the preceding sections was able to establish stability of a valence-bond order without full precision in determining how the lattice symmetries should be ultimately broken. It only produced information on which particular microstate is most frequently visited by the system (pattern of frustrated bonds in Fig.~\ref{EAHMstripeSP}); the other nearby microstates are not energetically suppressed, so that they are visited extremely often as well. We can now combine this information with the variational states. The assumption is that the correct ordering pattern must be ``synchronized'' with the state most frequently visited by the system. The Fig.~\ref{StripeComp} compares the entropically preferred stripe-like configuration of frustrated bonds with the only two compatible variational states. The scenarios (a) and (b) involve the stripe pattern of perfect hexagons. Intuitively, the case (a) does not seem likely, because the two stripe orientations of the compared states are different. In the case (b), the stripe orientations of the two compared states are the same, but the overlap period in the direction perpendicular to the stripes is relatively large. Perhaps the best match is accomplished in the scenario (c). The honeycomb pattern of perfect hexagons is also the most symmetric among all variational states, and the set of its symmetries is the closest to that of the Kagome lattice. Unfortunately, at this stage there is no unambiguous way of telling which match is the best, but we can also note that exactly this honeycomb pattern emerged from the analysis of the isotropic Heisenberg model \cite{KagVBC} as a likely ground-state. Based on all these arguments, we can propose the honeycomb variational state as the most likely qualitative description of the ordered ground state of the XXZ model.

Let us now turn to the disordered phase. Our goal is to show that conservation of the total Ising spin has profound consequences for the topological properties of disordered phases. Consider an arbitrary Hamiltonian that is invariant under the global spin-flip. It can be always expressed as a sum of local Hermitian operators that flip an equal number of ``up'' and ``down'' pointing spins. The eigenstates of all such operators either have any particular spin on the lattice fixed, or involved in a group of an even number of coherently fluctuating spins (for example, a group of two neighboring spins is a singlet or a symmetric triplet valence bond, appropriate for the pure XXZ dynamics). These eigenstates are somehow eventually superposed to give the ground state of the Hamiltonian. If all the spins fluctuate in the ground state so that there is no average magnetization on any site, then the ground state is a superposition of only the ``valence-group'' states in which \emph{every} spin belongs to a finite even-sized group of coherently fluctuating spins. This is a generalization of the valence-bond states (whose superpositions would yield the singlet ground states).

It is now possible to define topological sectors of these ``valence-group'' states. Choose an arbitrary pairing of spins within every group: every spin must be paired with one other spin (need not be a neighbor). Visualize the pairings by strings on the lattice that connect the paired spins (overlaps and shapes of the strings do not matter). A transition graph between any two string configurations can be constructed by overlapping them, in analogy to the hard-core dimer coverings. Then, any two ``valence group'' states from the superposition that forms the ground state will have the transition graph composed of finite closed loops (as long as the Hamiltonian has only local dynamics). If now the lattice is placed on a torus, there will be two topologically non-equivalent closed paths that go around the torus and intersect the bonds of the lattice. The topological sector of a string configuration is determined by the parities of the number of strings that each of these paths intersect. Two string configurations will belong to different topological sectors only if any of the paths intersects their transition graph an odd number times. Clearly, this can never happen if the transition graphs always consist of the finite closed loops: the ground state has a definite topology.

Therefore, any disordered state of the XXZ model is automatically a spin liquid, with four degenerate ground states on a tours. A characteristic feature of the Kagome lattice (and other corner-sharing lattices) is a manifestly weak dispersion in the far limit for which the spin liquid obtains. This indicates that the correlations in the spin liquid (away from the critical point) must be strictly short-ranged, virtually vanishing beyond a few lattice constants.


\section{Discussion}

We have explored two kinds of the Kagome lattice quantum Ising antiferromagnets. The first kind was endowed by spin dynamics that did not conserve the total Ising spin, and was represented by the transverse field Ising model (TFIM). The second kind conserved the total spin, and its simplest form was given by the XXZ model. Both TFIM and XXZ models contain only the shortest-range dynamical processes consistent with the required symmetries, acting as small perturbations to the pure Ising model. The considered extensions include further-neighbor and multiple-spin exchange dynamics, and thus they may reflect physics of the TFIM and XXZ models with stronger dynamical energy scale in comparison to the Ising interaction. The quantum phases found in these models are summarized in the Table ~\ref{PhDiag}.

\begin{table}[!b]
\begin{tabular}{|p{3cm}||p{2.5cm}|p{2.5cm}|}
\hline
dominant dynamical                    & simple               & multiple-spin and \\[-1.2mm]
processes:                            & short-ranged         & ring exchange\dots\\
\hline
\hline
do not conserve $\sum\limits_i S^z_i$ & disordered           & disordered        \\
\hline
conserve $\sum\limits_i S^z_i$        & valence-bond crystal & spin liquid       \\
\hline
\end{tabular}
\caption{\label{PhDiag}Quantum phases of the Kagome lattice Ising antiferromagnets with different kinds of spin dynamics.}
\end{table}

The disordered phase of the TFIM and related models was found to have trivial topology. Consistently, the table indirectly suggests that the same phase should be realized for all values of the transverse field. Our approach allowed us to gain some information about the probability amplitudes that various spin configurations have in the ground state superposition. Together with this information, the finding that excitations appear heavy or localized even for weak transverse fields suggested the following variational wavefunctions for many states: the eigenstates of the decoupled spins in transverse field should be projected to the manifold of minimized Ising frustration. It is evident that the corner-sharing structure of the Kagome lattice is responsible for the very weak dispersion or perhaps even localized nature of fluctuations.

Much richer physics is found when the total Ising spin is conserved. The XXZ and related models give rise to at least two non-trivial phases. The calculations indicated that the valence-bond order was most likely to be found for short-ranged and small dynamical perturbations, such as the one in the Heisenberg model with strong easy-axis anisotropy (simple XXZ model). Furthermore, a combination of arguments led to essentially two most probable ordered states, the stripe and honeycomb shaped patterns (Fig.~\ref{VarStates}). While no good arguments to rule one of them were provided, we suspect that the more symmetric honeycomb pattern is realized in most typical situations (no specially favored dynamical processes). This result is of potentially great importance, because the same type of lattice symmetry breaking has been proposed to occur in the ideal isotropic Heisenberg model, \cite{KagVBC} and accounted for the seemingly gapless band of singlet excitations observed in numerics. The physics of the ideal Heisenberg model is still largely mysterious: its ground state could be a spin liquid. Indeed, our calculations indicate that as the complexity of dynamical processes increases, which is similar to what happens when the amount of easy-axis anisotropy is reduced in the XXZ model, a phase transition into the spin liquid must occur. If our calculations could indeed qualitatively describe the anisotropy reduction, a question would arise whether the phase transition happens before or after the full isotropy is reached. In any case, both the valence-bond crystal and spin liquid phases found here are gapped (gap energy scale may be very small), and the same is expected to be true for the isotropic Heisenberg model, regardless of what phase it actually lives in (unless it sits at the critical point).

The Kagome lattice is a representative frustrated magnet suitable for learning more general lessons on the two-dimensional systems. One important question, driven by efforts to discover unconventional Mott insulators, is under what circumstances can disordered and spin liquid phases be found in the frustrated spin models. One mechanism that clearly emerges is adding sufficiently strong further-neighbor and multiple-spin exchange processes. This has been already indicated in various other cases. \cite{TriRing1, TriRing2, SaRe, KagFract} However, the corner-sharing lattices have been in focus due to a belief that even with the shortest-range dynamical processes one can still obtain the spin liquid physics. At least for the transverse field Ising model a disordered ground state is found in the Kagome system, making a sharp contrast to other usually studied systems \cite{KagIsing} (most dimer models, triangular and fully frustrated square lattice Ising models). This disordered phase is conventional, since the system behaves almost as completely decoupled. Apparently, the completely local transverse field dynamics is unable to bring up correlations in a poorly connected lattice. However, as soon as the transverse field is replaced by the next least correlated kind of dynamics (XXZ), a valence-bond ordered phase seems to emerge instead of a spin liquid. Since this happens in one of the most prominent systems for the short-range spin liquid, it is reasonable to speculate that the spin liquid in similar weakly perturbed quantum antiferromagnets quite generally requires further-neighbor, multiple-spin and ring exchange dynamics. Having a less connected lattice makes it easier for a trivial disordered phase to appear as a result of short-range dynamics, but perhaps not so much easier for the spin liquid phase.

Also, the arguments from the end of the section ~\ref{EAHMvar} indicate that the conservation of total Ising spin is a sufficient condition for non-trivial topology of disordered phases that have no net magnetic moment on any site. The SU(2) symmetric models, such as the Heisenberg model, are included as a special case, in agreement with the extension of the Lieb-Schultz-Mattis theorem to higher dimensions. \cite{LSMHastings} Clearly, the spin liquid can exist beyond this condition: as a stable phase, it can resist sufficiently weak spin non-conserving perturbations.

On the technical side, this paper demonstrated an alternative U(1) gauge theory to that of the Ref.~\cite{KagIsing}. Even though the present theory is more complicated, it provides different insight, and avoids some difficulties that otherwise might have been encountered for the XXZ problem. One of its advantages is ability to give information on the character of the ground (and excited) state wavefunctions, and a visual template for the kind of valence-bond orders that are possible in the Kagome XXZ models. We demonstrated a powerful analytical approach that successfully handles macroscopic degeneracy in the frustrated systems and extends the mean-field theories of the unfrustrated systems.

\section{Acknowledgements}

I am very grateful to T.~Senthil for suggesting this problem, numerous discussions and comments on the manuscript. I also wish to thank S.~Sachdev for a useful discussion. This research was supported by the National Science Foundation under the grant DMR-0308945, and the NEC Corporation.

\appendix

\section{Properties of the TFIM Lattice Theory}\label{AlgebraApp}

\begin{figure}
\includegraphics[width=1in]{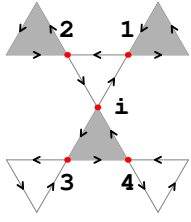}
\vskip -2mm
\caption{\label{ASPFig}Local neighborhood of a site $i$; $\varepsilon_{\bond{ij}}$ is $-1$ on the bonds of the shaded triangles, and $+1$ on the other bonds. Kagome bond orientations are also shown.}
\end{figure}

Here we derive two important properties of the lattice field theory ~(\ref{DHAction}) with regard to its saddle-points. The saddle-point vectors $\bx$ are given by ~(\ref{Vect}), where the bond variables $\xi_{\bond{ij}}$ describe the dimer coverings of the Kagome lattice with one dimer on every triangle, and an arbitrary even number of dimers on every hexagon. The value $\xi_{\bond{ij}}=1$ represents a dimer, while $\xi_{\bond{ij}}=0$ represents a vacancy. First, let us calculate the normalization of the saddle-point vectors $\bx$ (the $\xi_{\bond{ij}}$ variables have no time dependence, and we will drop the summation over time):
\begin{eqnarray}\label{SPVectNorm}
\bx^T\bx & \propto & \sum_i \Biggl\lbrack 
    \Biggl(\sum_{j \in i} \xi_{ij}\Biggr)^2 +
    \Biggl(\sum_{j \in i} \varepsilon_{\bond{ij}} \xi_{ij} \Biggr)^2 
    \Biggr\rbrack \\
  & = & \sum_i \Biggl\lbrack 2\sum_{j \in i}\xi^2_{ij} +
    \sum_{j_1,j_2 \in i}^{j_1 \neq j_2} 
    \Bigl(1 + \varepsilon_{\bond{ij_1}}\varepsilon_{\bond{ij_2}} \Bigr)
    \xi_{ij_1}\xi_{ij_2} \Biggr\rbrack \nonumber \\
  & = & \textrm{const.} + 4\sum_i \Bigl(\xi_{i1}\xi_{i2}+\xi_{i3}\xi_{i4} \Bigr)
    \ . \nonumber
\end{eqnarray}
In the last line we have used the notation from the Fig.~\ref{ASPFig}. Switching to the bond variables, we have:
\begin{eqnarray}\label{SPVectNorm2}
\bx^T\bx & \propto & \textrm{const.} -
    4\sum_i \Bigl( \xi_{\bond{i1}}\xi_{\bond{i2}}+\xi_{\bond{i3}}\xi_{\bond{i4}}
    \Bigr) \\
  & = & \textrm{const.} - 2\sum_{\triangle} 
    \Biggl( \sum_{\bond{ij}}^{\triangle} \xi_{\bond{ij}} \Biggr)^2 +
    2\sum_{\bond{ij}} \xi^2_{\bond{ij}} \nonumber \\
  & = & \textrm{const.} \ . \nonumber
\end{eqnarray}
We have used the facts that the sum of $\xi_{\bond{ij}}$ on every triangle is $1$, since every triangle holds one dimer ($N$ is the number of Kagome sites), and that the total number of dimers on the lattice is fixed (equal to the number of triangles). We see that all saddle-point vectors $\bx$ have the same normalization.

Now let us calculate how the coupling matrix $\bC$ from the action acts on the saddle-point vectors $\bx$. The quadratic parts of the expression ~(\ref{DHMatrAct}) reveal how the matrix $\bC$ acts on the height field vectors $\bh$ whose components are $\chi_i$ and $\lambda_i$. Substituting there $\sum_{j \in i}\xi_{ij}$ for every $\chi_i$, and $\sum_{j \in i}\varepsilon_{\bond{ij}}\xi_{ij}$ for every $\lambda_i$ reveals the action of $\bC$ on the saddle-point vectors $\bx$:
\begin{eqnarray}\label{ConXi}
(\bC\bx)_{\chi_i} & = & \sum_{j \in i} \omega_{(ij)} \ ; \\
(\bC\bx)_{\lambda_i} & = & \sum_{j \in i} \varepsilon_{\bond{ij}}\omega_{(ij)}
  \ , \nonumber
\end{eqnarray}
where $\omega_{(ij)}$ is neither a vector, nor a bond scalar:
\begin{equation}\label{omega}
\omega_{(ij)} = 4\xi_{ij} - \Biggl( \sum_{k \in j}\xi_{jk} +
  \varepsilon_{\bond{ij}} \sum_{k \in j} \varepsilon_{\bond{jk}} \xi_{jk}
  \Biggr) \ .
\end{equation}
Since we need to find the sums of $\omega_{(ij)}$ around a particular site, let us take a closer look at the neighborhood of a site $i$, and refer to the Fig.~\ref{ASPFig}:
\begin{eqnarray}\label{omega2}
\omega_{(ij)} 
  & = & 4\xi_{ij} - \sum_{k \in j}\xi_{jk}
    \Biggl(1 + \varepsilon_{\bond{ij}} \sum_{k \in j} \varepsilon_{\bond{jk}} \Biggr)
\\ 
\phantom{\omega_{(ij)}} & = & 4\xi_{ij} 
    - 2\delta_{j,1}(\xi_{ji}+\xi_{12}) - 2\delta_{j,2}(\xi_{ji}+\xi_{21}) \nonumber \\
  & & - 2\delta_{j,3}(\xi_{ji}+\xi_{34}) - 2\delta_{j,4}(\xi_{ji}+\xi_{43}) \nonumber \\
  & = & 6\xi_{ij} - 2\Bigl( \delta_{j,1}\xi_{12} + \delta_{j,2}\xi_{21}
                      +\delta_{j,3}\xi_{34} + \delta_{j,4}\xi_{43} \Bigr)
     \ . \nonumber
\end{eqnarray}
Then:
\begin{eqnarray}\label{ConXi2}
\sum_{j \in i}\omega_{(ij)} 
  & = & 6\sum_{j \in i}\xi_{ij} 
    -2 \Bigl( \xi_{12}+\xi_{21}+\xi_{34}+\xi_{43} \Bigr) \nonumber \\
  & = & 6\sum_{j \in i}\xi_{ij} \ ; \\
\sum_{j \in i}\varepsilon_{\bond{ij}}\omega_{(ij)}
  & = & 6\sum_{j \in i}\varepsilon_{\bond{ij}}\xi_{ij} 
        -2 \Bigl( \varepsilon_{\bond{i1}}\xi_{12}+\varepsilon_{\bond{i2}}\xi_{21} 
        \nonumber \\
  & & +\varepsilon_{\bond{i3}}\xi_{34}+\varepsilon_{\bond{i4}}\xi_{43} \Bigr)
      = 6\sum_{j \in i}\varepsilon_{\bond{ij}}\xi_{ij} \ . \nonumber
\end{eqnarray}
We see that the action of $\bC$ on a vector whose components are $\sum_{j \in i}\xi_{ij}$ and $\sum_{j \in i}\varepsilon_{\bond{ij}}\xi_{ij}$ simply reproduces those components, with an additional factor of 6. Therefore, all saddle-point vectors $\bx$ are degenerate eigenvectors of the coupling matrix $\bC$, with an eigenvalue 6.


\end{document}